\documentclass[preprint2,longabstract]{aastex}

\usepackage{epstopdf}

\slugcomment{\it To be submitted to the Astrophysical Journal}
\shortauthors{Fadda, D. , Yan, L. et al.}
\shorttitle{Ultra-deep Spitzer Mid-Infrared Spectroscopy of LIRGs and ULIRGs at $z \sim 1-2$ }

\def\deg{\ifmmode {^{\circ}}\else {$^\circ$}\fi}
\def\kms{\ifmmode {\rm\,km\,s^{-1}}\else
    ${\rm\,km\,s^{-1}}$\fi}
\def\ergcm2s{\ifmmode {\rm\,ergs\,cm^{-2}\,s^{-1}}\else
    ${\rm\,ergs\,cm^{-2}\,s^{-1}}$\fi}
\def\ergAcm2s{\ifmmode {\rm\,ergs\,cm^{-2}\,s^{-1}\,\AA^{-1}}\else
    ${\rm\,ergs\,cm^{-2}\,s^{-1}\,\AA^{-1}}$\fi}
\def\ergs{\ifmmode {\rm\,ergs\,s^{-1}}\else
    ${\rm\,ergs\,s^{-1}}$\fi}
\def\kmsMpc{\ifmmode {\rm\,km\,s^{-1}\,Mpc^{-1}}\else
    ${\rm\,km\,s^{-1}\,Mpc^{-1}}$\fi}

\def\nev{[\ion{Ne}{5}]$\lambda14.3\mu$m}
\def\oiv{[\ion{O}{4}$]\lambda 25.9\mu$m}

\def\spose#1{\hbox to 0pt{#1\hss}}
\def\simlt{\mathrel{\spose{\lower 3pt\hbox{$\mathchar"218$}}
     \raise 2.0pt\hbox{$\mathchar"13C$}}}
\def\simgt{\mathrel{\spose{\lower 3pt\hbox{$\mathchar"218$}}
     \raise 2.0pt\hbox{$\mathchar"13E$}}}

\def\gs{\mathrel{\raise0.35ex\hbox{$\scriptstyle >$}\kern-0.6em
\lower0.40ex\hbox{{$\scriptstyle \sim$}}}}
\def\ls{\mathrel{\raise0.35ex\hbox{$\scriptstyle <$}\kern-0.6em
\lower0.40ex\hbox{{$\scriptstyle \sim$}}}}

\newcommand{\um}{\,$\mu$m}

\newcommand{\spitz}{{\sl Spitzer }}

\begin{document}

\title{Ultra-Deep Mid-Infrared Spectroscopy of Luminous Infrared Galaxies at $z \sim$\,1 and $z$\,$\sim$\,2}

\author{Dario Fadda$^1$, Lin Yan$^1$, Guilaine Lagache$^3$, Anna Sajina$^2$, Dieter Lutz$^4$, Stijn Wuyts$^5$,David T. Frayer$^1$, Delphine Marcillac$^3$, Emeric Le Floc'h$^6$, Karina Caputi$^7$, Henrik W. W. Spoon$^8$,Sylvain Veilleux$^9$, Andrew Blain$^{10}$, George Helou$^1$\\
\affil{$^1$ IPAC, California Institute of Technology,  Pasadena, CA 91125}
\affil{$^2$ Haverford College, Haverford, PA 19041}
\affil{$^3$ Institut d'Astrophysique Spatiale, F-91405 Orsay, France; Univ. Paris-Sud 11 and CNRS (UMR 8617)}
\affil{$^4$ Max-Planck-Institut f\"{u}r Extraterrestrische Physik,  85741 Garching, Germany}
\affil{$^5$ W. M. Keck Postdoctoral Fellow, Harvard-Smithsonian Center for Astrophysics, 60 Garden Street, Cambridge, MA 02138}
\affil{$^6$ CEA Saclay, Service d'Astrophysique, 91191 Gif-sur-Yvette, France}
\affil{$^7$ SUPA Institute for Astronomy, University of Edinburgh, Royal Observatory, Blackford Hill, Edinburgh EH9 3HJ, UK}
\affil{$^8$ Dept. of Astronomy, Cornell University, 219 Space Sciences Building, Ithaca, NY 14853}
\affil{$^9$ Dept. of Astronomy, University of Maryland, College Park, MD 20742}
\affil{$^{10}$ Astronomy Department, California Institute of Technology, Pasadena, CA 91125}
\email{fadda@ipac.caltech.edu,lyan@ipac.caltech.edu}}

\begin{abstract}
We present ultra-deep mid-infrared spectra of 48 infrared-luminous
galaxies in the GOODS-South field obtained with the InfraRed
Spectrograph (IRS) on the \spitz Space Telescope. These galaxies are
selected among faint infrared sources (0.14\,--\,0.5\,mJy at 24\um) in
two redshift bins (0.76\,--\,1.05 and 1.75\,--\,2.4) to sample the
major contributors to the cosmic infrared background at the most
active epochs. We estimate redshifts for 92\% of the sample using
polycyclic aromatic (PAH) and Si absorption features obtaining, in
particular, 8 new redshifts difficult to measure from ground-based
observations. Only few of these galaxies (5\% at $z$\,$\sim$\,1 and
12\% at $z$\,$\sim$\,2) have their total infrared luminosity dominated
by emission from active galactic nuclei (AGN). The averaged mid-IR
spectrum of the $z$\,$\sim$\,1 luminous infrared galaxies (LIRGs) is a
very good match to the averaged spectrum of local starbursts. The
averaged spectrum of the $z$\,$\sim$\,2 ultra-luminous infrared
galaxies (ULIRGs), because of a deeper Si absorption, is better fitted
by the averaged spectrum of HII-like local ULIRGs. Combining this
sample with other published data, we find that 6.2\um\ PAH equivalent
widths reach a plateau of $\sim$\,1\um\ for $L_{24\mu
  m}$\,$\simlt$\,$10^{11}L_\odot$.  At higher luminosities,
$EW_{6.2\mu m}$ anti-correlates with $L_{24\mu m}$.  Intriguingly,
high-$z$ ULIRGs and sub-millimeter galaxies (SMG) lie above the local
$EW_{6.2\mu m}$-$L_{24\mu m}$ relationship suggesting that, at a given
luminosity, high-$z$ ULIRGs have AGN contributions to their dust
emission lower than those of local counterparts. A quantitative
analysis of their morphology shows that most of the luminous IR
galaxies have morphologies similar to those of IR-quiet galaxies at
the same redshift. All $z$\,$\sim$\,2 ULIRGs of our sample are
IR-excess BzK galaxies and most of them have $L_{FIR}/L_{1600\AA}$
ratios higher than those of starburst galaxies at a given UV
slope. The ``IR excess'' \citep{daddi07} is mostly due to strong
7.7\um\ PAH emission and under-estimation of UV dust extinction. On
the basis of the AGN-powered $L_{6\mu m}$ continuum measured directly
from the mid-IR spectra, we estimate an average intrinsic X-ray AGN
luminosity of $L_{2-10\,{\rm
    keV}}$\,=\,(0.1$\pm$\,0.6)$\times$\,$10^{43}$\,erg/s, a value
substantially lower than the prediction by \citet{daddi07}.
\end{abstract}

\keywords{infrared: galaxies  --
          galaxies: star formation --
          galaxies: high-redshifts --
          galaxies: evolution}

\section{Introduction}

The Cosmic infrared background (CIB) detected by {\it COBE}
\citep{puget96, fixsen98} peaks around 200\um\ with energy comparable
to the optical/UV background, implying that $\sim$\,50\%\ of the total
optical/UV emission is absorbed by dust and re-radiated at mid to
far-infrared. Using stacking methods, \citet{dole06} quantified that
the CIB emission at 70 and 160\um\ is primarily produced by
24\um\ sources with $S_{24\mu m}$\,$\sim$\,0.1\,--\,0.5\,mJy. At
longer wavelengths, a large part of the CIB is due to faint
24\um\ sources ($S_{24\mu m}$\,$\sim$\,0.06\,--\,0.3\,mJy) as shown by
cross-correlations with BLAST data at 350\um\ \citep{marsden09} and
SCUBA data at 450 and 850\um\ \citep{serjeant08}.  Going from far-IR
to sub-millimetric, the CIB becomes progressively dominated by more
luminous and higher redshift objects \citep{lagache05}. Indeed,
optical spectroscopy have shown that more than 70\% of these sub-mJy
24\um\ sources which are major contributors to the CIB lie beyond $z$
of 0.6, with a primary peak at $z\sim 1$ and a secondary one at $z\sim
2$, consistently with the modeling of deep 24\um\ number counts
\citep{karina07, emeric05, lagache04, desai07}.  These sub-mJy
24\um\ sources correspond to luminous infrared galaxies (LIRGs,
$L_{8-1000\mu m}$\,$\sim$\,$10^{11}$\,--\,$10^{12}L_\odot$) at
$z$\,$\sim$\,1 and to ultra-luminous infrared galaxies (ULIRGs,
$L_{8-1000\mu m}$\,$\sim$\,$10^{12}$\,--\,$10^{13}L_\odot$) at
$z$\,$\sim$\,2. The evolution of the IR population at
$z$\,$\sim$\,1\,--\,2 is clearly illustrated by the infrared
luminosity functions at these early epochs
\citep{emeric05,karina07,reddy08,magnelli09}. In essence, the knee of
the luminosity function, which marks the dominant contributors to the energy density,
moves to higher luminosity with redshift: from $L_{IR}$\,$\sim$\,
(3\,--\,5)$10^{11}L_\odot$ at $z$\,$\sim$\,1 to
(3\,--\,5)$10^{12}L_\odot$ at $z$\,$\sim$\, 2.  As shown by published
studies (see, e.g. \citet{emeric05}), $z$\,$\sim$\,1 LIRGs are
responsible for $70$\,$\pm$\,15\% of the total (UV+IR) luminosity
density, and, at $z$\,$\simgt$\,1, the contribution of ULIRGs becomes
increasingly dominant.

While deep 24\um\ surveys have revealed LIRGs and ULIRGs as primary
contributors to the global energy production at early epochs, the
fraction of their bolometric luminosity contaminated by AGN emission
is not well determined. Standard multi-wavelength criteria can be used
(using X-ray emission, optical spectroscopy, IRAC photometry) but none
of them is able to determine the relative AGN and starburst
contributions to the bolometric luminosity of IR-luminous galaxies
(that is mostly equal to the 3\,--\,1000\um\ luminosity). As of today,
a very efficient way to probe dust obscured AGN emission, is via deep
mid-IR spectroscopy. The difference between the mid-IR spectra of a
starburst and an AGN is striking. Starbursts are often characterized
by strong, low-excitation, fine-structure lines, prominent PAH
features and a weak 10\um\ continuum whereas AGN display weak or no
PAH features, plus a strong mid-infrared continuum (see, e.g.,
\citet{voit92}).  Mid-IR spectroscopic surveys have thus been
conducted on a number of low to high redshift samples of IR-luminous
galaxies with the Spitzer/IRS \citep{armus07, yan07, sajina08,
  farrah08, valiante07, spoon07, weedman06, eric09, veilleux09}. However,
these samples are limited to the bright-end of the IR luminosity
function. Previous deep mid-IR spectroscopic observations have been
attempted at the faint 24\um\ flux level of 0.1\,--\,0.5\,mJy, but the
number of observed sources is very small \citep{teplitz07,
  bertincourt09}.  To directly determine the AGN fraction at
the lowest IR luminosity possible, we carried out an observational
program to obtain low resolution, mid-IR spectra of 48 galaxies --- 24
at $z$\,$\sim$\,1 and 24 at $z$\,$\sim$\,2 --- with 24\um\ flux
densities $\sim$\,0.1\,--\,0.5\,mJy.  The main scientific goal of this
study is to provide the first observational data to characterize the
spectral properties of LIRGs and ULIRGs that are responsible for the
bulk on the CIB and output energy at high redshifts
($z$\,$\sim$\,0.7\,--\,2.5). Combining our deep mid-IR spectra with
the available optical and near-IR data in CDFS, we compare global
properties such as colors, morphologies, and stellar masses of our
sample galaxies with those of optically selected galaxies to provide
observational constraints on the mid-IR properties of optically
selected IR faint galaxies at $z$\,$\sim$\,1\,--\,2.


Throughout the paper, we adopt  $\Omega_M$\,=\,0.27,
$\Omega_\Lambda$\,=\,0.73\ and $H_0$\,=\,71\kmsMpc\ cosmology.
\begin{figure*}[!t]
\centering{\includegraphics[width=1.8\columnwidth]{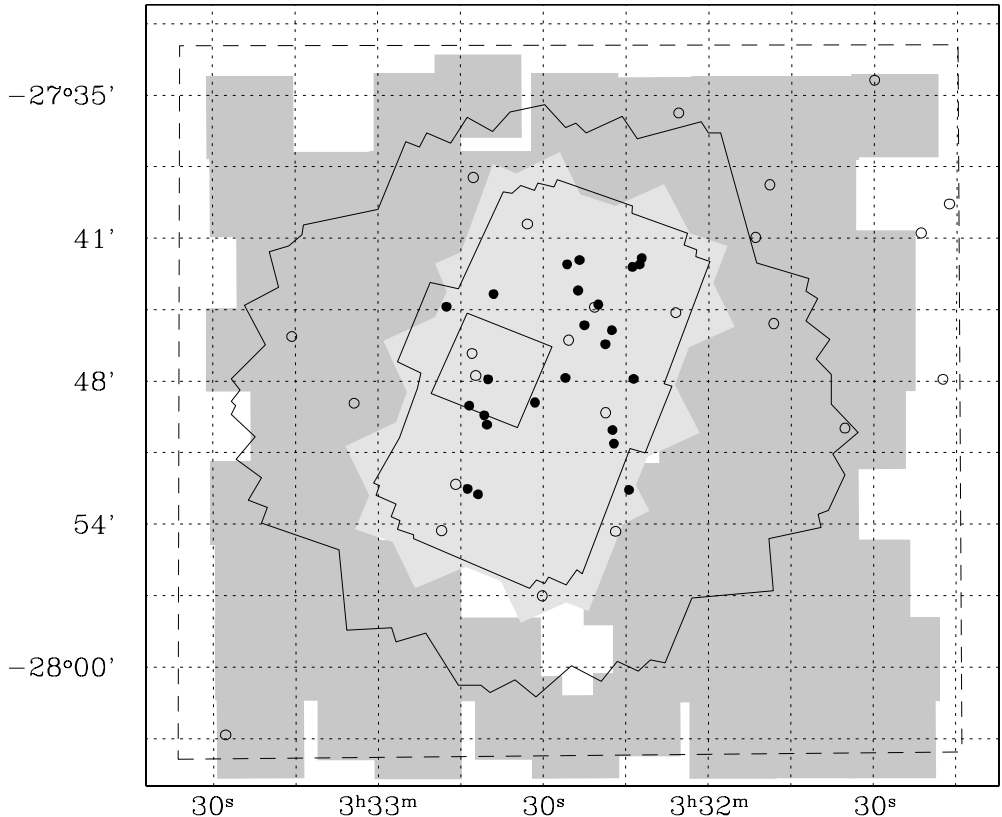}}
\caption{\footnotesize IRS targets overlapped on ACS images from GOODS
  (light gray shaded) and GEMS (gray shaded). Empty and full circles
  mark the $z\sim 1$ and $z\sim 2$ samples, respectively. The two
  internal contours mark the border of the 16$\mu$m imaging (deep and
  ultra-deep observations).  The external solid contour shows the
  coverage of deep Chandra imaging \citep{luo08}, while the external
  dashed contour marks the field covered by COMBO17 \citep{wolf04}
  .\label{fig:map}}
\end{figure*}

\section{Data Sample}

\subsection{Sample Selection}

LIRGs and ULIRGs are responsible for most of the integrated infrared
background. To study these two populations of galaxies, we selected
two samples of 24$\mu$m sources fainter than 0.5~mJy in two redshift
bins (0.76 -- 1.05 and 1.75 -- 2.2). In these two redshift bins, the
24\um\ fluxes translate to infrared luminosities roughly in the ranges
for LIRGs at z$\sim$1 and ULIRGs at z$\sim$2. This sample is therefore
crudely luminosity selected, and we do not apply any other
selections. The flux lower limit ($\sim 0.2$~mJy) is dictated by the
sensitivity of the \spitz InfraRed Spectrograph (IRS) \citep{houck04},
as demonstrated by the IRS instrument support team \citep{teplitz07}.
These targets were selected mostly using photometric redshifts
computed by \citet{karina06} since only 50\% and 10\% of them had
spectroscopic-$z$ in the $z$\,$\sim$\,1 and 2 bins, respectively.  We
note that all sources but two are actually in the chosen redshift
bins, as confirmed by their mid-IR spectroscopic redshifts (see
\S~\ref{sec:mirz}).  Our 48 galaxies were selected in the Chandra Deep
Field South (CDFS) for which a large sample of public ancillary data
is available.  The selected ULIRGs are {\it all} the 24\um\ sources in
the Chandra Deep Field South GOODS area with $S(24\mu
m)$\,$\sim$\,0.14\,--\,0.55\,mJy at $z$\,=\,1.75\,--\,2.4, excluding
three sources too close to a neighbor source to obtain unblended
spectra. The LIRGs have been selected among 24$\mu$m sources with
$S(24\mu m)$\,$\sim$\,0.2\,--\,0.5\,mJy at $z$\,=\,0.76\,--\,1.05 in
order to avoid a large-scale structure localized at z~0.734
\citep{szokoly04}.  To have a more representative sample of LIRGs,
spanning different luminosities within the LIRG range, we included in
our selection sources from the "extended CDFS" area.  The spatial
distribution of the targets relative to the available ACS images is
shown in Figure~\ref{fig:map}.  Table~\ref{obslog} and
Table~\ref{targets_data} list the 48 targets with J2000 coordinates,
on-target integration times for various slit modules, 8, 16, 24 and
70\um\ flux densities and redshifts.  \\
\begin{figure*}[!t]
\centering{\includegraphics[width=1.7\columnwidth]{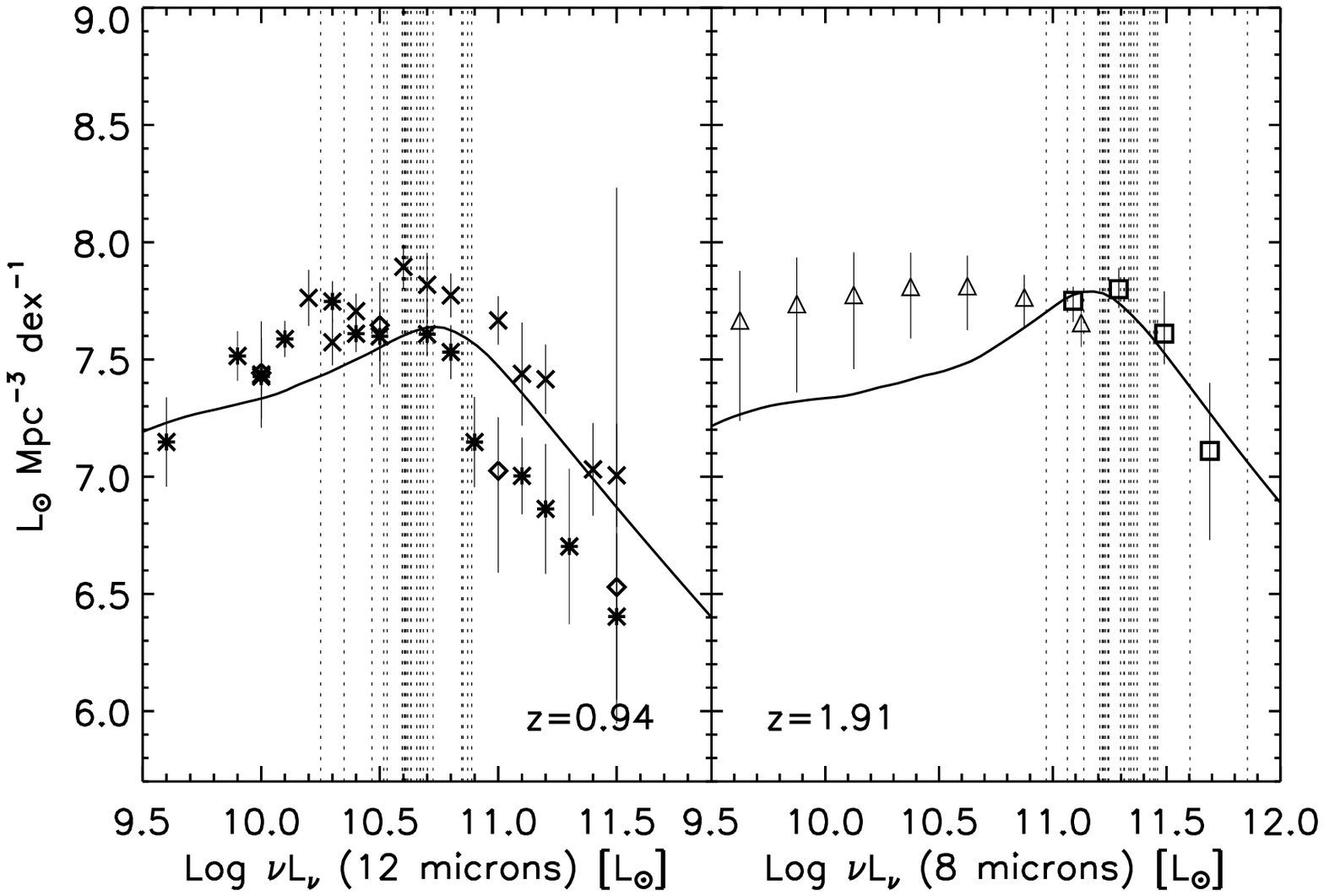}}
\caption{\footnotesize  Co-moving 12 and 8\um\ energy
  output per dex of 12 and 8\um\ luminosity (L dN/dLogL) at
  $z$\,=\,0.94 and $z$\,=\,1.91, corresponding to the mean redshifts
  of our two bins.  Continuous lines are from the model of
  \citet{lagache04}.  Diamonds, squares, and triangles are from
  \citet{emeric05,karina07, reddy08}, respectively.  At
  $z$\,$\sim$\,1, stars and crosses are derived using the $0.7<z<1$
  and $1<z<1.3$ 15\um\ luminosity functions from
  \citet{magnelli09}. Dotted lines represent the rest-frame 12 and
  8\um\ luminosities of our 48 targets computed from their
  24\um\ fluxes. No color correction has been applied.
 \label{fig:energy}}
\end{figure*}

\begin{figure*}[!t]
\centering{\includegraphics[width=1.5\columnwidth]{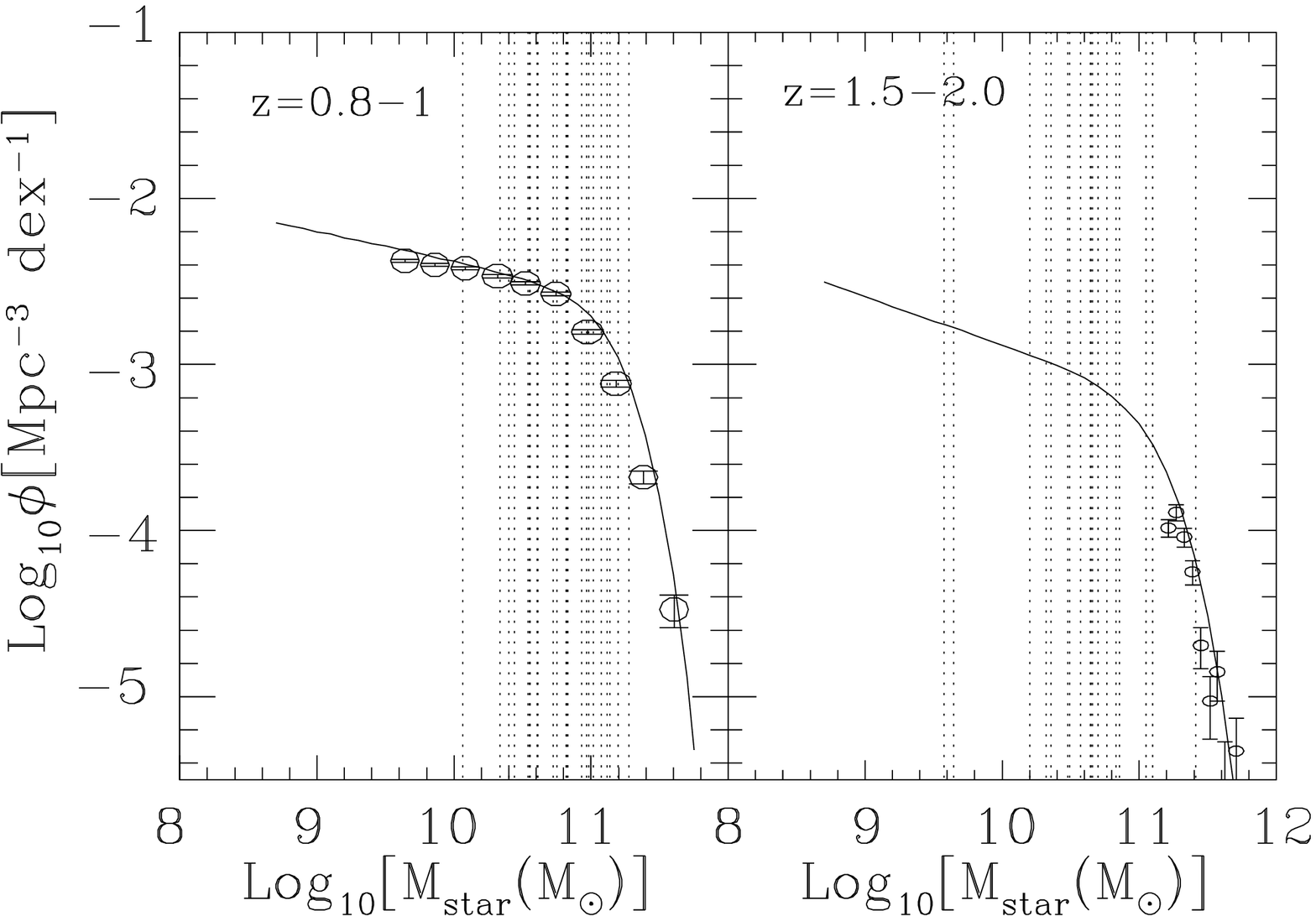}}
\caption{\footnotesize 
Stellar mass functions at $z$\,$\sim$\,1\,--\,2 published by \citet{olivier09} using the COSMOS data.  The dashed lines mark the estimated stellar masses of our sources estimated as in \citet{wuyts09}.  \label{fig:mass}}
\end{figure*}

One important question, which is also the basic driver of this
program, is how representative our sample is at the two redshift
slices.  Figure~\ref{fig:energy} quantitatively illustrates
that our targets represent populations of infrared luminous galaxies
which are energetically important in comparison with other sources.
We plot the contribution of 24\um\ sources to the co-moving
monochromatic 12\um\ and 8\um\ luminosity density per logarithmic
luminosity interval as a function of luminosity at $z$\,$\sim$\,1 and
$2$ respectively. Each of our targets is marked on the luminosity
scale with a dashed line.  The integrated co-moving luminosity density
is computed using both model (line) \citep{lagache04} and previously
published, measured infrared luminosity functions
\citep{karina07,emeric05,magnelli09,reddy08}.  It is interesting to
note that at $z\sim1$, the combined luminosity functions independently
derived by three different groups suggest that the infrared luminosity
density turns over around $L_{12\mu m}$\,$\sim$\,$10^{10.7}L_\odot$
\citep{emeric05,karina07,magnelli09}.  Although the knowledge of the
IR luminosity function is still poor at $z\sim2$, particularly at low
luminosities, the critical luminosity range $L_{8\mu
  m}$\,$\sim$\,$10^{11-11.5}L_\odot$ is fairly clearly highlighted in
in the right panel of Figure~\ref{fig:energy}.  We conclude that
our targets indeed probe the infrared populations which dominate the
energy productions at these two cosmic epochs.

Except having higher dust obscuration, our sample galaxies do not have
extremely deviant properties in the rest-frame UV/optical in
comparison with galaxies selected at observed optical/near-IR
band. For example, \S~\ref{sec:redcolor} discussed in detail that the
observed optical/near-IR colors of our sources are very similar to
those of extremely red galaxy populations selected by large area
K-band surveys.  Figure~\ref{fig:mass} shows the most recently
published stellar mass functions for redshift slices of 0.8\,--\,1.0
(black lines and circles) and 1.5\,--\,2.0 (red lines and circles) by
\citet{olivier09}, and the dashed, vertical lines mark the estimated
stellar masses for the galaxies in our sample.  The method used to
derive the stellar masses is described in \citet{wuyts09}. It assumes
a Kroupa's initial mass function \citep{kroupa01}, stellar population
synthesis models by \citet{bruzual03} and the reddening law of
\citet{calzetti00}.  The same set of assumptions are also used by
\citet{olivier09}.  This figure illustrates clearly that our
$z$\,$\sim$\,1 LIRGs mostly have stellar masses near or a bit lower
than $M^*$, the turn-over of the mass function, thus probing the most
representative and important mass scales at this redshift slice.  The
$z$\,$\sim$\,2 ULIRGs reaches much lower masses,
$\simlt10^{11}M_\odot$, than what was published in \citet{olivier09}
using COSMOS data. This is because the complete catalog used to
construct these mass functions is cut at $i_{AB}\simlt 25.5$.  This
suggests that many of our $z$\,$\sim$\,2 ULIRGs have fainter optical
magnitudes ($i_{AB} > 25.5$).

\subsection{IRS Observations and Ancillary Data \label{sec:obs}}

\begin{figure}[!t]
\plotone{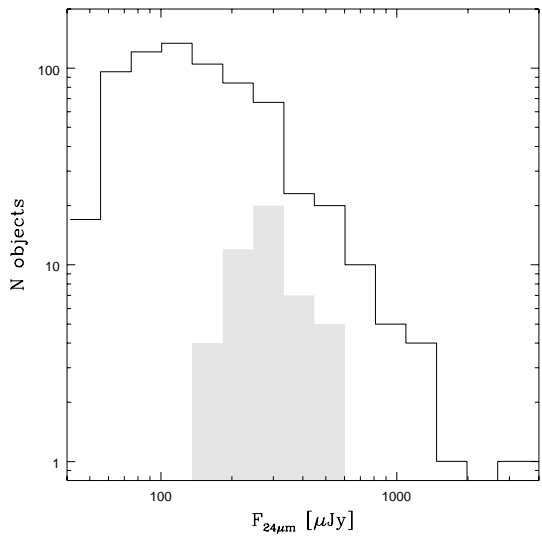}
\caption{\footnotesize 24\um\ fluxes of our IRS
  targets (shaded histogram) compared with the flux distribution of
  all the 24\um\ sources in the GOODS-CDFS field (solid
  histogram). \label{fig:histoflux}}
\end{figure}

As shown in Figure~\ref{fig:histoflux}, our targets have 24\um\ fluxes
of 0.14\,--\,0.55\,mJy in comparison with the distribution of all
24\um\ galaxies within the GOODS-South field (approximately 0.05 sq
degs).  For these ultra-deep, low resolution observations, we use the
IRS in mapping mode, with 4 dithered positions along the slit for each
target.  All observations were scheduled during each campaign just
after the dark calibration observations.  This method has been shown
to be effective in reducing potential latencies left by previous
observations of bright sources \citep{teplitz07}.  Our IRS spectra
will sample important PAH features at 6.2\um\ and 7.7\um, with the
observed coverage of 7\,--\,21\um\ for LIRGs and the 14\,--\,35\um\ range
for ULIRGs.

The CDFS field is remarkable for the great amount of ancillary data
publicly available. In this paper, we make use of this rich dataset,
including Chandra X-ray data, deep HST images at optical/UV,
ground-based near-IR data, and the \spitz images taken with MIPS
(24\um\ and 70\um), IRS (16\um), and IRAC (8.0\um).  We reduced the
\spitz data by ourselves, and directly took the reduced HST images
from the GOODS \citep{giavalisco04} and GEMS surveys \citep{rix04}, as
well as the recent FIREWORKS \citep{wuyts08} and
MUSYC\footnote{www.strw.leidenuniv.nl/fireworks, www.yale.edu/MUSYC}
multi-wavelength photometric catalogs \citep{musyc06}.  Most of the
sources of our IRS sample are covered by the 2\,Ms Chandra Deep-Field
South Survey \citep{luo08} (see Figure~\ref{fig:map}).  We use the
X-ray catalog from \citet{luo08} for our analysis in
\S~\ref{sec:agnfrac}.

\subsection{Optical Spectra \label{sec:optspec}}

\begin{figure*}[!t]
\centering{\includegraphics[width=1.8\columnwidth]{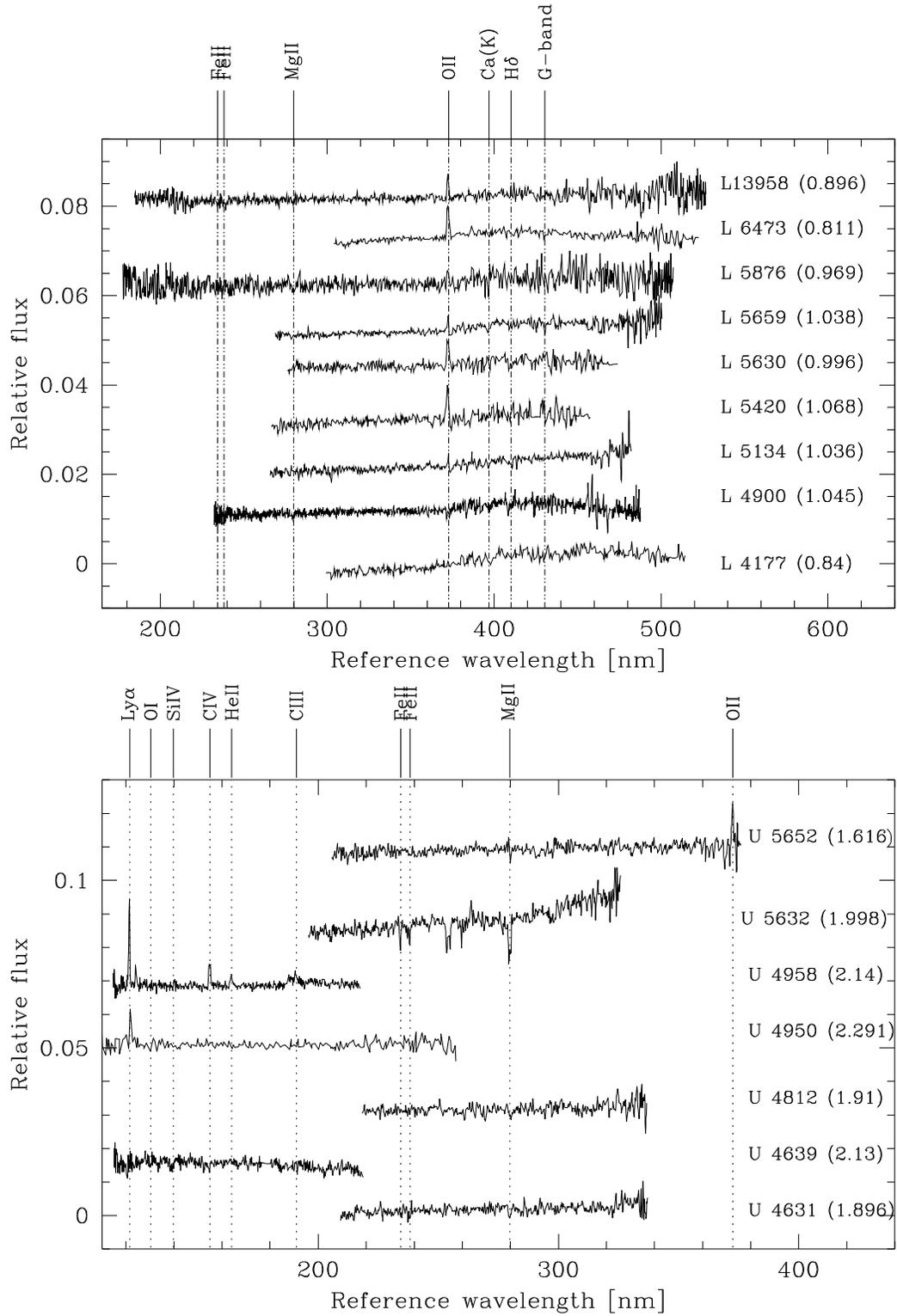}}
\caption{\footnotesize Optical spectra available for LIRGs (top) and
  ULIRGs (bottom) of our sample. The main absorption and emission
  features are marked.  Estimated optical redshifts are reported between
  brackets.
\label{fig:optspec}}
\end{figure*}

Published optical spectra exist for 17 ULIRGs and 9 LIRGs of our
sample. The redshift estimate is usually easy for the $z$\,$\sim$\,1 sample
because of the large number of features in the spectra ([OIII], Ca,
G-band) and becomes more complex in the case of $z$\,$\sim$\,2 sources. In
fact, there are good optical spec-$z$ estimates for 9 LIRGs while only 7
ULIRGs have optical spectra with enough features to give reliable redshifts
(see Table~\ref{targets_data}).  Remarkably, all these spectra are
available as FITS files and we were able to inspect each of them and
confirm or correct the redshift estimates of the original
publications. In the case of L4900, the published $z_{spec}(\rm opt)$
of 0.978 \citep{popesso08} was revised to 1.045 which is now
perfectly consistent with the estimate derived from the IRS spectrum.
All the spectra with good redshift estimates (according to the
published references) are reported in Figure~\ref{fig:optspec}.  The
only exception is the spectrum of U5795 (from \citet{mig05}) whose
FITS file is corrupted. The published optical redshift is not reported in
Table~\ref{targets_data} since we were not able to confirm it.  On the
other hand, we include the spectrum of U4631 in the figure and table
although it has been classified as unsure since it has clear MgII
and FeII lines and its optical spec-$z$ is similar to that of estimated from the mid-IR spectrum. 
The optical spectral lines will be used in the discussion
on the AGN fraction in \S~\ref{sec:agnfrac}.

\section{Mid-IR Data Reduction}

The reduction of the IRS data was carried out with a new software
written by the first author specifically for low resolution IRS
spectra. A detailed description of the technique used is given in the
appendix.

We reduced and measured fluxes of our targets in the 8\um\ IRAC,
16\um\ IRS, 24 and 70\um\ MIPS images.  For the IRAC data, we started
from the \spitz BCDs applying artifact corrections before
mosaicking. The artifact corrections include column pull-downs,
muxbleed, optical banding, and background droops.
We mosaicked the image with MOPEX with a pixel size of $0.6^{''}$. The
photometry within an aperture of $3.8^{''}$ in radius was measured
using SExtractor \citep{bertin96}.  The final IRAC 8\um\ flux is
computing using a flux conversion factor of 0.2021\,MJy/sr/(DN/s), and
aperture corrections of 1.84 (see SWIRE technical
report\footnote{swire.ipac.caltech.edu/swire/astronomers/publications},
page 31).  For the IRS 16\um\ images, we obtained a superflat by
stacking the original BCDs.  Each BCD is then corrected with this
superflat.  Because we are interested in only point sources, we
subtracted the median background from each BCD before mosaicking them
with MOPEX with a 0.9\,arcsec pixel. We extracted the sources using PSF
fitting with the code StarFinder \citep{diolaiti00} inside an aperture
of 9.45\,arcsec and applied an aperture correction of 1.1626 (computed
with a theoretical PSF obtained with
STinyTim\footnote{ssc.spitzer.caltech.edu/archanaly/contributed/stinytim
}).  The flux conversion used is 0.0117\,MJy/sr/(DN/s).
In the case no counterpart was found, we computed also 5$\sigma$ flux upper limits.
\begin{figure*}[!t]
\includegraphics[width=2.15\columnwidth]{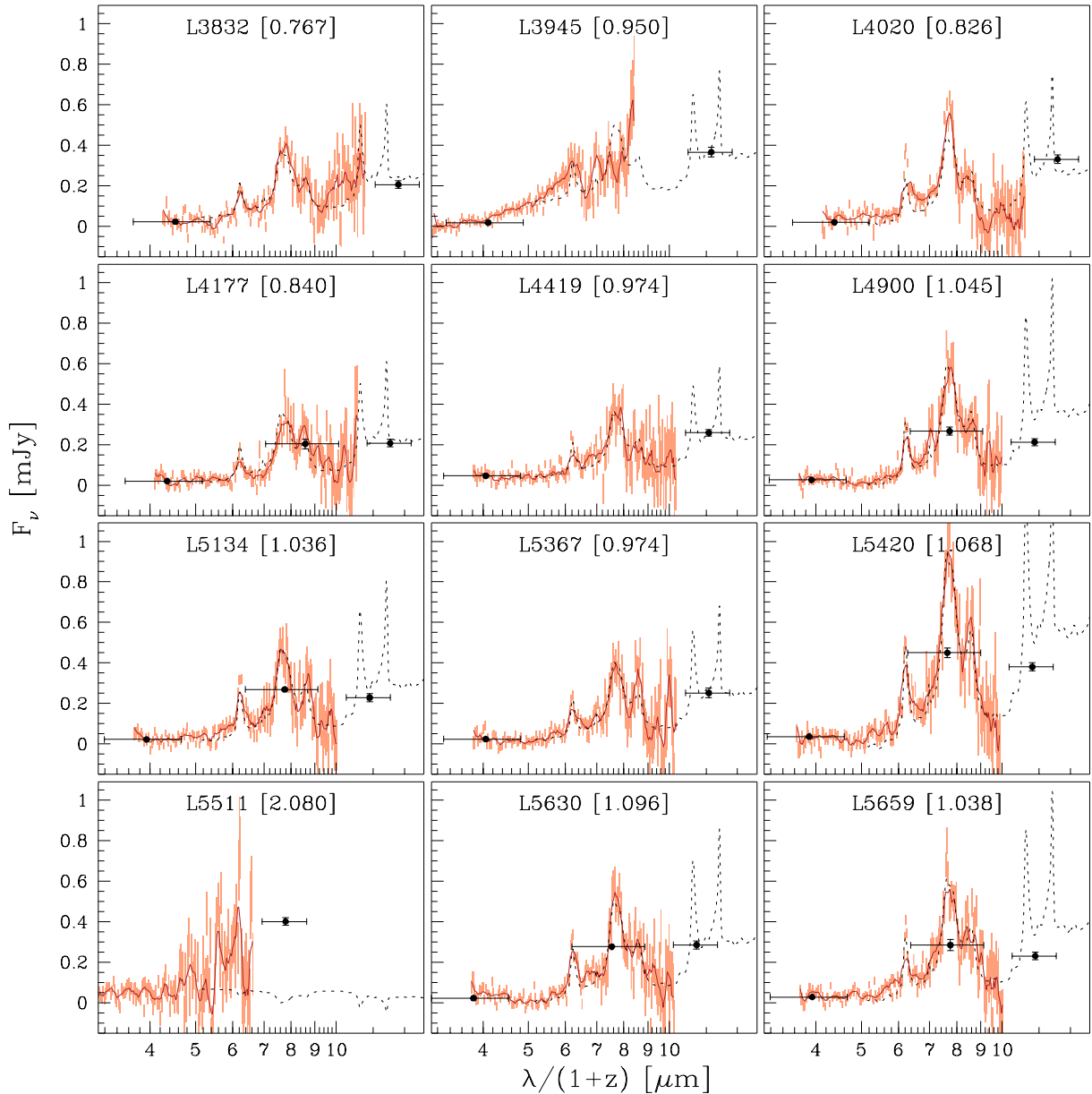}
\caption{ \footnotesize Observed spectra for our sample of 24
  LIRGs. The solid line, corresponding to the data with a smoothing of
  5 pixels, is on the top of the shaded 1$\sigma$ error bars for each
  pixel.  Broad band fluxes are overplotted. Dotted lines are either
  the averaged local starburst template from \citet{brandl06} or the
  hot dust dominated spectrum from \citet{spoon07} rescaled to the
  data. Source names and redshifts appear on the of each panel.
\label{spec2}}
\end{figure*}
\addtocounter{figure}{-1}
\begin{figure*}[!t]
\includegraphics[width=2.15\columnwidth]{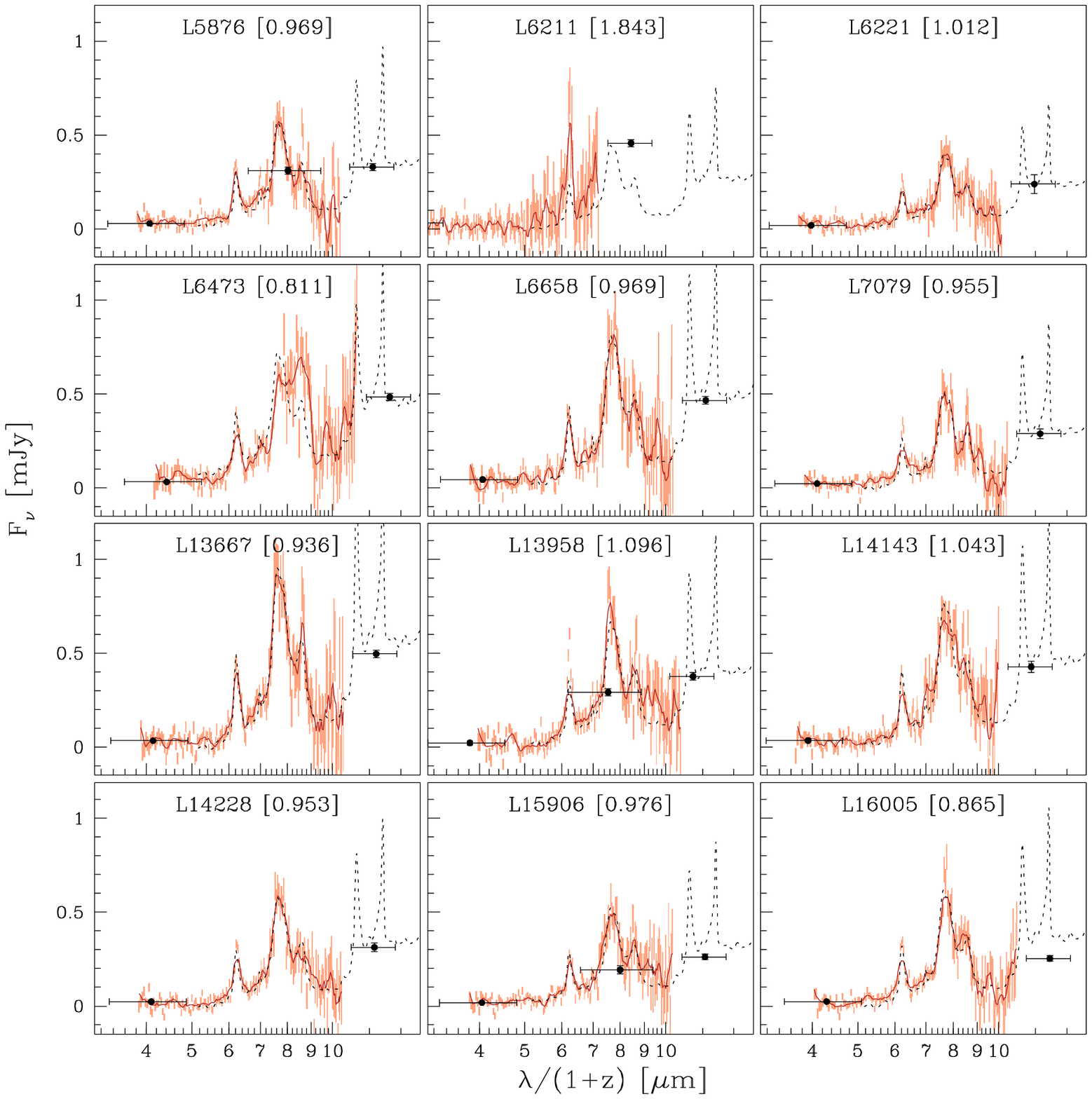}
\caption{ \footnotesize continued.
\label{spec2}}
\end{figure*}
\begin{figure*}[!t]
\includegraphics[width=2.15\columnwidth]{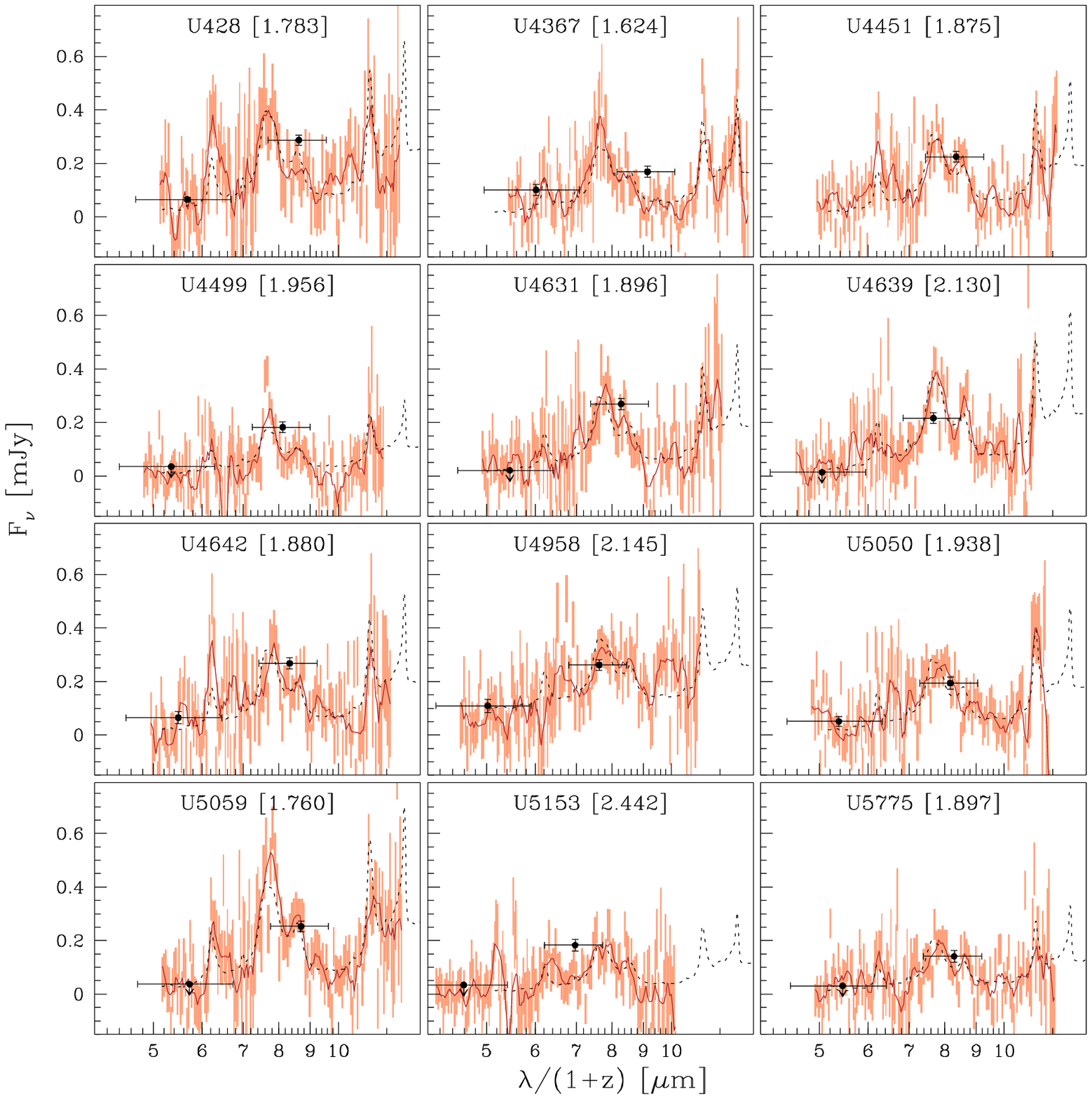}
\caption{ \footnotesize Observed spectra for our sample of 24 ULIRGs.
  The solid line, corresponding to the data with a smoothing of 5
  pixels, is on the top of the shaded 1$\sigma$ error bars for each
  pixel.  Broad band fluxes are overplotted. Dotted lines are either
  the averaged local starburst template from \citet{brandl06} or the
  hot dust dominated spectrum from \citet{spoon07} rescaled to the
  data. Source names and redshifts appear on the top of each panel.
\label{spec}}
\end{figure*}
\addtocounter{figure}{-1}
\begin{figure*}[!t]
\includegraphics[width=2.15\columnwidth]{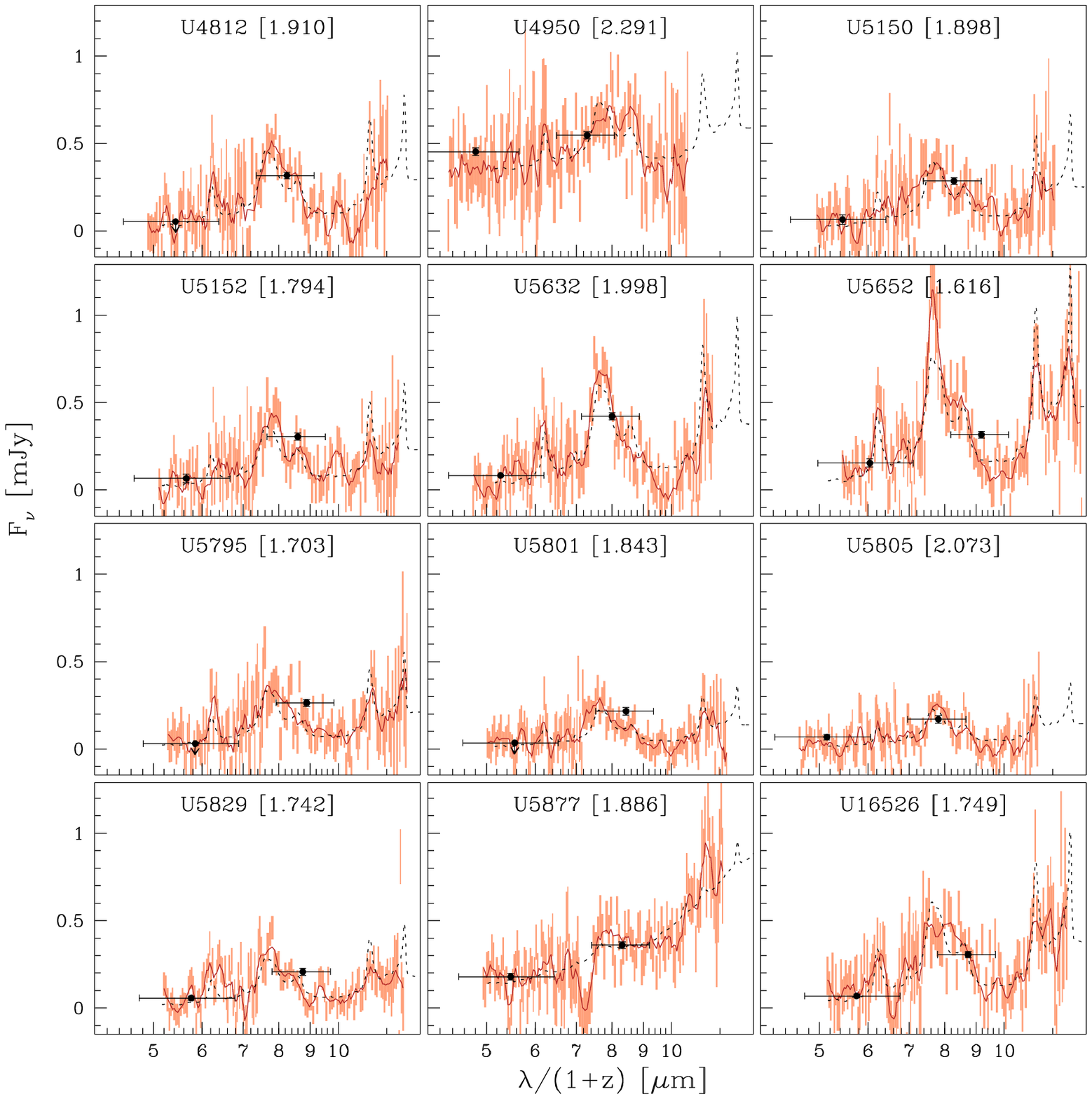}
\caption{ \footnotesize continued.
\label{spec}}
\end{figure*}
\begin{figure*}[!t]
\includegraphics[width=2.15\columnwidth]{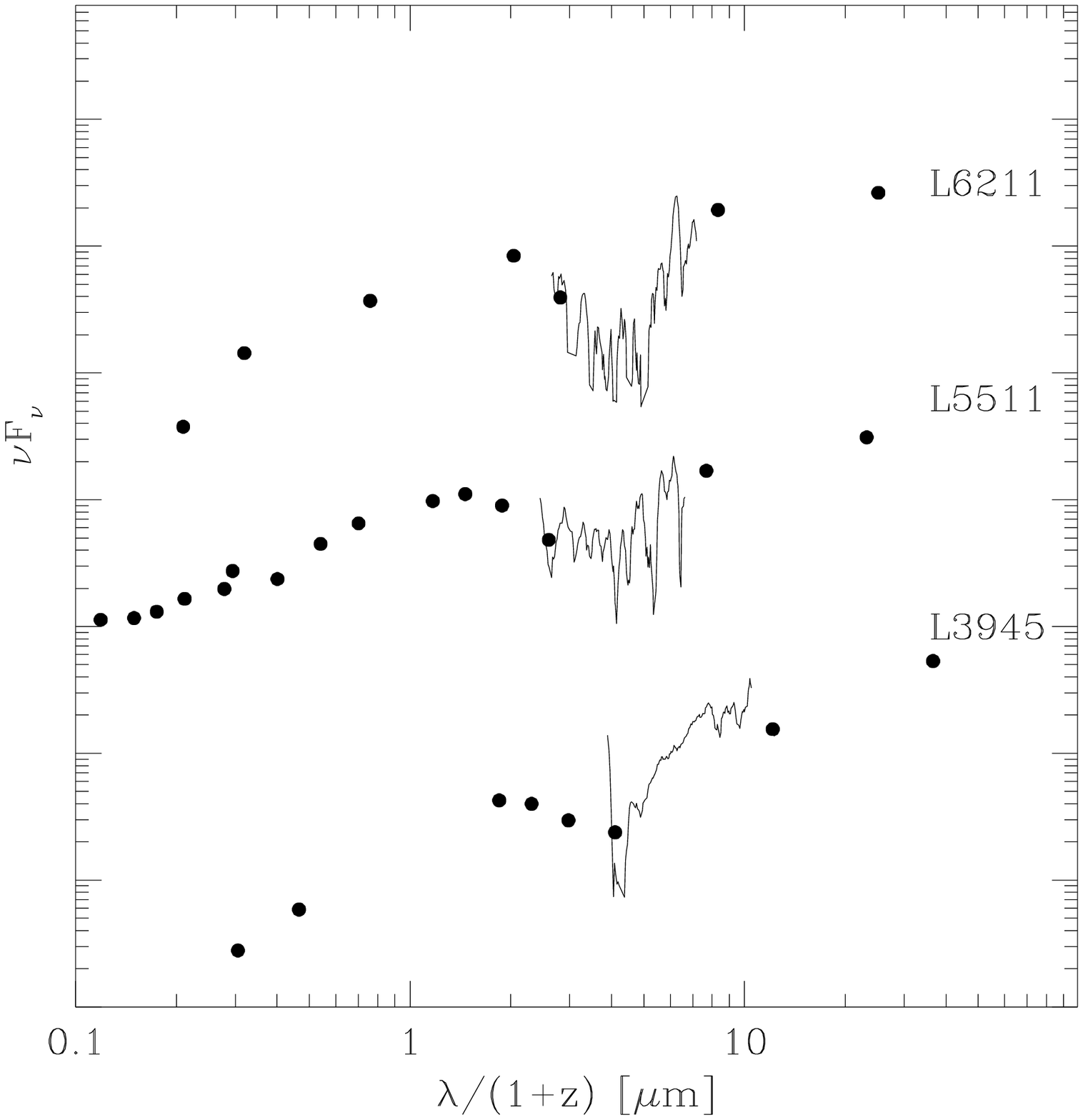}
\caption{ \footnotesize Available broad-band fluxes and IRS spectra
  for three LIRGs without strong PAH features. The photometric
  redshifts used in the sample selection were 0.754 and 0.776 for
  L5511 and L6211, respectively \citep{karina06}.  The proposed
  redshifts of 2.08 and 1.84 for L5511 and L6211, respectively,
  correctly place the valley between the parts of the SED dominated by
  stellar and dust emission as well as a strong feature visible in the
  L6211 IRS spectrum (identified as 6.2\um\ PAH).
\label{fig:SED}}
\end{figure*}

The 24\um\ image was reduced using an improved version of the method
described in \citet{fadda06}, which now corrects the long-term
transient and the regular short-term baseline variations.
The long-term is treated as a multiplicative effect since it has been
proved in the case of MIPS 70 and 160\um\ for which we have an
internal calibration that it is due to a variation in the responsivity of
the detector.  The short-term regular variation of the baseline is due
to distortions introduced by the different positions of the cryogenic
mirror assumed to freeze the image during a scan. This variation is
corrected by taking into account these distortions.
We made the mosaic image with MOPEX and extracted sources using
StarFinder with an aperture of $14^{''}$, and applying the aperture
and color correction of 1.158 and 0.961 respectively
\citep{fadda06}\footnote{The MIPS photometric system is tied to
  spectra with constant $\nu f_{\nu}$, see MIPS data
  handbook\footnote{http://ssc.spitzer.caltech.edu/mips/dh}, page 31}.
The flux conversion is 0.0454\,MJy/sr/(DN/s).

Finally, the 70\um\ imaging data is the deepest available, from the
Far-Infrared Deep Extragalactic Legacy Survey
(FIDEL)\footnote{http://ssc.spitzer.caltech.edu/legacy/all.html}.  We
start our reduction from the raw data using the off-line \spitz
software (GeRT\footnote{http://ssc.spitzer.caltech.edu/mips/gert}).
The pipeline corrects the calibration drifts, non-linearity, and
latencies due to stimflashes, and applies the flux conversion before
computing the slopes of the ramps. We remove any streaking in the
final images by using filtering on the first-pass images with bright
sources masked out.  The final mosaic with a pixel size of $4^{''}$
was obtained with MOPEX and sources were extracted with StarFinder
with an aperture of $42^{''}$ in diameter.  We apply the aperture
correction of 1.1665 (computed with STinyTim) and a color correction
of 0.918 to tie the photometric system to spectra with constant $\nu
f_{\nu}$.  The flux conversion is 702\,MJy/sr/(DN/s).  In the case no
counterpart was found, we computed also 5$\sigma$ flux upper limits.
The fluxes in these bands are reported in Table~\ref{targets_data} and
in Figures~\ref{spec2},~\ref{spec}, and~\ref{fig:sed}.  The flux
limits in these two figures are in $5\sigma$ for all bands.  All the
flux conversion factors reported are the conversion factors applied by
the \spitz pipeline used at the time of our data reduction. We report
these factors, as well as the aperture corrections used, to allow a
comparison with older or future reduction of other Spitzer data.

\section{Results \label{results}}

\subsection{Observed Mid-IR Spectra \label{sec:mirspec}}

Figures~\ref{spec2} and \ref{spec} present the observed mid-IR spectra
for the 24 LIRGs at $z$\,$\sim$\,1 and 24 ULIRGs at $z$\,$\sim$\,2. In each
panel, the smoothed spectrum in the rest frame wavelength is shown as a
red solid line, a local spectral template as a dashed line, and the
unsmoothed spectrum with errorbars as an orange histogram.  Each panel
indicates the broad band fluxes at 8\um, 16\um, and 24\um, the
object name, the redshift based on the mid-IR spectrum, or optical
redshifts when the mid-IR spectra do not yield redshifts (which is only
the case for 3 LIRGs). Downward arrows mark 5$\sigma$ flux upper limits.
The spectral template shown in dashed line in each panel is either the
averaged starburst spectrum obtained by \citet{brandl06} based on the
\spitz\ spectra of 22 local starburst galaxies, or the AGN template 1A
\citep{spoon07} if the observed spectrum has no obvious PAH
emission. Redshifts are estimated from the mid-IR spectra by
cross-correlating these two templates to the data.  On
Figures~\ref{spec2} and \ref{spec}, the overlapped local spectral
templates allow a simple characterization of the observed spectra of
our high-z LIRGs and ULIRGs.

\subsection{Mid-IR Spectroscopic Redshifts \label{sec:mirz}}
\begin{figure}[!t]
\plotone{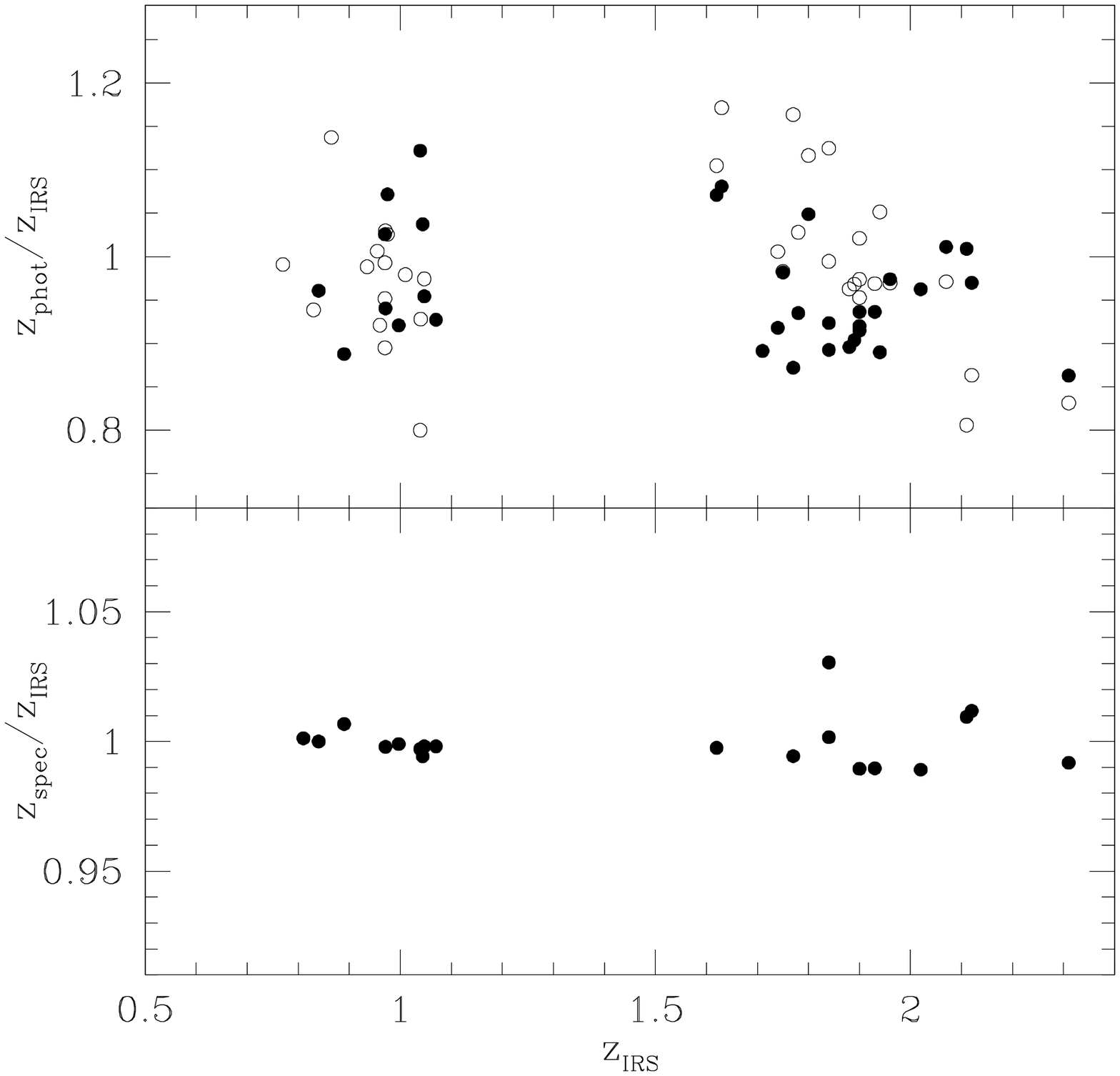}
\caption{ \footnotesize Comparison between mid-IR spectroscopic
  redshifts and optical redshifts (spectroscopic and photometric). In
  the top panel, photometric redshifts are from \citet{wuyts08} (full
  circles) and \citet{karina06} (empty circles).
\label{compz}}
\end{figure}

Of our 48 sources, we were able to estimate redshifts for 44 of them
from their mid-IR spectra based on the PAH emission features and
silicate absorption. We estimated the redshifts by cross-correlating
the averaged starburst template of \citet{brandl06}. 
For three sources in the LIRG sample, we were not able to estimate
spectroscopic redshifts from the IRS data because of the lack of
recognizable features. When plotting their spectra with the broad band
photometry points (see Figure~\ref{fig:SED}), it is clear that two of
them (L5511 and L6211) are at redshifts higher than those estimated by
\citet{karina06} (0.754 and 0.776, respectively). They are actually
ULIRGs at $z$\,=\,2.08 and $1.84$, respectively.  Furthermore,
Figure~\ref{spec2} shows that the spectrum of L6211 has fairly strong
6.2\um\ PAH emission at $z$\,=\,1.84, with the rest of the PAH
spectrum not sampled by the observations (LIRG observations used only
LL1, not LL2).  The wrong estimates are due to wrong
identifications with optical sources at $\sim$\,1.5\,arcsec from their
positions. The real counterparts are faint sources very close to these
bright galaxies and usually blended with them in ground images which
are not reported in the COMBO17 catalog. The faint optical
counterparts appear very clearly in the HST images (see
Figure~\ref{morphl}).

  
Figure~\ref{compz} compares mid-IR spectroscopic redshifts with two
sets of photometric redshifts in the top panel and with the optical
spectroscopic redshifts in the bottom panel.  The mid-IR spectroscopic
redshifts are consistent with the available optical spectroscopic and
photometric redshifts within 20\%, and the photometric redshifts have
larger uncertainties than that of mid-IR spectroscopic redshifts.  The
photometric redshifts from \citet{wuyts08} have a slightly lower
dispersion than those from \citet{karina04} since they are based on
optical and infrared data. Table~\ref{targets_data} reports
preferentially photometric redshifts from the former study.  We
conclude that our sample selection based on initial photometric
redshifts has proven effective.

Our IRS spectra have provided 8 new spectroscopic redshifts used in
\citet{wuyts09} for which ground-based optical/near-IR spectroscopy is
difficult.  Particularly, at $z$\,$\sim$\,1.7\,--\,2.0, $H_{\alpha}$
falls between the H- and K-band atmospheric windows, making
ground-based near-IR spectroscopy impossible.  For 24\um\ galaxies in
this redshift range, their \spitz mid-IR spectra provide useful
distance measurements.

\subsection{AGN fraction \label{sec:agnfrac}}

\begin{figure}[!h]
\plotone{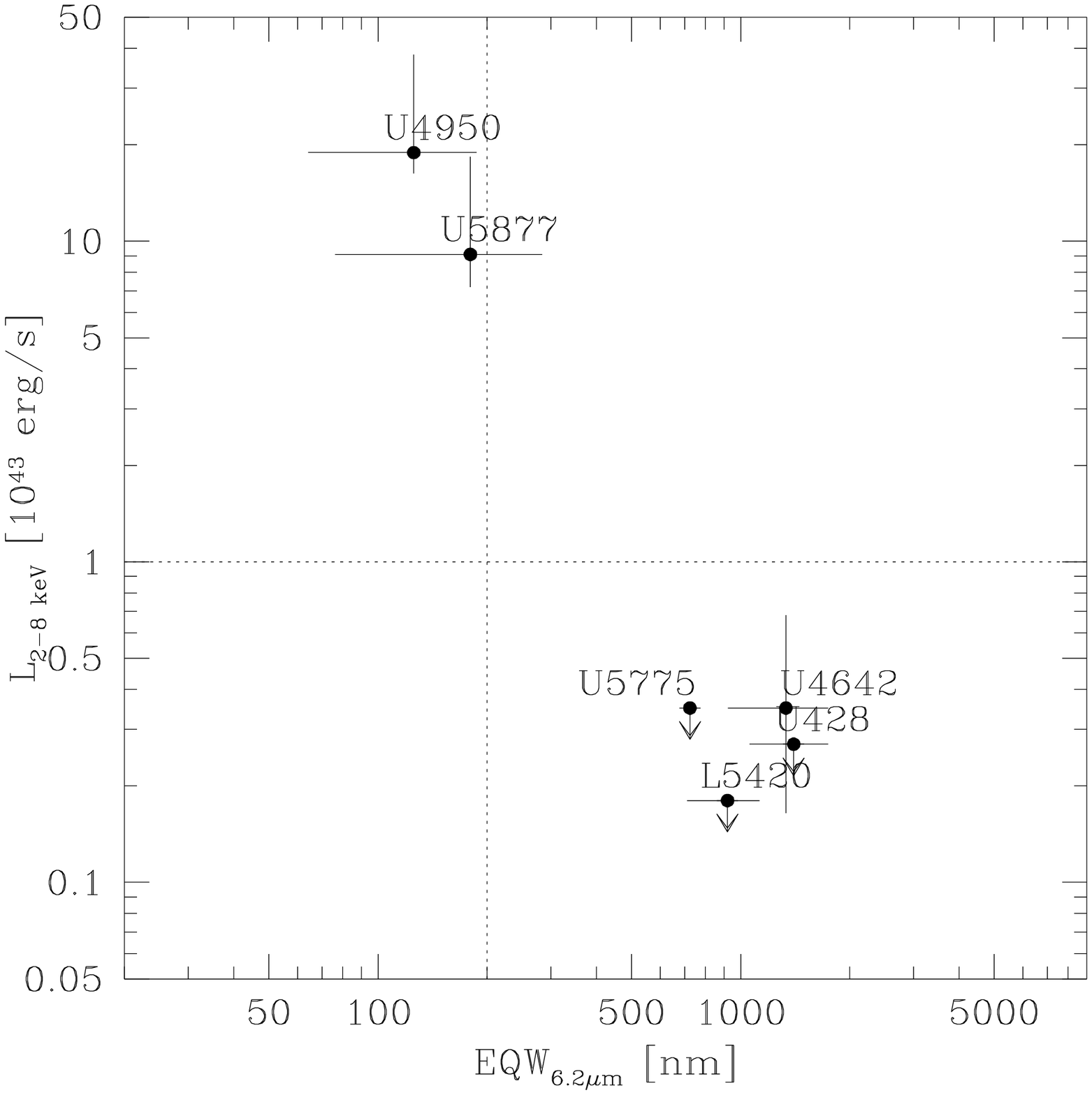}
\caption{ \footnotesize Hard X-ray 2-8\,keV luminosity versus
  equivalent width of the 6.2\um\ PAH feature for the IRS sources with
  X-ray counterparts. The infrared output of the objects on the left
  of the vertical line is dominated by AGN emission \citep{armus07,
    veilleux09}.  Objects above the horizontal line are typically AGN
  dominated sources (see, e.g., \citet{fadda02}). The dashed vertical
  line marks $EW_{6.2\mu m}$\,=\,0.2\um, below which galaxies are AGN
  dominated. See \S~\ref{sec:agnfrac} for the detailed discussion of
  this value.
\label{lxeqw}} 
\end{figure} 

\begin{figure}[!h]
\plotone{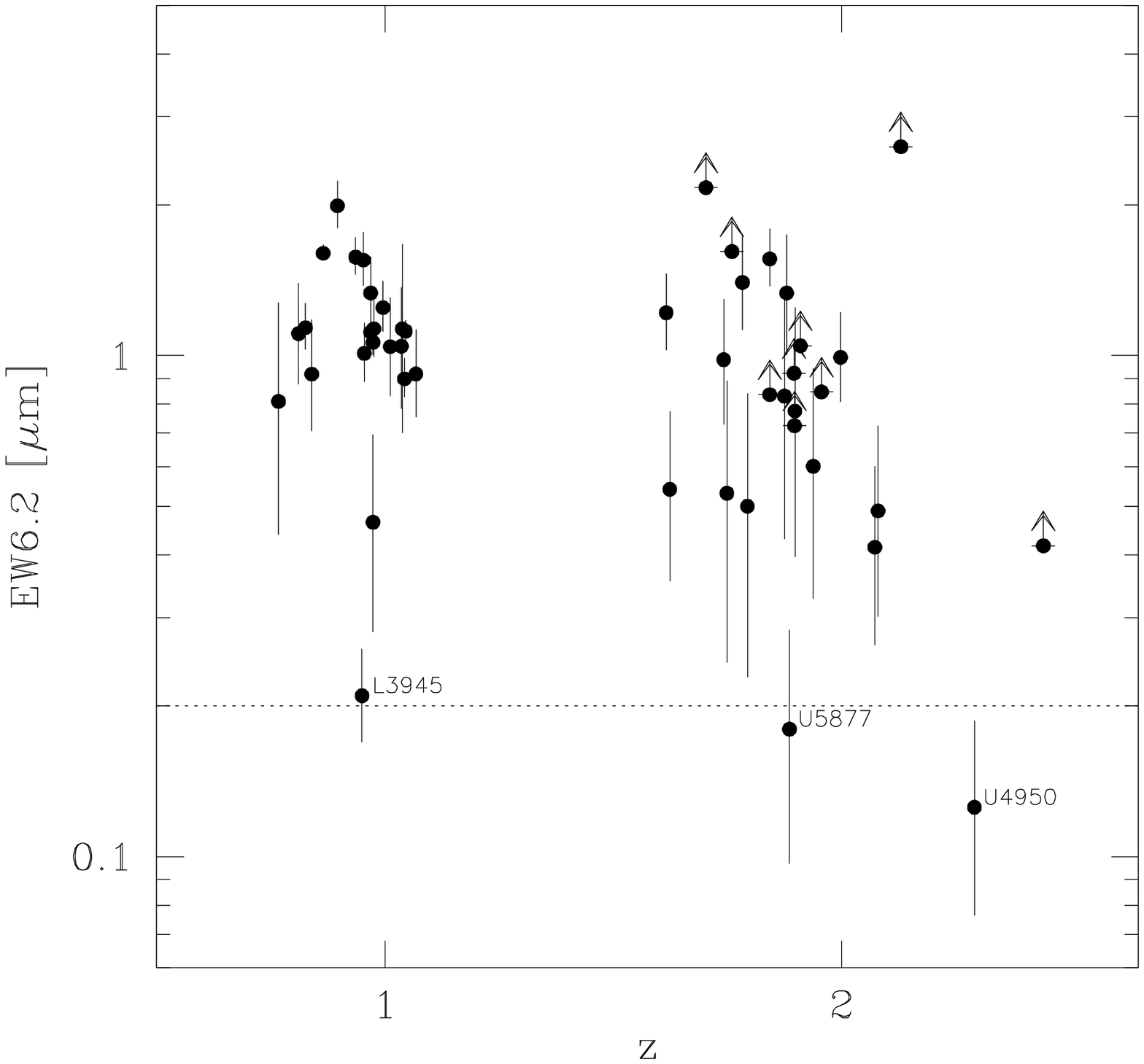}
\caption{ \footnotesize 6.2\um\ equivalent width versus
  redshift. The lower limits come from upper limits to the continuum
  from 16\um\ observations. The horizontal line corresponds to the
  limit between AGN and star formation dominated systems. $EW_{6.2\mu m}$\,=\,0.2\um\ is derived from the combination of \citet{armus07} and \citet{veilleux09}. 
\label{eqwz}}
\end{figure} 

One motivation for this program is to use mid-IR spectral diagnostics
to determine the contribution of AGN to the total energy output among
LIRGs at $z$\,$\sim$\,1 and ULIRGs at $z$\,$\sim$\,2, particularly the obscured
AGN which may not be detected by X-rays. It is important to understand
that, although in some cases AGN can completely dominate the
bolometric luminosity, there is an entire population of galaxies with
enshrouded AGN whose bolometric luminosities are primarily from
star formation.  To objectively classify AGN, we use several
indicators such as: broad lines and high ionization lines in optical
spectra, lack of a 1.6\um\ stellar bump in the SED, X-ray bright
sources, low mid-IR 6.2\um\ PAH equivalent width, and finally their
optical morphology.  We note that our samples at $z$\,$\sim$\,1 and $z$\,$\sim$\,2 now have 22 and 26 sources respectively, because two galaxies
initially assumed to be LIRGs at $z$\,$\sim$\,1 have now been determined to
be ULIRGs at $z$\,$\sim$\,2 (see \S~\ref{sec:mirz}).

\begin{figure*}[!t]
\includegraphics[width=2.2\columnwidth]{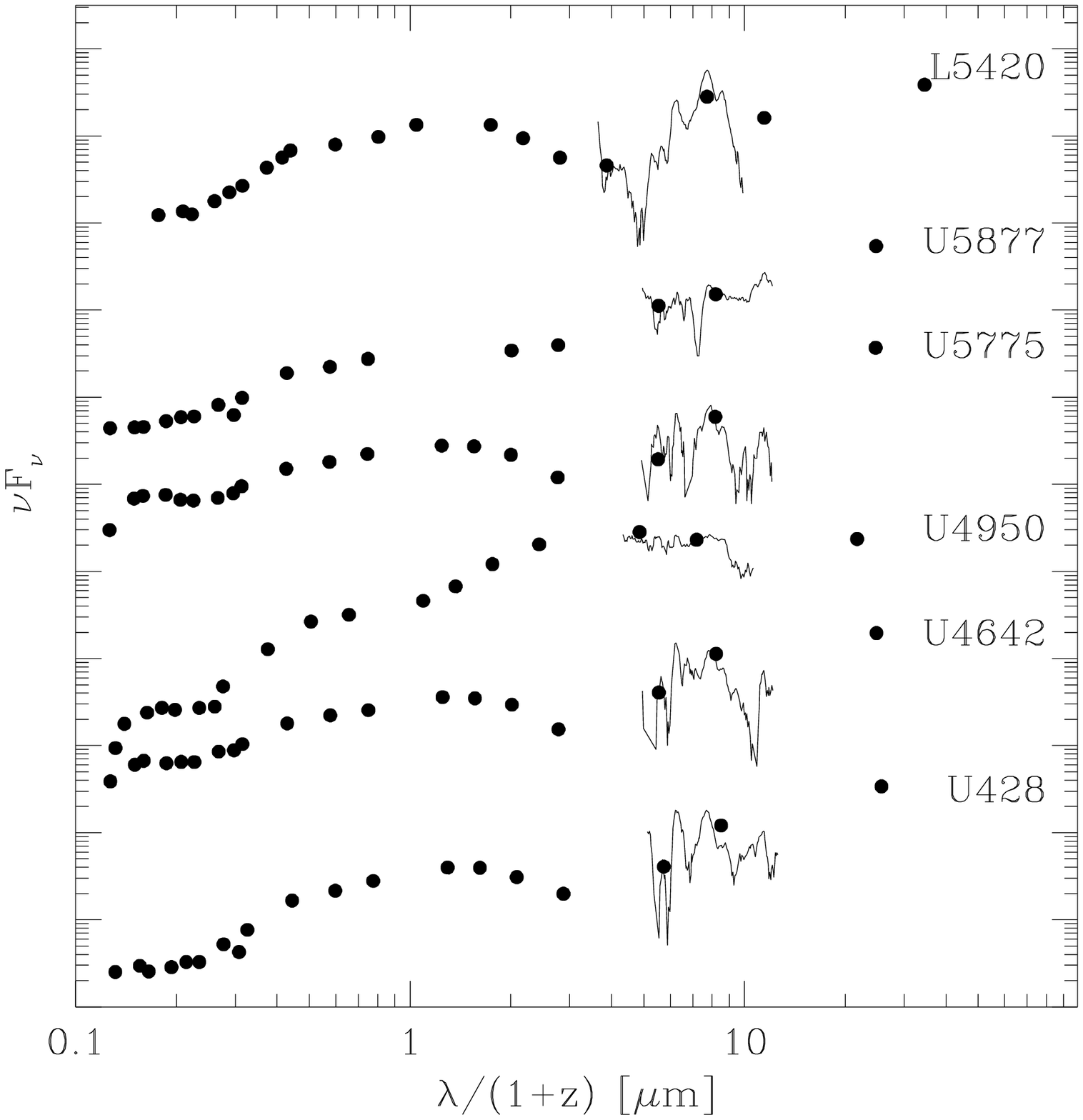}
\caption{ \footnotesize Broad band photometry and IRS spectra for the
  six IRS sources with X-ray counterparts.   Only two of
  them (U4950 and U5877) have SED and spectral features compatible
  with AGN dominated sources.
\label{fig:xray}}
\end{figure*}

The 2\,Ms Chandra X-ray survey covers all the ULIRGs and most of the
LIRGs of our sample. Six of our sources have X-ray counterparts from
the catalog of \citet{luo08} (see Table~\ref{xray}).  Two of them,
U4950 and U5877, are clearly detected in the (2\,--\,8)\,keV hard X-ray
band.  Another one (U4642) is detected at the limit of sensitivity of
the survey.  Two more ULIRGs (U5775 and U428) are detected only in the
soft X-ray and one LIRG (L5420) is detected at a low-$\sigma$ level in
the soft X-ray (secondary catalog). A cross-match with the more
extended X-ray observations of \citet{lehmer05} does not yield any
further counterpart.  Figure~\ref{fig:xray} compiles all of the broad
band photometry and mid-IR spectra for these 6 sources.  U4950 and
U5877 have IRS spectra compatible with a power law and also their
broad band photometry is typical of an AGN, i.e. no 1.6\um\ stellar bump
and a power-law spectrum over  the entire wavelength range.
We note that the HST morphologies of U4950 and U5877 show compact,
unresolved nuclei, supporting the AGN classification from both X-ray
and mid-IR spectra.  The other source detected in the hard X-ray
(U4642) has, on the contrary, a more complicated morphology with
multiple clumps. So, one clump can host the AGN, but the AGN does not
dominate the optical-IR emission. All the other
sources (U5775, U428 and L5420) have a softer X-ray emission, the
mid-IR spectra have obvious PAH emission (see the quantitative numbers
below) and their broad-band SEDs shown in Figure~\ref{fig:xray} reveal
a strong stellar bump at rest-frame wavelength 1.6\um, suggesting that
although these three sources may have low-luminosity AGN, their host
galaxies dominate the stellar emission in optical to near-IR. The
X-ray luminosity of the fainter X-ray sources is also compatible with
star formation using the relationships from \citet{ranalli03}.

The optical spectrum of U4958 (see Figure~\ref{fig:optspec}) shows
a broad CIII line and strong NV and CIV emission lines. Considering 
 that the infrared spectrum is rather featureless, we can classify
this source as AGN dominated.

\begin{deluxetable}{lrrrrr}
\tablecaption{Cross-correlation with X-ray survey \label{xray}}
\tablewidth{0pt}
\tablehead{
\colhead{Name} & \colhead{Distance} & \multicolumn{3}{c}{Flux [$10^{-16}\ erg/s/cm^2$]} & \colhead{$L_{2-8keV}$}\\
               & \colhead{(arcsec)} & \colhead{(0.5-8 keV)} & \colhead{(0.5-2 keV)} & \colhead{(2-8 keV)} & \colhead{[$10^{43}\ erg/s$]}\\}
\startdata                                                              
      \multicolumn{6}{l}{primary catalog:}\\
       U428   &   0.29 &  0.76  &       0.35 &   $<$1.26  &$<$0.27\\
       U4642  &   0.12 &  2.05  &       0.74 &      1.41  &0.35\\
       U4950  &   0.18 & 60.90  &      13.00 &     47.00  &18.9\\
       U5775  &   0.48 &$<$0.97 &       0.42 &   $<$2.39  &$<$0.35\\
       U5877  &   0.18 & 36.40  &       0.73 &     36.40  &9.1\\
      \multicolumn{6}{l}{secondary catalog:}\\
       L5420  &   0.40 &$<$2.07 &       0.63 &   $<$2.91  &$<$0.18\\
\enddata
\end{deluxetable}
The strength of PAH features has been widely used as a star formation
indicator \citep{lutz96,genzel98,armus07,veilleux09}. Particularly,
the 6.2\um\ PAH is relatively isolated, and its rest-frame equivalent
width can be measured cleanly and used to classify starburst-dominated
systems.  This is supported by the fact that $EW_{6.2\mu m}$ is shown
to broadly anti-correlate with flux ratios of [NeV]/[NeII] and
[OV]/[NeII] \citep{lutz96,genzel98,armus06,veilleux09}, where the
detections of \nev\ and \oiv\ indicate the excitation of the ionized
gas by black hole accretion due to their extremely high ionization
potentials (97.1\,eV and 59.4\,eV respectively).  We used the code
PAHFIT \citep{smith07} to fit lines and dust features together with a
continuum from dust and stellar emission to our IRS spectra. During
our analysis, we discovered two bugs in the routine that computes
equivalent widths.  The routine was incorrectly applying Gaussian
profiles instead of Drude profiles in the computation and, more
importantly, using the flux of the line corrected for extinction and
not corrected continuum. These bugs have been reported to the author
of the code and will be fixed for the next release.  To measure the
equivalent width of the 6.2\um\ PAH, we used a Chiar-Tielens silicate
profile and added an absorption component to take into account the
water ice and hydrocarbon features at 5.7\,--\,7.8\um.  Following
\citet{veilleux09}, we used the profile taken from observations of
F00183-7111 \citep{spoon04} for this absorption feature. The addition
of the water ice and hydrocarbon absorption features at
5.7\,--\,7.8\um\ is important mainly in strongly absorbed local ULIRGs
(see \S~\ref{sec:starformation}). Neglecting this component in our
sample spectra and other high-z spectra does not change significantly
the $EW_{6.2\mu m}$ estimates.

In the case of LIRGs, the signal-to-noise ratio of the spectra is high
enough to measure directly the equivalent width of the 6.2\um\ PAH
feature. In the case of ULIRGs, the low continuum combined with the
higher noise of the spectra lead to large errors in the estimates of
$EW_{6.2\mu m}$. In particular, the error is dominated by the
uncertainty in the estimate of the continuum from the spectrum. To
obtain firmer estimates for individual sources, we used 16\um\ fluxes
as estimates of the continuum since they have a much lower
uncertainty. We have 16\um\ fluxes for 14 ULIRGs and upper limits for
an additional 9.  Although, in general, most of the 16\um\ flux is due
to the continuum, the contribution of the 6.2\um\ PAH is increasingly
important at redshifts lower than 1.7.  To compensate for this effect,
we applied a multiplicative correction to the 16\um\ flux equal to the
ratio of the 5.8\um\ flux and the expected 16\um\ of the averaged
starburst spectrum moved at the redshift of the ULIRG.  This
correction ranges from 0.6 at $z$\,$\sim$\,1.6 to 0.9 at
$z$\,$\sim$\,2 and does not vary significantly by using different
templates as the average HII-like ULIRG spectrum which better
describes ULIRGs around the Si absorption.  We were able to measure
the $EW_{6.2\mu m}$ for all the sources with the exception of U4958
which we already classified as AGN on the basis of its optical
spectrum.
\begin{figure*}[!t]
\includegraphics[width=2.2\columnwidth]{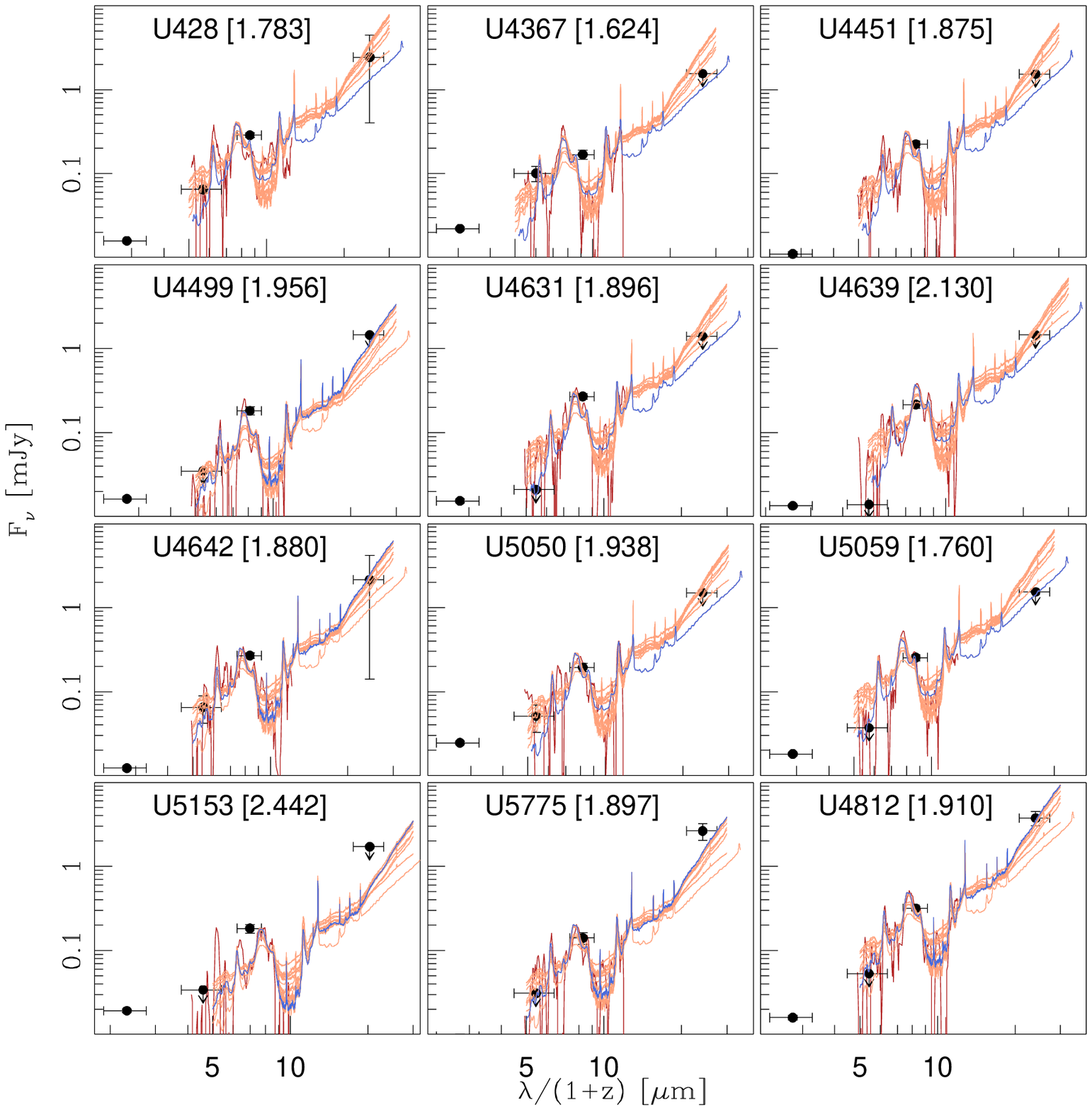}
\caption{ \footnotesize Templates of average starburst
  \citep{brandl06} and ULIRG composites \citep{veilleux09} rescaled to
  the smoothed IRS spectra (red line) of non AGN-dominated ULIRGs. The
  best fit compatible with the 70\um\ flux is shown in blue.
\label{fig:sed}}
\end{figure*}
\begin{figure*}[!t]
\addtocounter{figure}{-1}
\includegraphics[width=2.2\columnwidth]{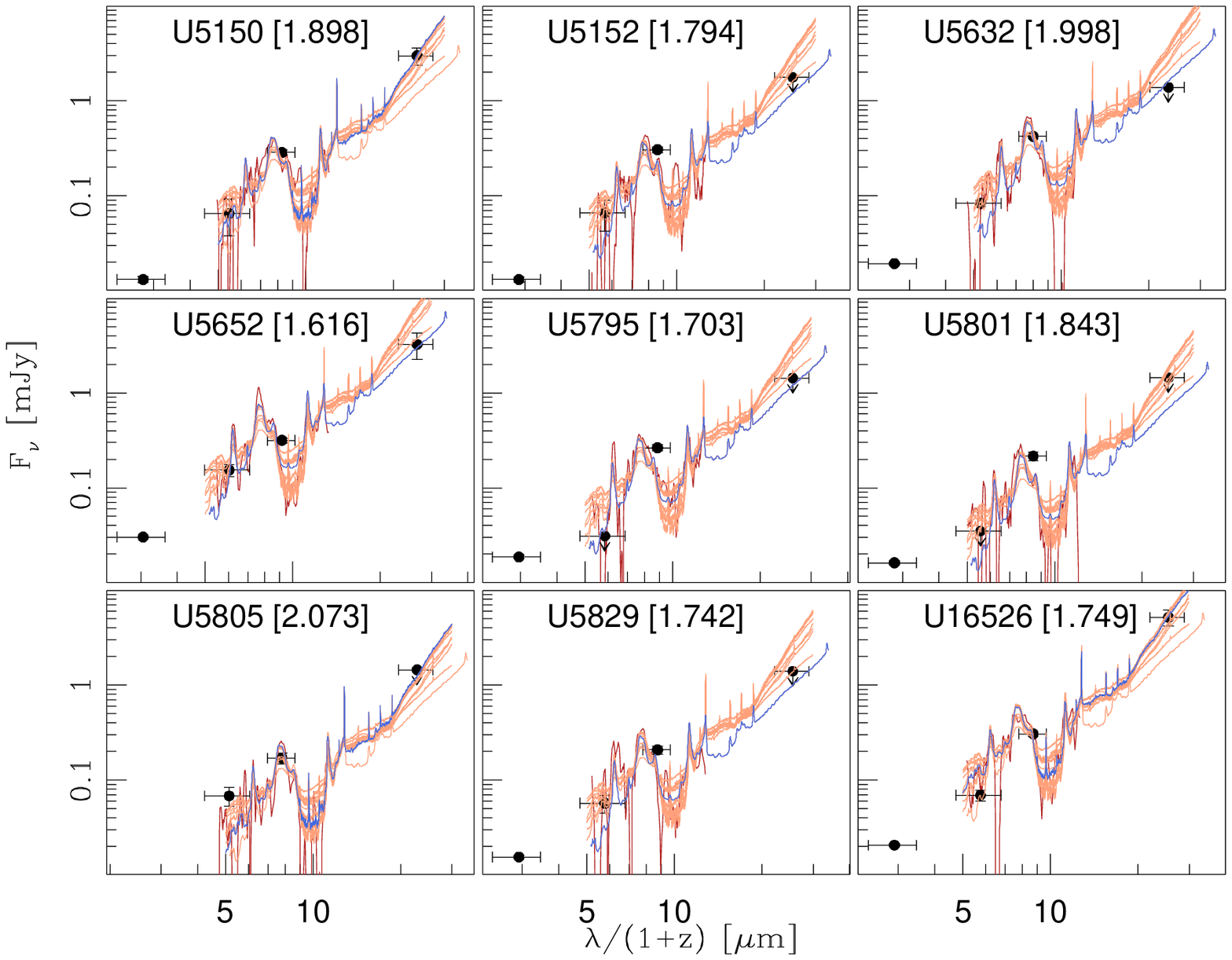}
\caption{ \footnotesize continued.
}
\end{figure*}

The 6.2\um\ equivalent widths of the X-ray emitting sources are
reported in Figure~\ref{lxeqw} and support our previous discussion.
Studies of local LIRG and ULIRG spectra have shown that the percentage
of AGN contribution to the total energy output correlates with
$EW_{6.2\mu m}$. \citet{armus07} estimates that $EW_{6.2\mu m} <
0.3$\um\ are AGN-dominated, where \citet{veilleux09} put the limit
around 0.1\um\, while at 0.3\um\ the AGN contribution is around 30\%.
The two results are not in contradiction since \citet{veilleux09} compare 
AGN fractional contributions to the bolometric luminosities by
applying bolometric corrections to their numbers (see Table~10 in  \citet{veilleux09} and discussion in the Appendix). \citet{armus07} compare instead
AGN fractional contributions to the [NeII] and mid-IR luminosities.
Nevertheless, according to other criteria used by \citet{veilleux09},
also sources with $EW_{6.2\mu m}$ up to 0.2\um\ can be AGN-dominated (see Fig.~\ref{eqw}).
In the following, we assume 0.2\um\ as a conservative limit.

Figure~\ref{lxeqw} shows that the mid-IR 6.2\um\ PAH equivalent
criteria has yielded the same conclusion as we stated above, that 2
X-ray sources U4950 and U5877 are strong AGNs, and the rest of four
have their infrared emission dominated by star formation.

Since many LIRGs are outside of the X-ray field or in the less sensitive
parts and because of the limited hard X-ray sensitivity,  it
is useful to analyze the 6.2\um\ equivalent widths of our entire sample.
Figure~\ref{eqwz} shows the 6.2\um\ equivalent width versus redshift
for all 48 sources with measurable 6.2\um\ PAH. The infrared emission of almost all
the LIRGs and the majority of the ULIRGs is dominated by star
formation.  The LIRG with low 6.2\um\ equivalent width
(L3945) lie in the less sensitive and external regions
of the X-ray survey and therefore it is not a surprise that it is not
detected. 
\begin{figure*}[!t]
\includegraphics[width=1.\columnwidth]{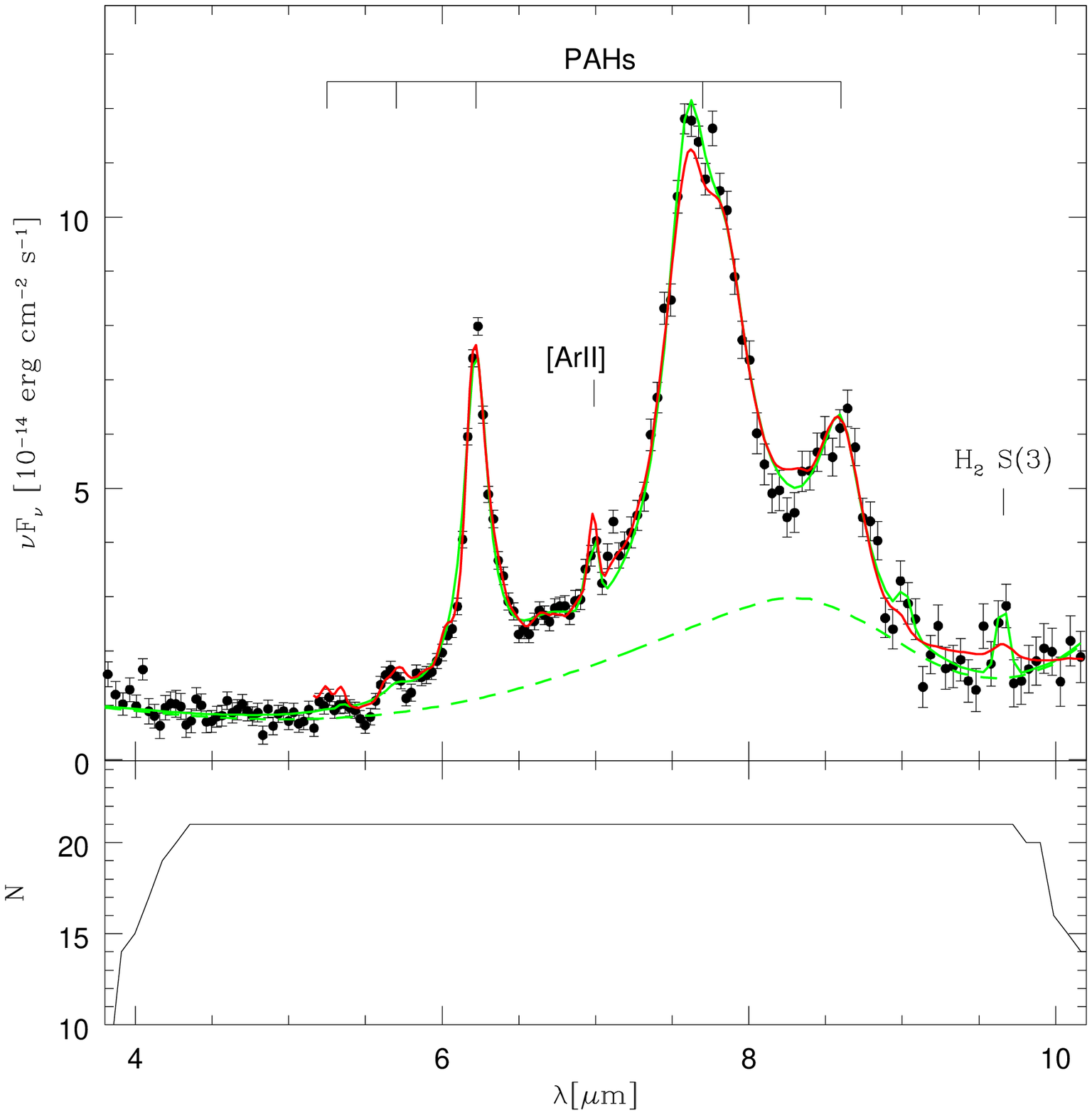}
\includegraphics[width=1.\columnwidth]{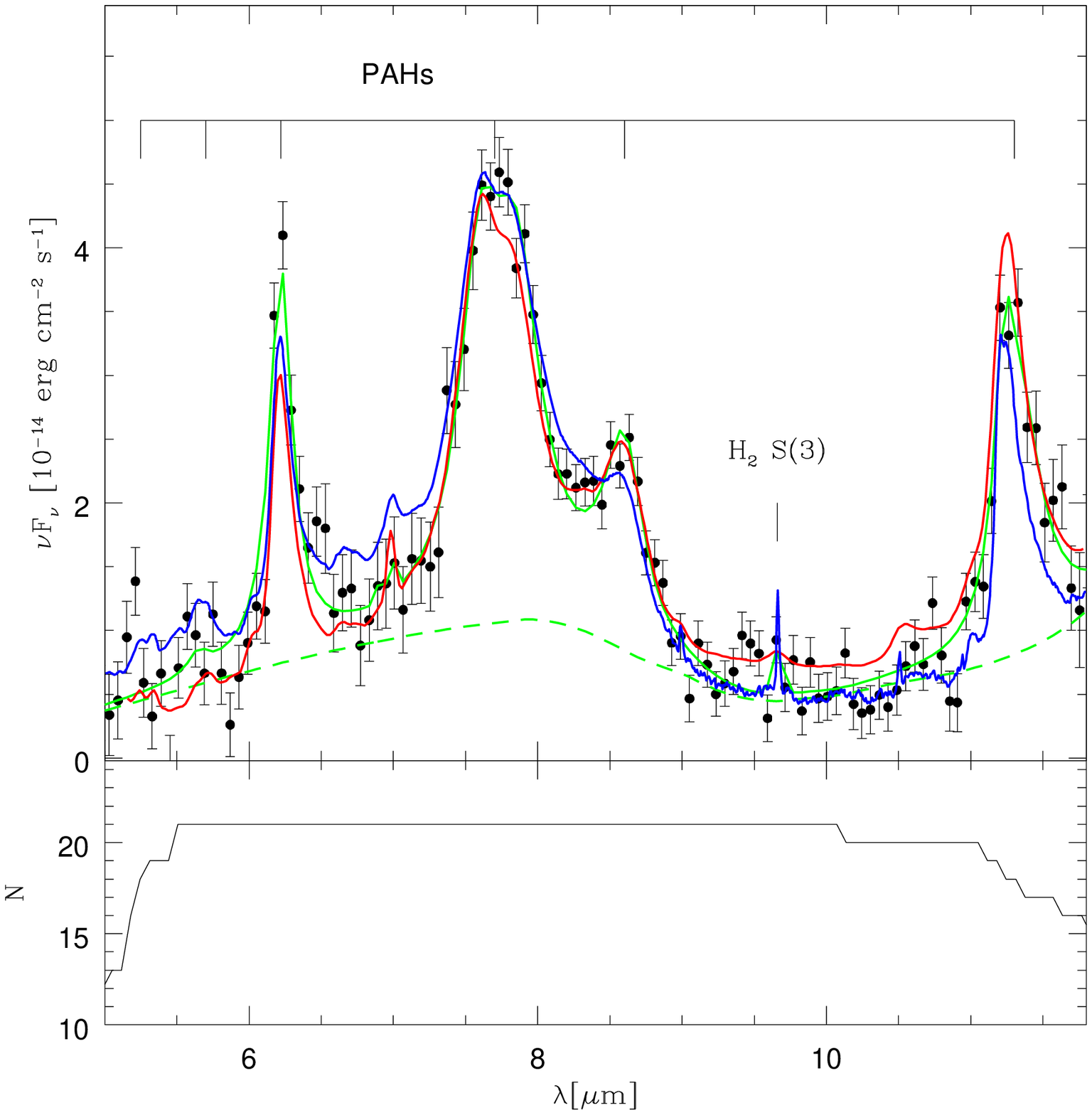}
\caption{ \footnotesize The averaged $z\sim1$ LIRG (left) and $z\sim2$
  ULIRG spectrum (right) obtained by stacking all the spectra which
  are not AGN-dominated. The spectra have been normalized to the
  24\um\ rest-frame luminosity before stacking them. The lower panel
  shows the number of spectra used for each wavelength. The continue
  and dashed green lines correspond, respectively, to the total flux
  and continuum fitted with PAHFIT. The red and blue lines correspond,
  respectively, to the averaged starburst from \citet{brandl06} and
  the average local HII ULIRG from \citet{veilleux09} normalized to
  the stacked spectra.
\label{stacking}}
\end{figure*}

Among LIRGs, therefore only one source is dominated by the AGN emission (5\% of the
sample). In the case of ULIRGs, we exclude L5511 (reclassified as
ULIRG) from the sample because its spectrum does not have enough
coverage to yield usable 6.2\um\ equivalent width measurement.  Thus,
of the total 25 ULIRGs at $z$\,$\sim$\,2, only 12\%\ (3/25) have a dominant
AGN.

These numbers are significantly lower than those for bright ($S_\nu$\,$
\simgt$\,1\,mJy) 24\um\ samples in the same redshift range.  Based on
the mid-IR spectral features as well as IR SEDs, the fraction of AGN
dominating the IR luminosities are close to
(50\,--\,75)\%\ \citep{yan05,yan07,sajina07,sajina08}.  Including weak AGNs
or starburst/AGN composite systems, the majority ($\simgt$\,75\%) of
bright 24\um\ samples contain AGNs \citep{sajina08}. Studies of local
ULIRGs have found that the AGN fraction becomes significant at $L_{IR}$\,$
\simgt$\,$10^{12.3}L_\odot$ \citep{lutz98}, although the exact numbers
suffer from many uncertainties. There are speculations (see
\S~\ref{sec:starformation}) that at high-$z$, this AGN-starburst
transitional luminosity moves to higher values. One supporting
evidence has been SMGs, whose $L_{IR}$ is on average $\simgt$\,$
10^{12.3}L_\odot$, and it is commonly accepted that bolometric
luminosities of SMGs are mostly from starbursts (see, e.g.,
\citet{alexander03,valiante07}).  One result from this paper is that
the AGN fraction for $z\sim2$ galaxies is small for $L_{IR}$\,$\sim$\,$
10^{12.6}L_\odot$, which implies that at high-$z$, the AGN-starburst
transitional luminosity must be higher than $L_{IR}$\,$\sim$\,$
10^{12.6}L_\odot$.


\subsection{Stacked Mid-IR Spectra \label{sec:stackspec}}

How do the relative PAH strength of high-$z$ LIRGs and ULIRGs compare
with those of local infrared galaxies?  As shown in Figures~\ref{spec2}
and \ref{spec}, our individual spectra are very similar to the
averaged $z\sim 0$ starburst spectrum \citep{brandl06}.  We note that
our spectra are typically not deep enough to individually probe the silicate
absorption feature at 9.7\um.  In this section, we make this
comparison by producing stacked spectra at $z$ of 1 and 2 redshift
bins.
This technique has been widely used in recent years to detect and
quantify certain weak features which are not detected in the data of
individual sources.

\begin{figure*}[!th]
\centering{\includegraphics[width=1.9\columnwidth]{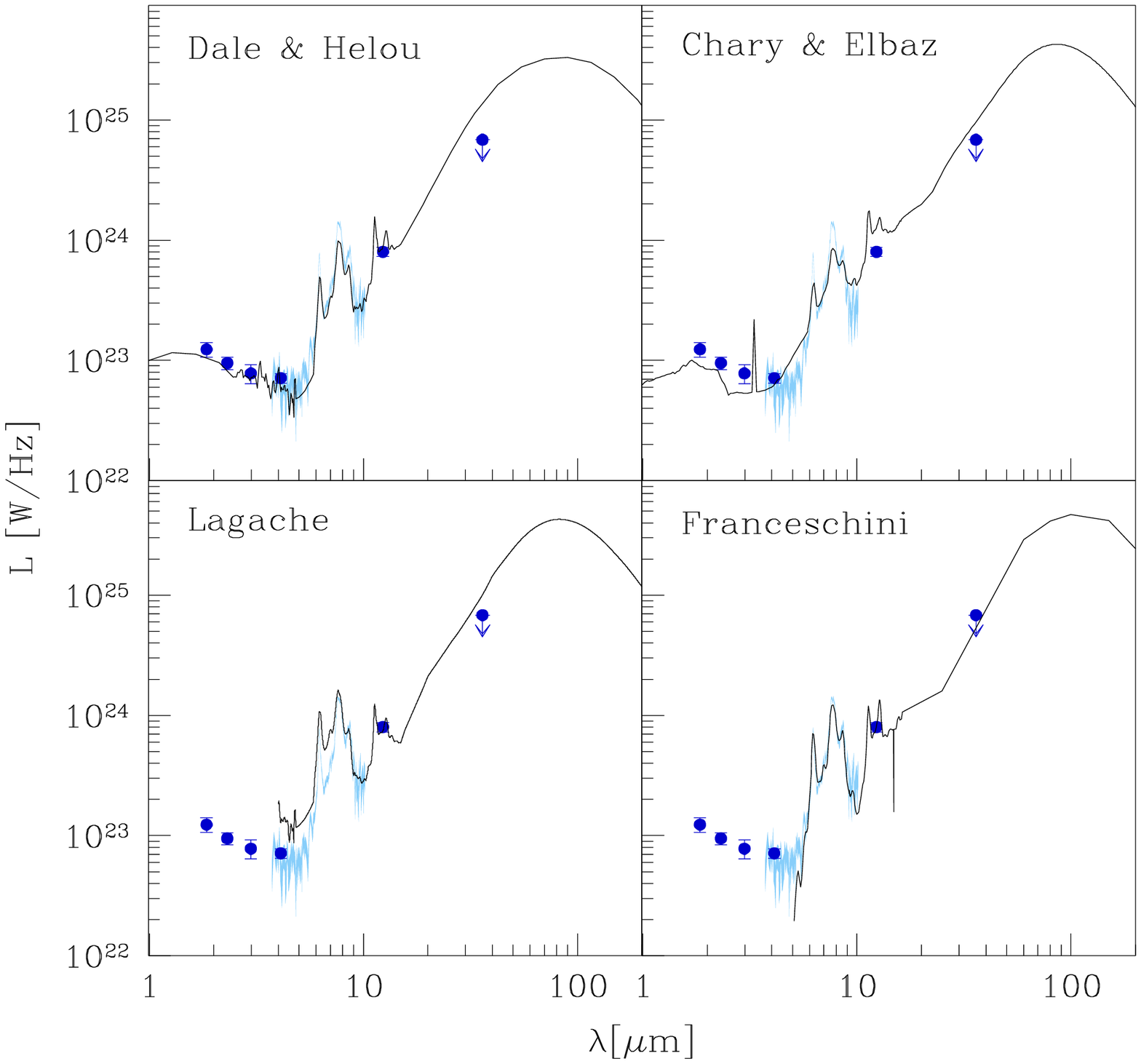}}
\caption{
\footnotesize
Composite LIRG spectrum and IRAC and MIPS fluxes compared to four popular
theoretical templates \citep{franceschini09, lagache04, dale02, chary01}.
\label{lirg_templates}}
\end{figure*}
\begin{figure*}[!t]
\centering{\includegraphics[width=1.9\columnwidth]{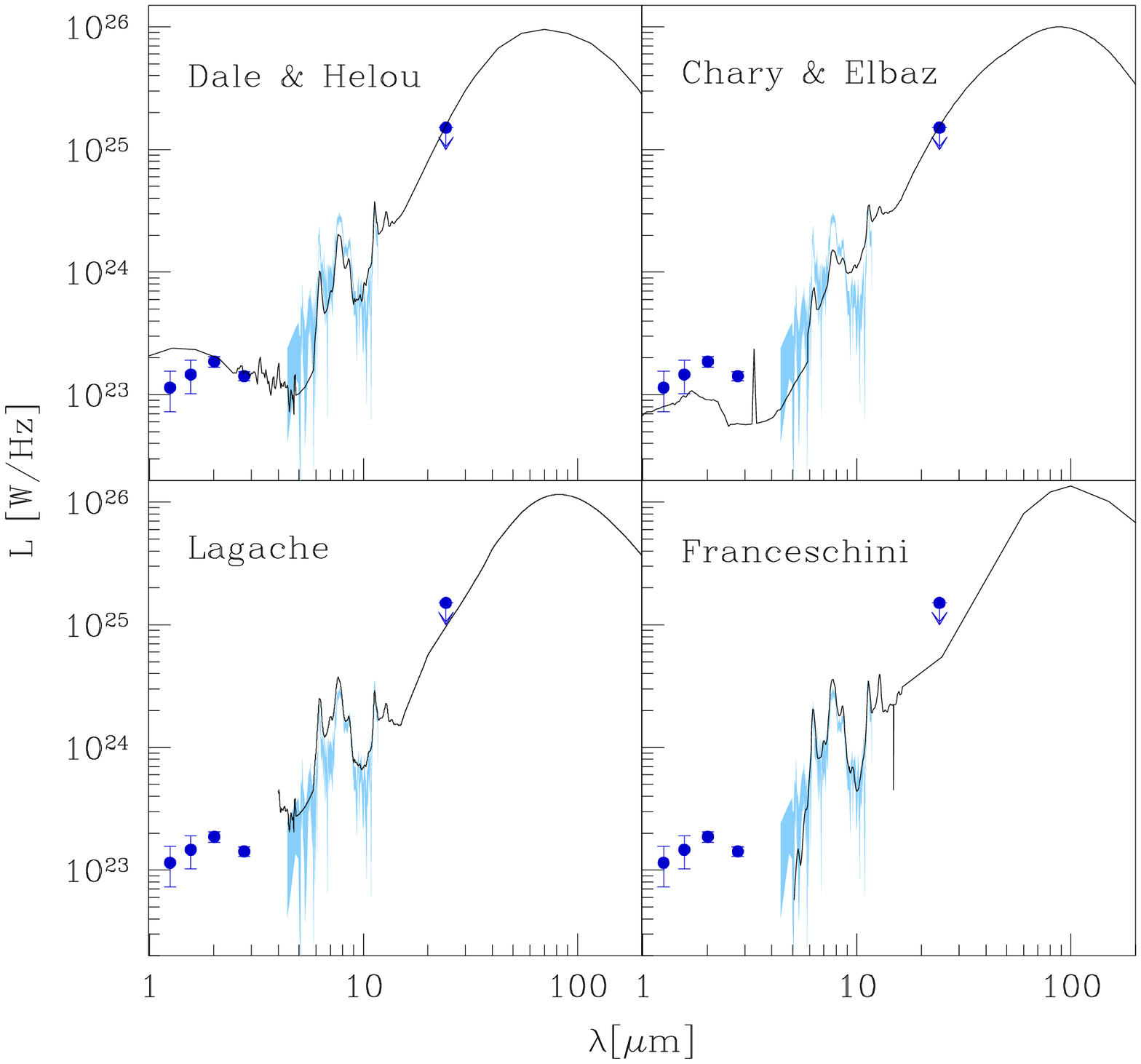}}
\caption{
\footnotesize
Composite ULIRG spectrum and IRAC and MIPS fluxes compared to four popular
theoretical templates \citep{franceschini09, lagache04, dale02, chary01}.
\label{ulirg_templates}}
\end{figure*}

\begin{figure*}[!t]
\centering{\includegraphics[width=1.9\columnwidth]{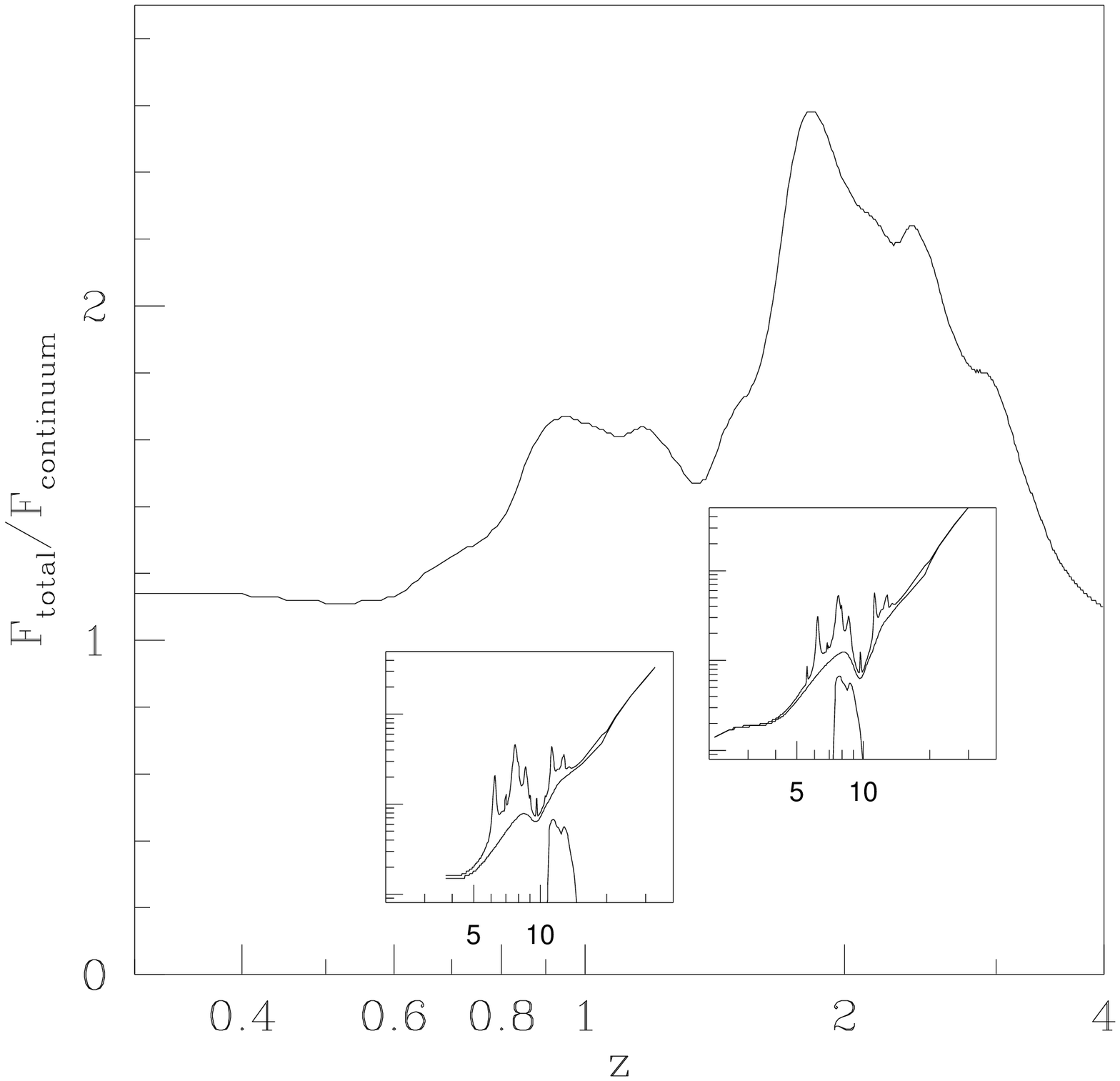}}
\caption{
\footnotesize
Fraction of the total over continuum 24\um\ flux for luminous galaxies in the range
between redshift 0 and 4, based on our composite LIRG and ULIRG spectra.
The presence of PAH features boosts the detection of 24\um\ sources around
$z$\,$\sim$\,0.9 and $z$\,$\sim$\,1.8, the median redshifts of our two subsamples.
The insets show the decomposition of the two composite spectra in PAH plus continuum
and the locations of the 24\um\ filter at the two sensitivity peaks.
\label{pah_boost}}
\end{figure*}

We exclude from the stacking the few sources whose emission is
AGN-dominated: three ULIRGs (U4958, U4950, U5877) and one LIRG
(L3945), as explained in \S~\ref{sec:agnfrac}.  We choose to
normalize the spectra to their 24\um\ rest-frame luminosity which
correlates with their total infrared luminosity \citep{rieke09}.
Here, the rest-frame 24\um\ luminosity is calculated differently for
the two samples. For $z$\,$\sim$\,1 LIRGs, we simply interpolated the
8, 24, and 70\um\ fluxes in the $\log(\lambda)$ versus
$\log(\nu\ F_{\nu}$) plot. We do not consider the 16\um\ flux since it
is severely affected by the 7.7\um\ PAH complex.  In the case of
$z$\,$\sim$\,2 ULIRGs, the 70\um\ flux is very close to the rest-frame
24\um\ flux. Unfortunately, for many ULIRGs we have only upper
limits of the 70\um\ fluxes. In these cases, we estimate the
rest-frame 24\um\ fluxes by scaling a series of templates, including
the averaged starburst from \citet{brandl06} and the composite ULIRGs
from \citet{veilleux09}, to fit the spectrum. The best fitting 
template compatible with the 70\um\ flux or upper limit is then considered
to estimate the rest-frame 24\um\ flux. The rescaled templates are
shown in Figure~\ref{fig:sed}.  In
Table~\ref{targets_data}, we list the estimated total infrared
luminosities.

The result of the stacking is shown in Figure~\ref{stacking}.  We have
resampled the original spectra on a common wavelength grid as close as
possible to the original data. At each wavelength, we computed the 
average values weighted by the errors on the single spectra. To check
if outliers create artificial strong features, we also obtained
spectra by using the biweight mean and we did find only negligible 
variations between the two methods.

The averaged spectra show clearly the characteristics of starburst
dominated galaxies with weak dust continuum at 4\,--\,14\um\ and
strong PAH emission at 6.2,7.7,8.6, and 11.3\um. Weak PAH features at
5.25 and 5.70\um\ are also visible, especially in the LIRG spectrum.
For comparison, we report in Figure~\ref{stacking} also local average
spectra normalized to our data (by minimizing the $\chi^2$ of the
residuals).  The average LIRG is surprisingly close to the average
starburst spectrum from \citet{brandl06}.  In the case of the average
ULIRG, we compare it to the average starburst spectrum from
\citet{brandl06} and to several average local ULIRG spectra from
\citet{veilleux09}. The best fit among the \citet{veilleux09} spectra
is obtained using the H II-like ULIRG average spectrum.  Although the
average starburst spectrum fits well to the relative PAH strength in
the averaged spectrum of $z$\,$\sim$\,2 ULIRG, the local average ULIRG
reproduces better the 9.7\um\ silicate absorption.  This implies that
high-$z$ ULIRGs have more dust extinction than lower luminosity LIRGs.

\subsection{Comparison with Empirical IR SED Templates}

Empirical IR SED templates used to model counts from deep surveys and
the cosmic infrared background are mostly based on local galaxies. One
important question is how these empirical SEDs compare with our
observed SEDs, including broad band photometry and spectra.  Here we
considered all the spectra which are not dominated by AGN emission and
obtained composite spectra by weighting the individual spectra by
their total infrared luminosities. The same procedure has been applied
to the IRAC and MIPS fluxes of the sources.  The composite SED plus
IRS spectra are compared to four different models, among the most
popular, in Figures~\ref{lirg_templates} and \ref{ulirg_templates}.  For each
model, we chose the one with closest total infrared luminosity and
rescaled it to the total infrared luminosity of the composite
spectrum.  In the case of the \citet{dale02} models, which give spectra as
a function of the $F_{60}/F_{100}$ color, we used the relationship
between this color and the total infrared luminosity in \citet{chapin09}.

Two models extend down to near-IR fluxes \citep{dale02, chary01},
while the other two \citep{lagache04, franceschini09} are limited to
wavelengths longer than 4\um.  It is interesting to note that of the
four empirical SED template models, only the Franceschini model
templates fit the mid-IR spectral features, although all four models
more or less in agreement with the overall observed SEDs for both
LIRGs and ULIRGs.  However, the  few observational
constraints in the far-IR part are
consistent with all four templates.  For LIRGs, only the
\citet{franceschini09} model SED template agrees with the observed
spectral features and is able to reproduce the far-IR limit for the
observed LIRG composite.

Using the composite spectra plus the IRAC and MIPS fluxes, we tried
also to estimate how much the presence of PAH features boosts the detection of
infrared galaxies in the 24\um\ filter as a function of redshift.  As
shown in Fig.~\ref{pah_boost}, we fitted the composite LIRG and
ULIRG with PAHFIT to estimate the continuum. Then, we computed the
ratio of the flux in the 24\um\ filter for the total spectrum and
for the continuum only. The ratio of these two fluxes has two clear
peaks approximately at $z$\,=\,0.9 and $z$\,=\,1.8 which corresponds to the
redshifts where the 24\um\ filter includes the two main PAH complexes
(11.3 and 7.7\um).  For comparison, the median redshifts of our two
samples are 0.95 and 1.87.  Figure~\ref{pah_boost} suggests that at
$z$\,$\sim$\,2, 24\um\ broad band filter selection will preferentially
select galaxies with strong PAH emission rather than power-law
sources, and this bias can be as high as a factor of 2.

\subsection{6.2$\mu$m equivalent width versus far-IR luminosity
\label{sec:starformation}}

In this section, we explore the relationship between two quantities
which are mainly linked to star formation: the rest-frame
24\um\ monochromatic luminosity $(\nu L_{\nu})_{24\mu m}$ and the
equivalent width of the 6.2\um\ PAH feature.  The rest-frame $(\nu
L_{\nu})_{24\mu m}$ is related to the total infrared luminosity
$L_{IR} = L_{8-1000\mu m}$ \citep{rieke09} which is an indicator of
star formation.  On the other hand, PAH features are created in
star-forming regions and destroyed around AGN, due to their intense
ionization fields ( see, e.g., \citet{voit92, lutz98}). We
expect, therefore, that in presence of an AGN component, $(\nu
L_{\nu})_{ 24\mu m}$ is proportional to the sum of the AGN and
starburst luminosities.  On the other side, $EW_{6.2\mu m}$ is the
ratio between the strength of the line (due only to star-formation
activity) and the underlying continuum (created by AGN and
star-formation activity). These two quantities should behave like:
\begin{eqnarray}
\label{eq:agn}
(\nu L_{\nu})_{ 24\mu m}  \propto (L_{AGN}+L_{SB}),\\
EW_{6.2\mu m} \propto \frac{L_{SB}}{L_{AGN}+L_{SB}},
\end{eqnarray}
where $L_{AGN}$ and $L_{SB}$ are the AGN and starburst luminosities, respectively.

\begin{figure*}[!t]
\centering{
\includegraphics[width=1.9\columnwidth]{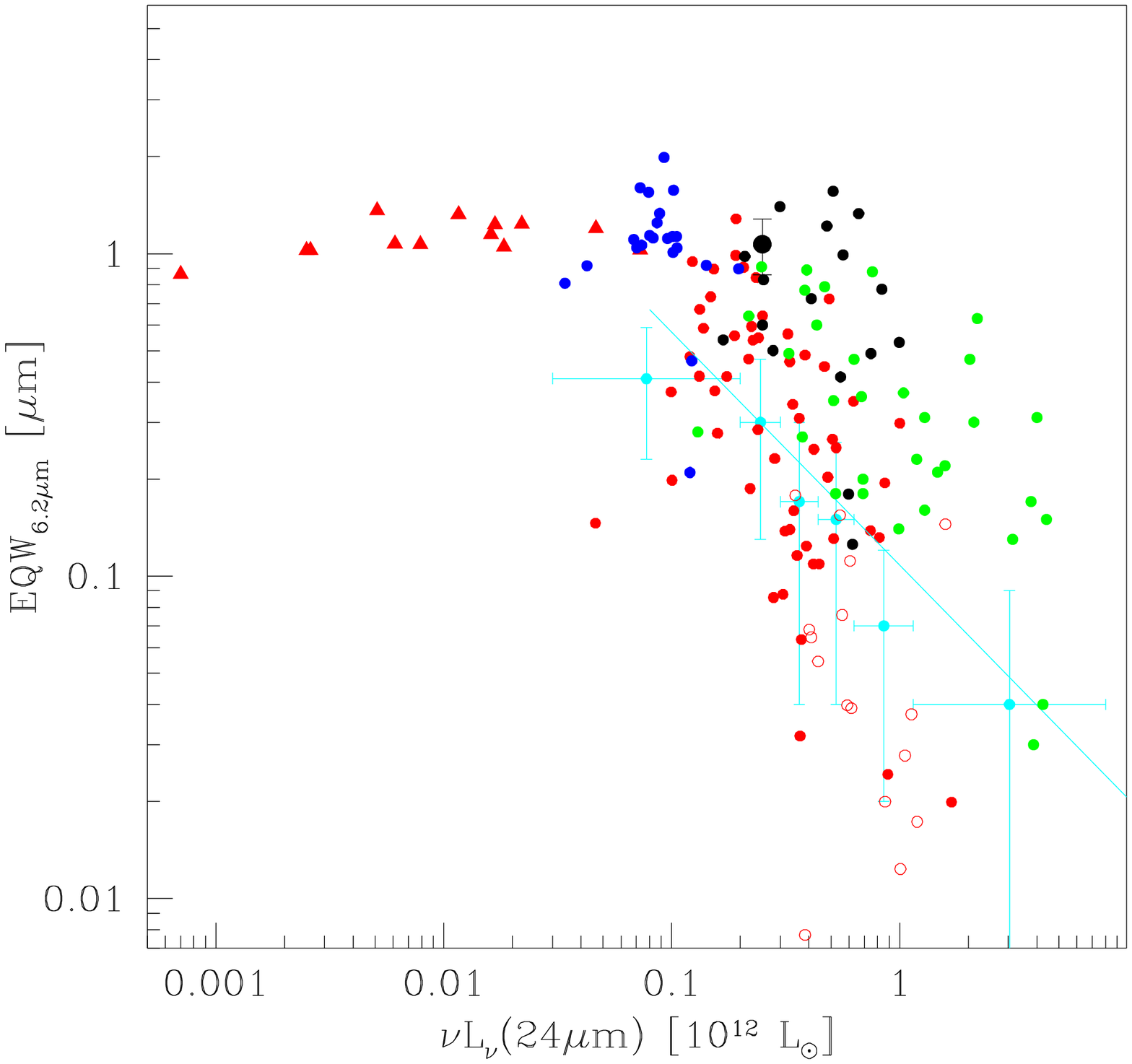}
}
\caption{ \footnotesize Equivalent width of the 6.2\um\ PAH feature
  versus the rest frame luminosity at 24\um for galaxies of our
sample ($z$\,$\sim$\,1 and $z$\,$\sim$\,2 galaxies are marked with blue and
black dots, respectively) compared to local and other high-z galaxies. We note that we re-measure $EW_{6.2\mu m}$ and $L_{24\mu m}$ for all objects in this figure. 
The black dot with error bars is  for the average $z$\,$\sim$\,2 ULIRG.
The red triangles correspond to local starbursts from \citet{brandl06},
while the red circles are ULIRGs from the local 1\,Jy
  sample by \citet{veilleux09}. The empty circles are classified as
at least 50\% AGN dominated by \citet{veilleux09}. Finally, the green
dots are high-redshift infrared sources from literature
\citep{sajina07,menendez07,pope08,valiante09}.
All the equivalent widths have been remeasured in the same way as 
the one from our sample. Cyan points and line are from \citet{desai07}. 
\label{eqw}}
\end{figure*}

In Figure~\ref{eqw}, we report the 6.2\um\ equivalent widths measured
for the $z$\,$\sim$\,1 LIRGs (blue dots) and ULIRG at $z$\,$\sim$\,2
(black dots) as a function of their rest-frame 24\um\ luminosities. We
also plot the $EW_{6.2\mu m}$ of the composite ULIRG (black dot with
errorbar).  The stacked ULIRG spectrum has a high enough
signal-to-noise ratio to directly measure its mid-IR continuum and
therefore obtain an average $EW_{6.2\mu m}$.

In Figure~\ref{eqw}, we compare local starbursts \citep{brandl06},
1\,Jy ULIRGs \citep{veilleux09} with a collection of high-$z$ \spitz
ULIRGs and SMGs \citep{sajina07,menendez07,pope08,valiante09}.  We
report also the average points from \citet{desai07} computed from
local galaxies.  To make this comparison meaningful, we re-measured
$EW_{6.2\mu m}$ and $L_{24\mu m}$ for all of the sources included in
Figure~\ref{eqw}.  We measured the $EW_{6.2\mu m}$ using our revised
version of PAHFIT (see \S~\ref{sec:agnfrac}).  The values of
$EW_{6.2\mu m}$ for highly extincted ULIRGs agree very well with those
computed by \citet{veilleux09}. On the contrary, for less extincted
galaxies we have systematically lower values ($\sim$~70\%) probably
due to the fact that \citet{veilleux09} do not fit directly the lines
but uses templates to fit the spectra whose actual relative PAH
strength may differ from that of templates.
\begin{figure*}[!t]
\centering{\includegraphics[width=1.8\columnwidth]{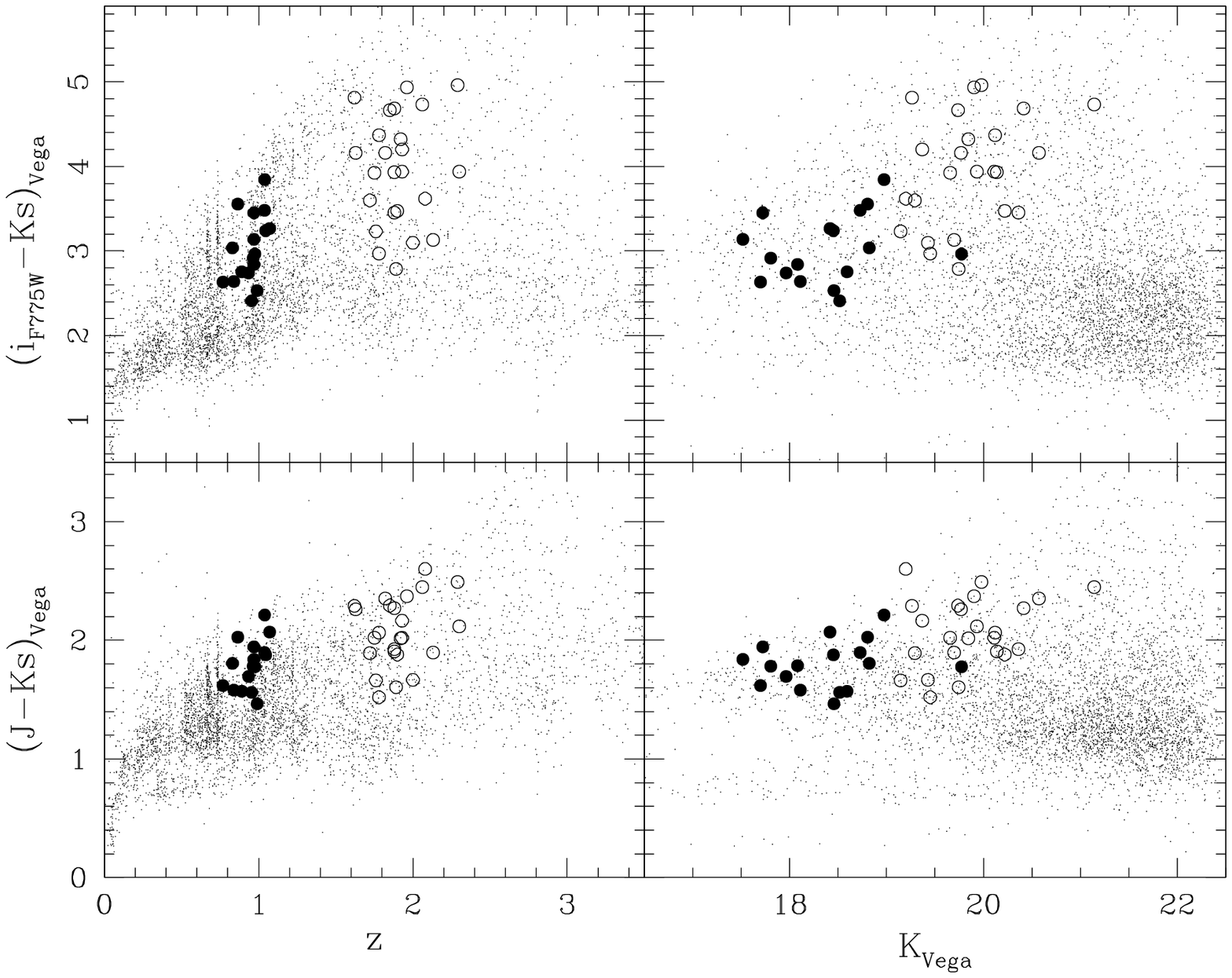}}
\caption{
\footnotesize
Color-color plots for galaxies in the GOODS field (dots) and galaxies of our
sample. Full and empty circles mark the LIRGs and ULIRGs of our sample, respectively.
\label{colordiag}}
\end{figure*}

One striking result shown in Figure~\ref{eqw} is the maximum
$EW_{6.2\mu m}$ value, $\sim$\,1\um, an upper limit for all starburts, LIRGs
and ULIRGs, independent of redshifts.  This maximum limit is expected
from Equation~\ref{eq:agn}, indicating that when
$L_{AGN}$\,$<<$\,$L_{SB}$, $EW_{6.2\mu  m}$ should be constant. We
have no explanation why this constant is $\sim$1\um.  
At $L_{24\mu m}$\,$\simgt$\,$10^{11}L_\odot$, $EW_{6.2\mu
  m}$ and $L_{24\mu m}$ are clearly anti-correlated with a slightly
different slope for local and high-redshift sources. The relation is
broad not only because of errors in the measurement of the $EW_{6.2\mu
  m}$ but also due to AGN contributions that dilute the
$EW_{6.2\mu m}$ intrinsically broadening the correlation. As a matter
of fact, the dispersion of pure local starbursts (for which
the AGN contamination is negligible) is much smaller than that of the
local ULIRGs (that are contaminated by AGN).

Among high luminosities sources ($L_{24\mu m}\, > 10^{11}L_{\sun}$),
at a given luminosity, high-redshift sources have PAH emission
stronger than that of local ULIRGs.  This can be interpreted in at
least two different ways.  Either high-redshift sources are forming
stars more efficiently than local counterparts or local ULIRGs have a
higher AGN contribution to their infrared continua. Supporting
evidence for the first hypothesis is that SMGs at $z$\,$\sim$\,2,
which are the majority of the high-z luminous galaxies in our plot,
have more gas, higher star formation rates and more efficient star
formation than local ULIRGs \citep{greve05,tacconi06}. On the other
hand, if the AGN contribution to the mid-IR continuum is smaller for
$z$\,$\simgt$\,1 ULIRGs than that for local ULIRGs at a fixed
luminosity, high-$z$ ULIRGs would have bigger $EW_{6.2\mu m}$ due to
less AGN boosting of the mid-IR continua.  The good fit of the
composed spectra of our sample with the averaged starburst template
and the low estimated AGN contribution to their total emission (see
also next section) tend to support this last hypothesis.
\begin{figure*}[!t]
\includegraphics[width=1.1\columnwidth]{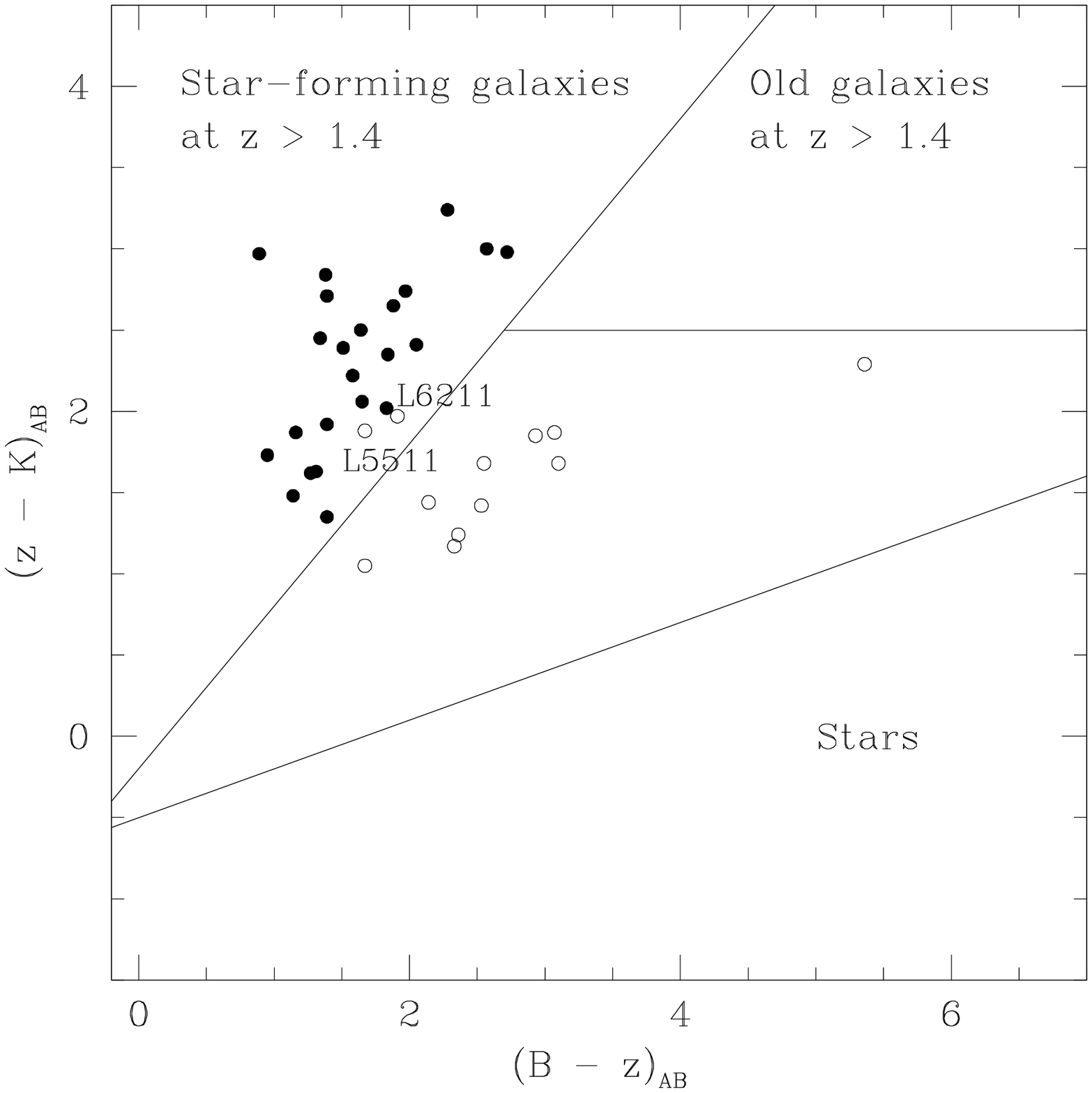}
\includegraphics[width=1.1\columnwidth]{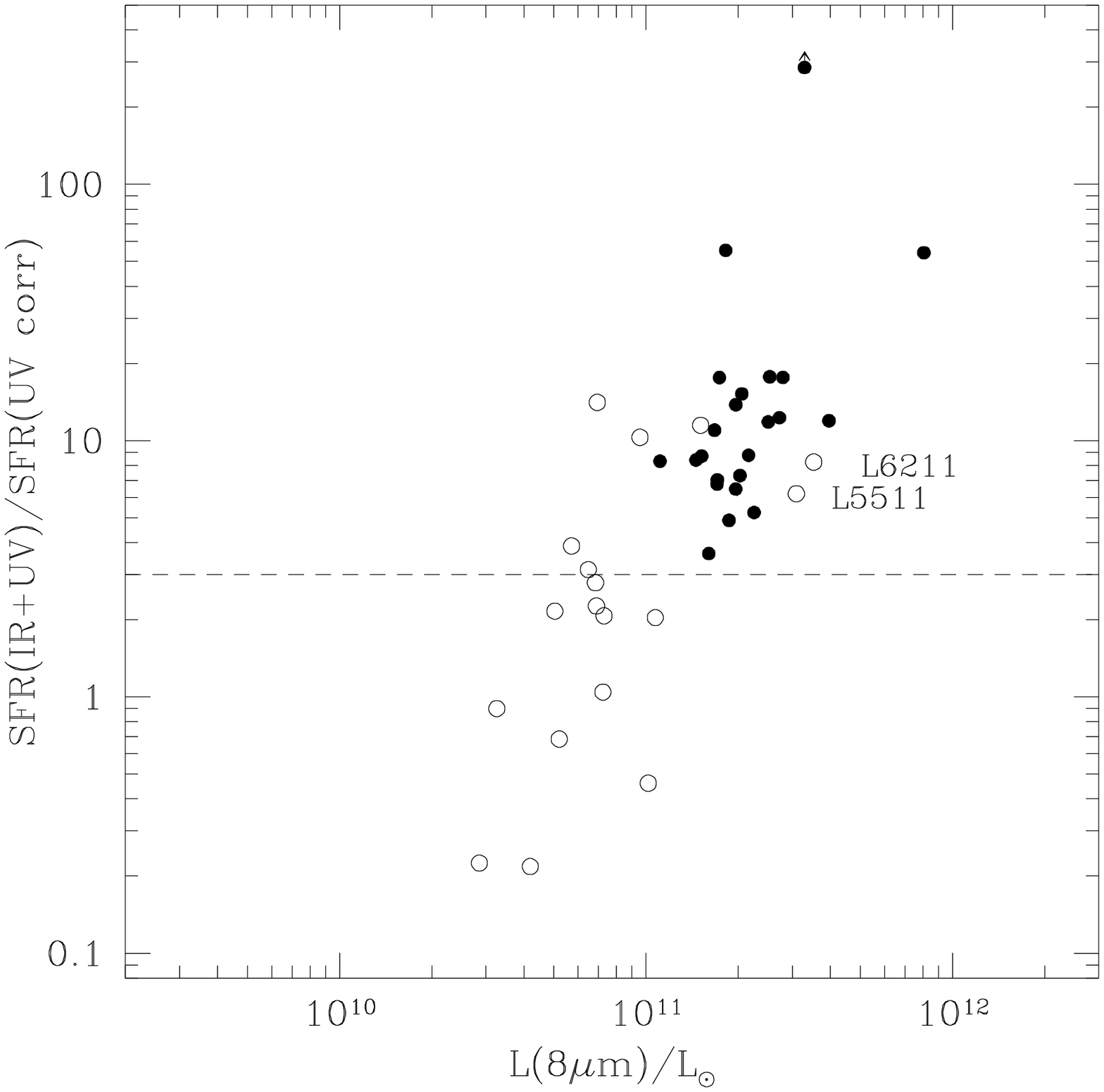}
\caption{ \footnotesize {\it Left:} optical/near-IR color plot; {\it Right:} IR
  excess vs rest-frame 8$\mu$m luminosity. In the two figures, sources
  from the ULIRG and LIRG samples are marked with filled and empty
  circles, respectively. The names indicate the two sources from the
  LIRG sample which are actually at redshift $\sim$\,2.
\label{bzk}}
\end{figure*}

\subsection{Red Optical Colors\label{sec:redcolor}}

Many of our sources have very red optical SEDs.  As shown in
Figure~\ref{colordiag}, with definitions for Extremely Red Galaxies
(ERG) and Distant Red Galaxies (DRG) of $(i-K)_{\rm Vega}$\,$\simgt$\,4.0
and $(J-K)_{\rm Vega}$\,$\simgt$\,2.3, respectively
\citep{yan00,yan05,franx03,vandokkum04}, we find that of the 24
$z$\,$\sim$\,2 ULIRGs, 63\%\ (15/24) are ERGs and 33\%\ (8/24) are DRGs.
These near-IR color selected DRGs have very faint optical magnitudes,
and their redshifts are difficult to obtain with optical/near-IR
spectroscopy \citep{kriek08}.  Our IRS spectra have provided 8
spectroscopic redshifts which do not have optical spec-$z$ and have been used 
for the analysis of the reliability of
DRG color selections \citep{wuyts09}.  Our finding of 8 $z$\,$\sim$2 ULIRGs having very red $(J-K)$ colors suggest that a significant fraction of DRGs selected by K-band surveys are dusty star forming galaxies, not passively evolving stellar populations at $z$\,$\sim$\,2\,--\,3.  This is consistent with previously published studies, which found as much as 50\%\ of color selected DRGs being dusty starbursts \citep{wuyts07,labbe05,kriek08,wuyts09}.


Furthermore, all of the $z\sim2$ ULIRGs also satisfy the $(z-B)$
vs. $(K-z)$ color-color selections (BzK) \citep{daddi04}, as shown in
Figure~\ref{bzk}. These BzK color cuts shown as lines in
Figure~\ref{bzk} have been used to select star forming and passively
evolved galaxies at $z$\,$\simgt$\,1.4 \citep{daddi04}.  These $z$\,$\sim$\,2
ULIRGs fall into the star forming BzK region on the (z-B) vs. (K-z)
color-color diagram.  As expected, their red BzK colors are mostly due
to the dust absorption the UV photons from O, B types of young stars
produced by recent starbursts.

\begin{figure*}[!t]
\centering{\includegraphics[width=1.8\columnwidth]{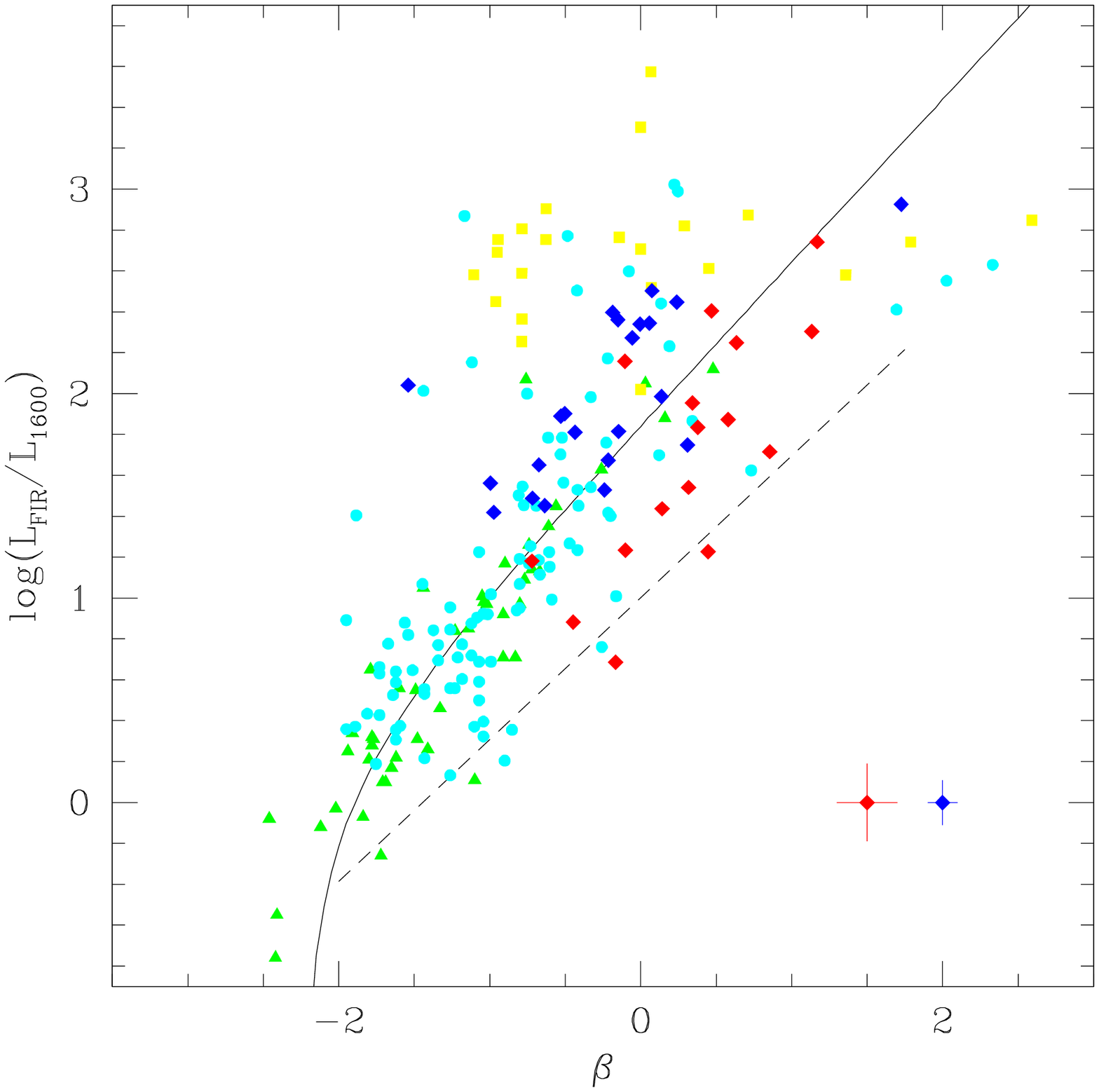}}
\caption{ \footnotesize Ratio of far-IR to UV flux at 1600{\rm \,\AA}
  compared to the UV spectral slope $\beta$ for our galaxies and some
  published comparison galaxies.  The solid line shows the linear fit
  to the A$_{1600{\rm \,\AA}}$-$\beta$ relationship by
  \citet{meurer99}.  The dashed line show the relation for normal
  star-forming galaxies \citep{cortese06}.  The samples from
  \citet{meurer99, reddy06}, and \citet{chapman05} are marked with
  green triangles, cyan dots, and yellow squares, respectively.  Our
  $z\sim 1$ and $z\sim 2$ sample are marked with red and blue
  diamonds, respectively. The typical error of the points is shown in
  the bottom right corner of the plot.  Most of our $z\sim 2$ ULIRG
  lie above the curve, in the region occupied by SMGs and HLIRGs.
\label{uvir}}
\end{figure*}

In Figure~\ref{bzk}, we report also the IR excess as a function of
$L_{8\mu m}$ computed as in \citet{daddi04}. All our $z\sim 2$ ULIRGs
and a few LIRGs are IR excess galaxies according to the definition of
\citet{daddi04}.  Although it is a known fact that optical and IR
surveys select galaxies with different amounts of dust, the question here
is if our IR selected galaxies follow the same dust extinction scaling
relationship as optically selected objects with low dust content.
Figure~\ref{uvir} shows the far-infrared to
1600\AA\ monochromatic luminosity ratio as a function of UV spectral
index $\beta$, where $f_{\lambda}$\,$\sim$\,$\lambda^\beta$,
1000\AA\,$<$\,$ \lambda$\,$<$\,2500\AA.  The ratio
$L_{FIR}/L_{1600\AA}$ measures how much energy is re-distributed from
UV to infrared, and $\beta$ measures the UV SED slope, {\it i..e} UV
reddening.  Studies of local UV selected starbursts have shown that
dust absorption is correlated with UV reddening
\citep{meurer95,meurer99}.  This correlation provides a powerful
empirical tool, particularly for high-$z$ UV/optical surveys, to
recover the total, dust-absorption corrected, UV fluxes, using UV
photometry alone.  Figure~\ref{uvir} compares the location of our
sample with those of local galaxies \citep{meurer99} and high-redshift
galaxies \citep{chapman05,reddy06} in the plot $L_{FIR}/L_{1600\AA}$
versus $\beta$. In the case of our sample, we used the available
optical magnitudes from U to R to compute $L_{1600\AA}$ by
interpolating in $\nu F_{\nu}$, while $\beta$ was computed using the
relationships from \citet{papovich06}.  The $L_{FIR}$ is computed from
the rest-frame 24$\mu$m luminosity using the relationship from
\citet{rieke09} to compute the total IR luminosity and from
\citet{elbaz02}, to pass from total IR luminosity to far-IR
luminosity.  In the case of SMG galaxies from \citet{chapman05}, we
used the available B and R magnitudes to compute $\beta$ and
$L_{1600\AA}$. The relationship from \citet{papovich06} were used
after applying median corrections to the B and R magnitudes to
transform them in the appropriate bands.  For the $z$\,$\sim$\,1
sample, the UV luminosity and $\beta$ are poorly constrained since we
do not have UV observations. These
 sources appear to scatter 
between the normal star-forming galaxies and the UV-selected starbursts 
relations, which as discussed above can be attributed to a range of UV extinction 
slopes.
For our $z$\,$\sim$\,2 ULIRGs, we note that they largely lie above the
Meurer's relationship, although usually not quite so far as the SMGs
and HLIRGs (hyper-luminous IR galaxies).
Overall, the large scatter in this diagram, and the strong dependence on the 
UV-slope of the unknown extinction curve serve as caveats in applying standard 
UV dust-corrections. \citet{boquien09} caution that applying the wrong extinction law 
to the UV would results in over- or under- estimating the true SFR by factors of up to $\sim$\,7. 
We return to this point in the following section. 

\subsection{``IR Excess'' and Obscured AGN Luminosity
\label{sec:agnlum}}

As discussed in \S~\ref{sec:redcolor}, the BzK method has been used to
study 24\um\ galaxies in the GOODS field \citep{daddi07}.  This paper
found that as much as (20\,--\,30)\%\ of 1.4\,$<$\,$z$\,$<$\,2.5 BzK selected galaxies
with $\rm K_{\rm Vega}$\,$\simlt$\,22 have infrared excess, in the sense that
$\rm SFR_{IR}+SFR_{UV,uncorrected}$\,$\simgt$\,3.16\,$\times$\,
$\rm SFR_{UV,corrected}$.
Using the X-ray stacking method on these 24\um\ galaxies with ``IR
excess'', this study found that these galaxies contain highly
obscured, Compton thick AGNs, and the predicted unobscured X-ray
luminosity $L_{2-8\,keV}$\,$\sim$\,(1\,--\,4)\,$\times$\,$10^{43}$\,ergs/s.  All our
$z$\,$\sim$\,2 ULIRGs are BzK galaxies satisfying the IR excess criteria
(see Figure~\ref{bzk}).

In principle, if we know the correct relationship to estimate $L_{IR}$
from rest-frame $L_{8\mu m}$, the correct UV extinction, and if the
AGN contribution is negligible, we should get $SFR_{UV,
  corrected}$\,=\,$SFR_{IR}+SFR_{UV, uncorrected}$. The fact that all
our $z$\,$\sim$\,2 ULIRGs have ''IR excess'' could be due to either
(a) non-negligible AGN contributing to total luminosities , or (b)
wrong UV extinction correction, or (c) incorrect estimate of $L_{IR}$.
\citet{daddi07} has argued that this ``IR excess'' is primarily due to
Compton-thick AGN heating of mid-IR luminosities, causing the
over-estimation of $L_{IR}$, thus ``excess'' of $SFR_{IR}+SFR_{UV,
  uncorrected}$.  Our mid-IR spectra suggest that most $z$\,$\sim$\,2
ULIRGs (92\%) have strong PAH emission, and AGN contribution to their
$L_{IR}$ is small. However, the rest-frame $L_{8\mu m}$ estimated from
observed broad band 24\um\ photometry will be over-estimated because
of broad, strong PAH emission among starbursts. As already note by
\citet{eric09}, the relationship used by \citet{daddi07} to compute
$L_{IR}$ from $L_{8\mu m}$ over-estimates $L_{IR}$. For our ULIRGs,
the median ratio between SFR estimated from $L_{8\mu m}$ and $L_{24\mu
  m}$ is 3.1.  Lastly, another big factor causing this ``IR excess''
is the under-estimated UV dust extinction correction.  This is further
supported by Figure~\ref{uvir}, where the $z$\,$\sim$\,2 ULIRGs are
systematically offset above the local relationship defined by low
luminosity, low extinction star-forming galaxies.  It is not
surprising that UV dust extinction of heavily obscured ULIRGs is only
a lower limit because, for a mixed distribution of stars and dust,
increasing amounts of dust will further absorb stellar emission, but
the reddening of the integrated colors of a galaxy will saturate as
the most obscured spectral regions have less and less weight in the
integrated SED \citep{wuyts09b}.

Our IRS spectra illustrate that the ``IR excess'' feature in BzK
galaxies is really due to strong broad PAH emission, particularly
7.7\um\ feature, produced by recent starbursts. This excess is not due
to dust being heated by obscured, Compton thick AGNs as stated in
\citet{daddi07}.  Below, we will use the IRS spectra to qualitatively
estimate obscured AGN luminosities.
\begin{figure*}[!t]
\centering{
\includegraphics[width=2.\columnwidth]{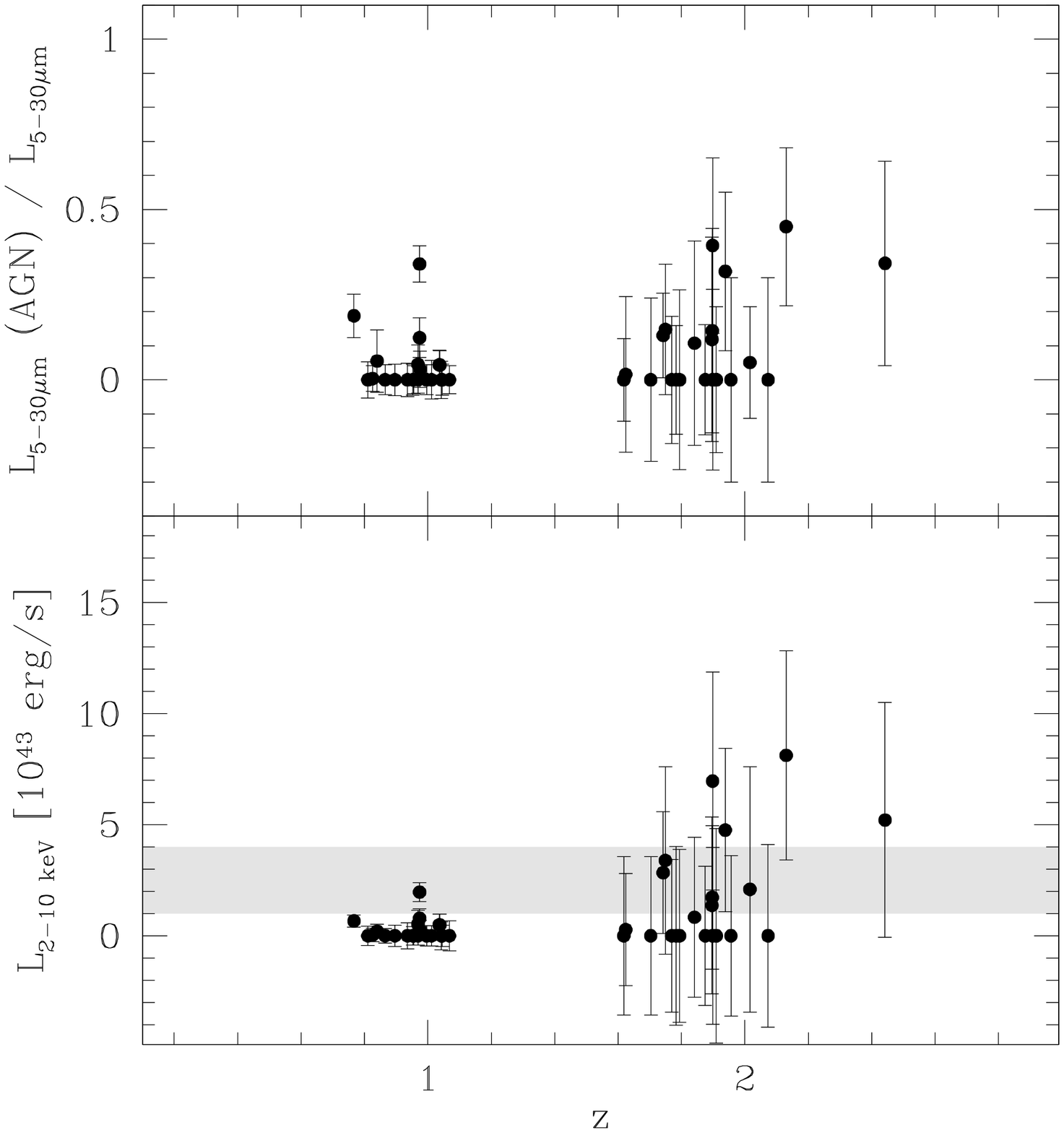}
}
\caption{ \footnotesize Fraction of the mid-infrared emission due to
  the AGN component and its X-ray luminosity (top and bottom panels,
  respectively) for the LIRGs and ULIRGs of our sample which are not
  dominated by AGN emission. The grey band marks the expected X-ray
  luminosity of obscured AGN according to
  \citet{daddi07}. \label{agnfrac}}
\end{figure*}

Many mid-IR spectroscopic studies have shown that galaxies with
obscured AGN tend to have excess mid-IR continuum at 4\,--\,6\um,
which is thought to be produced by small, hot dust grains heated by UV
photons from AGN \citep{laurent01}. There is a narrow correlation between
the AGN luminosity at 5.8\um, $\rm L_{5.8\mu m}$, and its intrinsic X-ray
luminosity $\rm L_{2-10\,keV}$ for a local samples of Seyferts and QSOs:
\begin{equation}
\log L_{5.8\mu m} = 1.209\ \log L_{2-10\,keV} - 8.667,
\end{equation}
with luminosities expressed in $erg/s$ \citep{lutz04,bauer09}.

To estimate the continuum due to the AGN emission at 5.8$\mu$m, we fit
to each spectrum a combination of a linear component plus the scaled
average starburst template from \citet{brandl06} considering for the
fit the 5.5-6.85$\mu$m interval. This direct approach provides results
consistent to the approach of \citet{lutz04} and \citet{valiante09}
who used M82 as starburst template and adjusted for differences
between M82 and an average starburst galaxy.  To compare our results
to the work of \citet{daddi07}, we follow his criteria excluding
galaxies detected in the 2\,--\,8\,keV band in the 1~Ms Chandra survey
and those without a clear decreasing in their SED after
1.6\um. Therefore, we do not consider the two ULIRGs whose mid-IR
spectra show very little PAH emission, have a power-law SED and are
clearly dominated by nuclear emission.  We also exclude U4958 since
its optical spectrum shows the presence of an AGN and the mid-IR
spectra is almost featureless.  Since several of the LIRGs of our
sample are also ``IR excess'' galaxies, we apply the same analysis to
this sample excluding three objects (L3945, L5511, L6211) which are
not in the correct redshift range or have little or no PAH emission in
their spectra.

To estimate the AGN contribution to the total mid-IR emission, we
proceed as \citet{eric09} by scaling the SED of Mrk231 to the
5.8\um\ AGN continuum and fitting the residual SED with the average
starburst from \citet{brandl06}. As shown in Figure~\ref{agnfrac}, the
emission from most of the ULIRGs and almost all LIRGs is completely
dominated by star formation. This confirms and extends the analysis of
\citet{eric09}, since our study is done on a bigger, fainter, and more
homogeneous sample.  Moreover, our analysis allows us to directly
estimate the X-ray luminosity of the AGN harbored in the galaxies of
our sample.  In the bottom panel of the same Figure~\ref{agnfrac}, we
report the estimated X-ray luminosities of the AGN component of the
LIRGs and ULIRGs of our sample compared to the predictions based on
stacked X-ray analysis from \citet{daddi07}.  Also in this case, the
majority of ULIRGs have AGN with X-ray luminosities lower than these
values. A straight average of the values found would give a value of
$L_{2-10\,keV}$\,=\,(1.8$\pm$0.6)$\times$\,$10^{43}$\,erg/s which
would be compatible with the analysis of \citet{daddi07}.  A robust
average, using a biweight estimator, gives a value of
$L_{2-10\,keV}$\,=\,(0.1$\pm$0.6)$\times$\,$10^{43}$\,erg/s.  This
means that a few galaxies with bright hidden AGN dominate the stacking
of the faint infrared sources in the aforementioned analysis,
highlighting the intrinsic danger of any stacking analysis.  We can
conclude that the infrared emission from the majority of our ULIRGs
(21/24, i.e. 88\%\ of our sample) is dominated by star formation and
the analysis of our sample of ULIRGs sets a low upper limit to the
presence of Compton thick AGN in luminous infrared galaxies at
$z$\,$\sim$\,2, with respect to the analysis of \citet{daddi07}.

\begin{figure*}[!t]
\includegraphics[width=2.2\columnwidth]{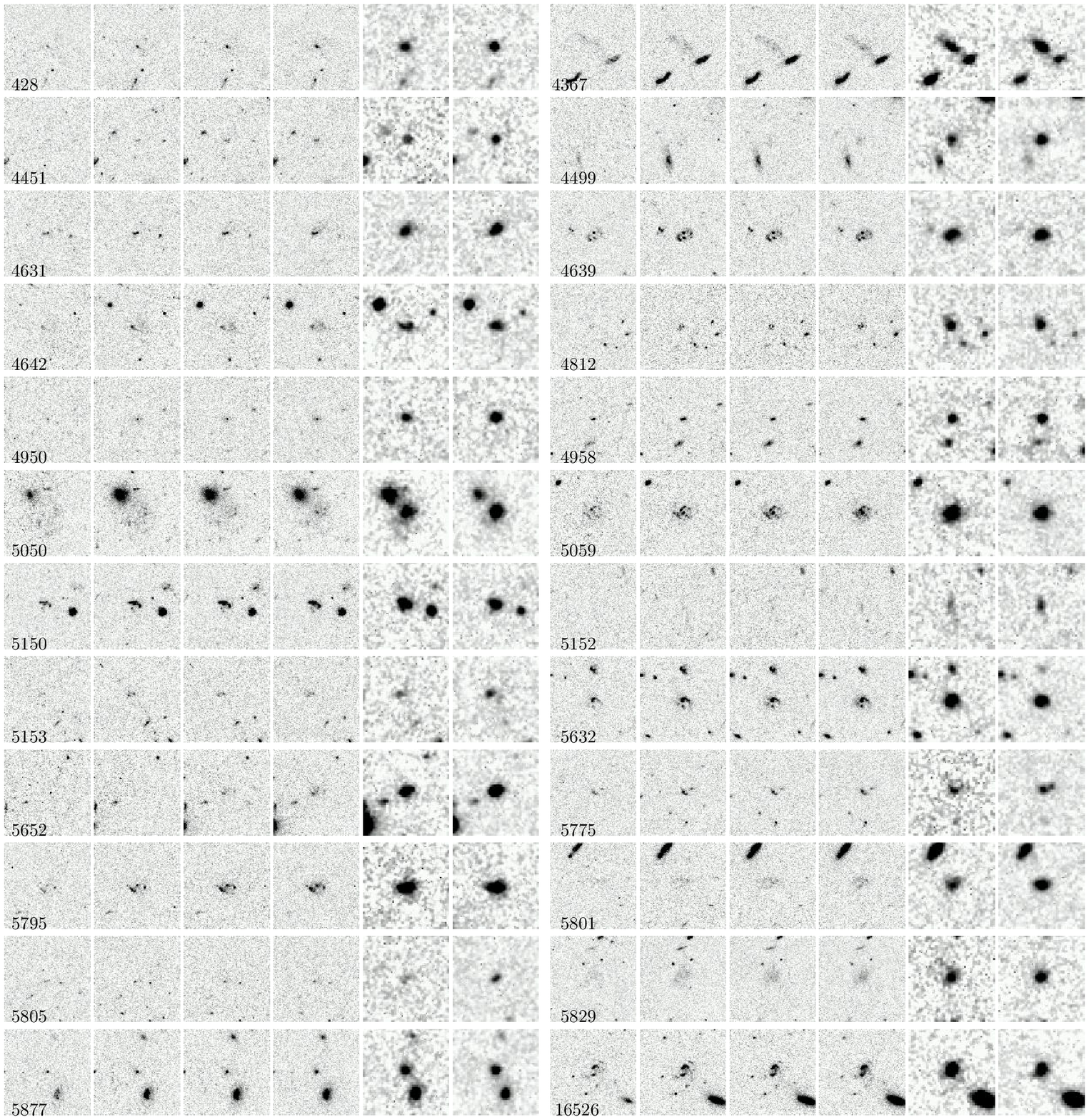}
\caption{ \footnotesize The image cutouts for each ULIRG are in the
filters F435W, F606W, F775W, F850LP from {\it HST} ACS images of the
GOODS treasury survey. $J$ and $Ks$ images are from the public EIS
survey done with ISAAC on the VLT telescopes. The object ID is
labeled on the F435W image.  Each image stamp is
$10^{''}\times10^{''}$. \label{morphu}}
\end{figure*}
\begin{figure*}[!t]
\includegraphics[width=2.2\columnwidth]{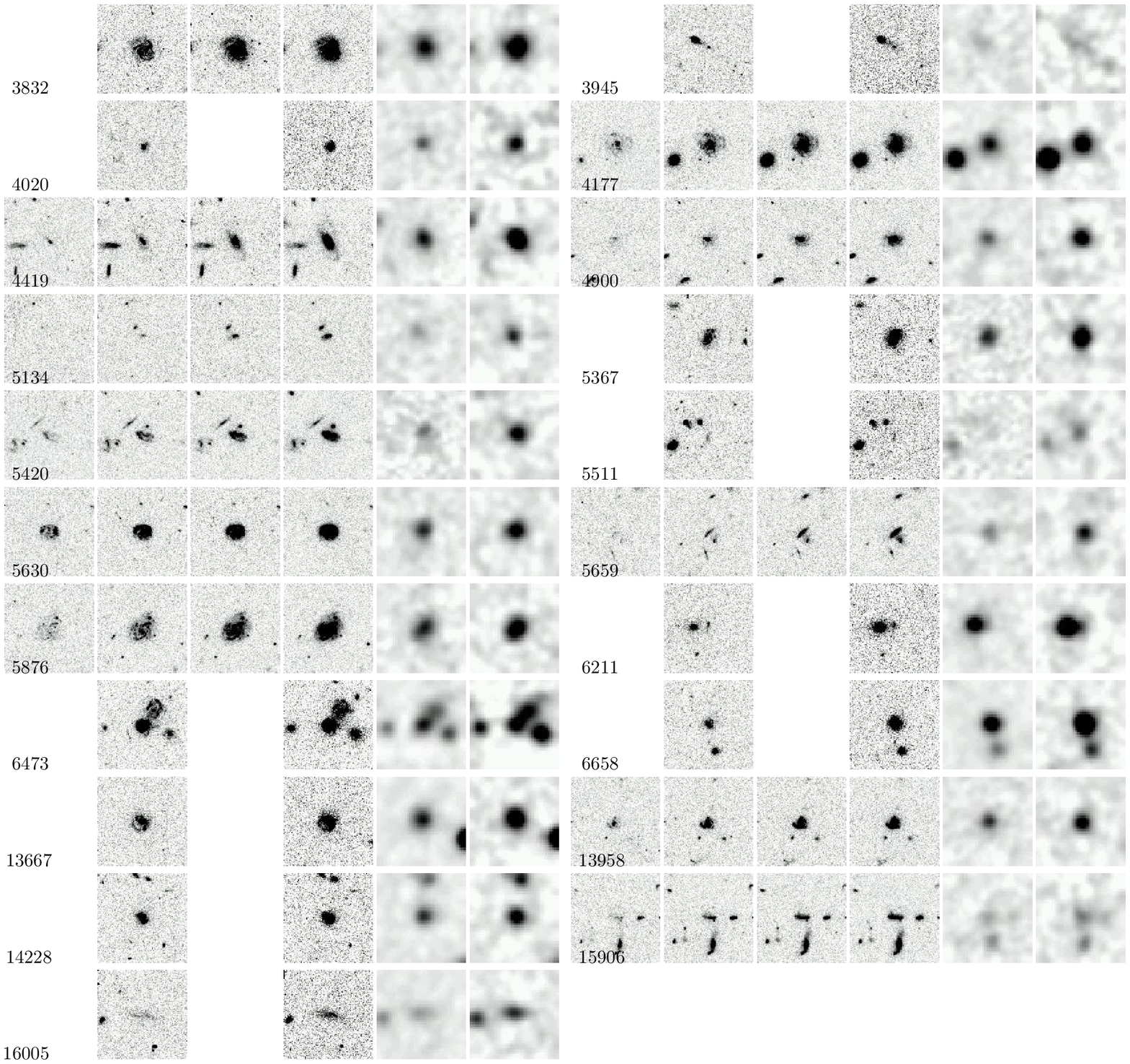}
\caption{
\footnotesize
Similar to Figure~\ref{morphu}, this plot shows the image
cutouts for 21 LIRGs at $z\sim1$.  We used all of the available
observations in HST/ACS F606W and F850LP filters for LIRGs within the GEMS 
survey, and F435W, F606W, F775W and F850LP for LIRGs within the GOODS
region. For objects in the GOODS region, $J$ and $Ks$ images are from
the EIS survey. Otherwise, they  are from the MUSYC survey
\citep{musyc06}. \label{morphl}}
\end{figure*}
\begin{figure*}[!t]
\centering{
\includegraphics[width=2.2\columnwidth]{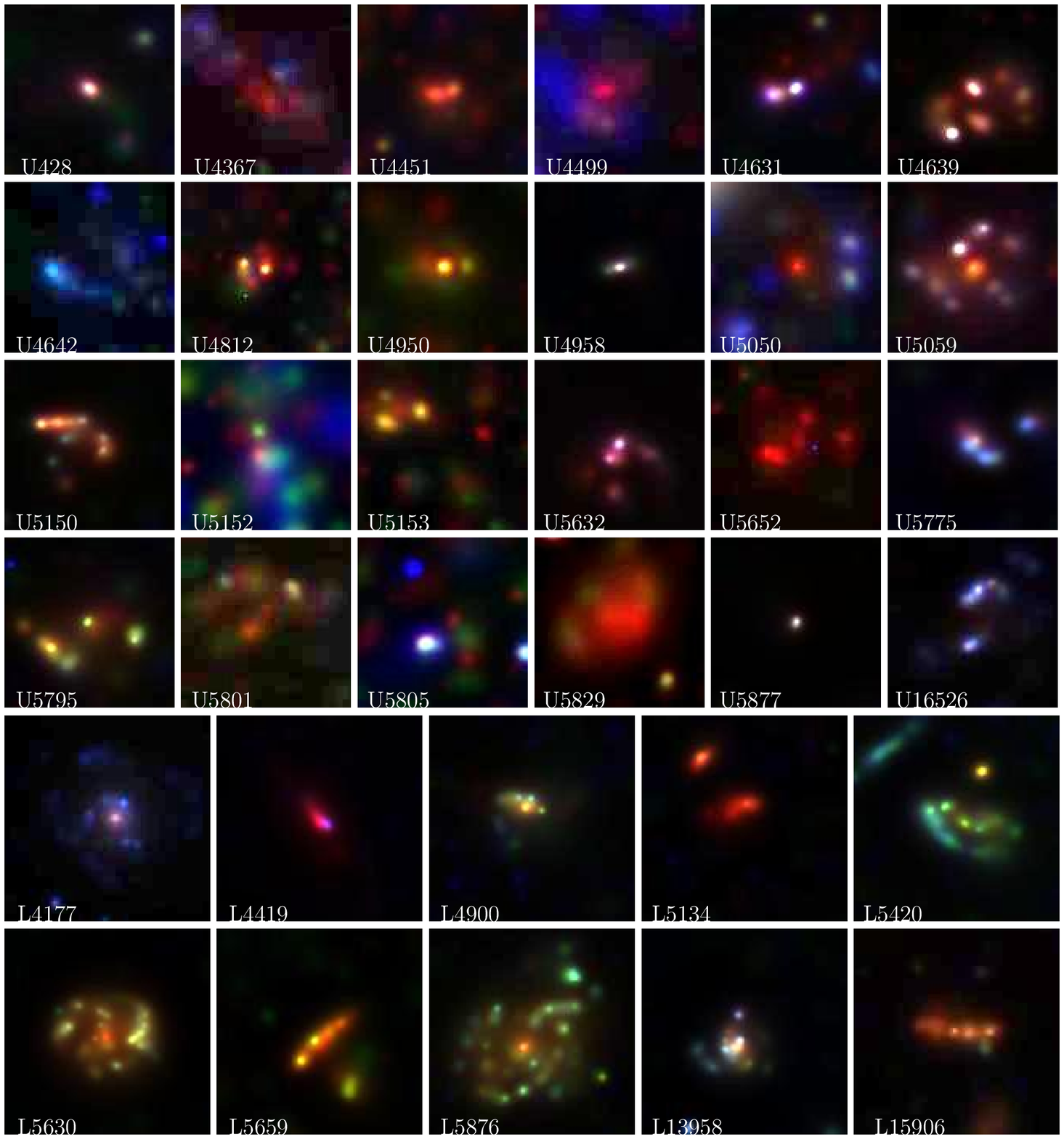}}
\caption{
\footnotesize
Three color images for 24 ULIRGs and 10 LIRGs of our sample
with F435W, F606W, and F850LP ACS coverage. The field of view
is  $3{''}\times3^{''}$ and $4^{''}\times4^{''}$ for ULIRGs
and LIRGs, respectively. 
The composed image was obtained by giving the same weight to each band.
 \label{colormorph}}
\end{figure*}
\begin{figure*}[!t]
\centering{
\includegraphics[width=1.8\columnwidth]{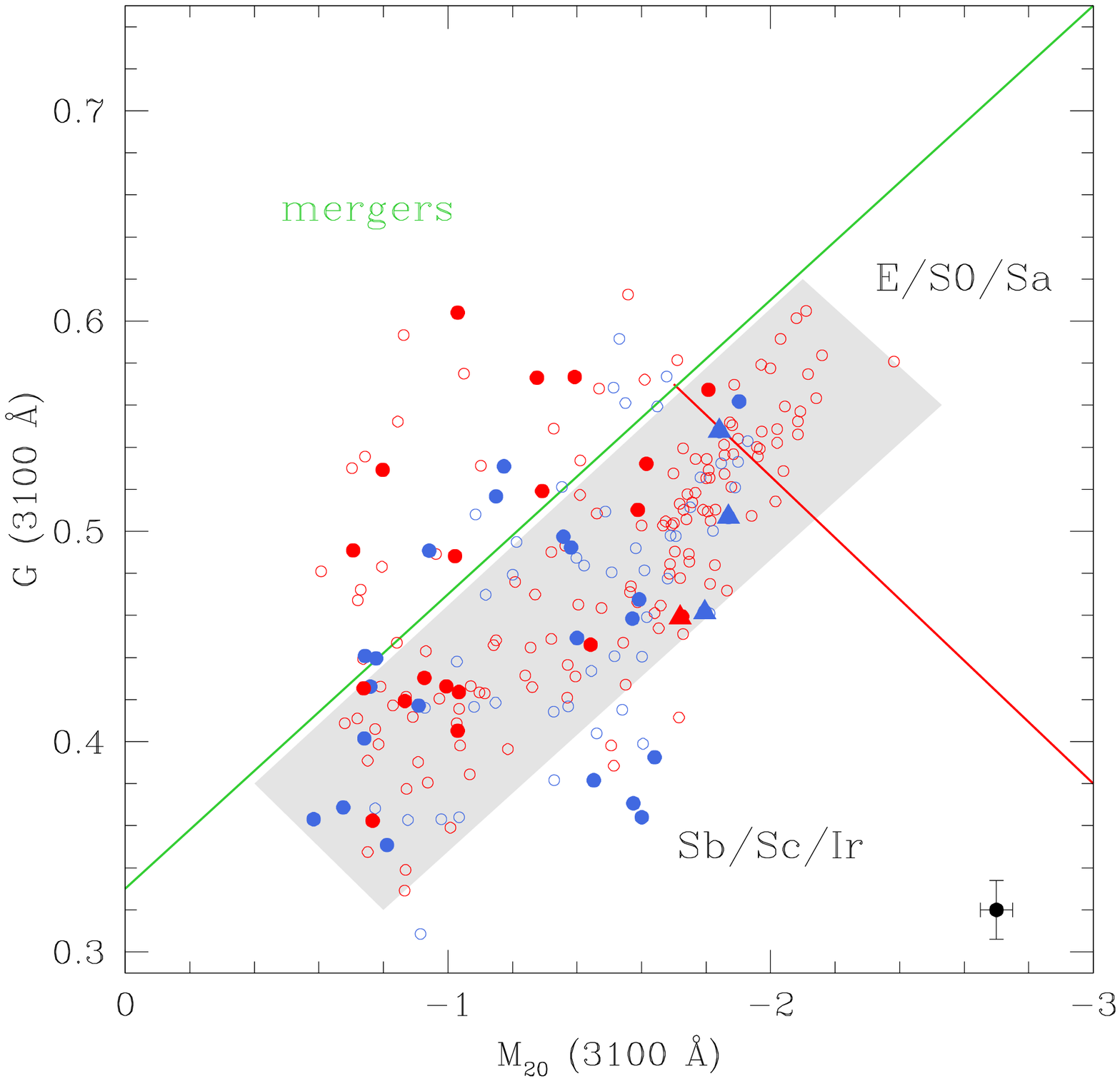}}
\caption{ \footnotesize Gini coefficient and M$_{20}$ statistic for
  the sample of LIRGs (red filled dots) at z$\sim$1, ULIRGs (blue
  filled dots) at z$\sim$2 and faint infrared sources in the redshift
  range z$\sim$1 and z$\sim$2 (red and blue empty dots,
  respectively). The triangles correspond to the sources in our sample
  whose mid-IR emission is AGN-dominated. The grey shaded region
  highlights the Gini-M$_{20}$ sequence which, locally, is populated
  by ``normal'' galaxies while local ULIRGs lie above the
  sequence. The three regions defined in \citet{lotz08} are marked
  with solid lines. The statistics have been computed from ACS images
  in the bands F606W and F850LP for galaxies at z$\sim$1 and z$\sim$2,
  respectively, to be approximately at the rest-frame wavelength of
  3100{\rm \,\AA}.
\label{fig:gini}}
\end{figure*}
\begin{figure*}[!t]
\centering{
\includegraphics[width=2.1\columnwidth]{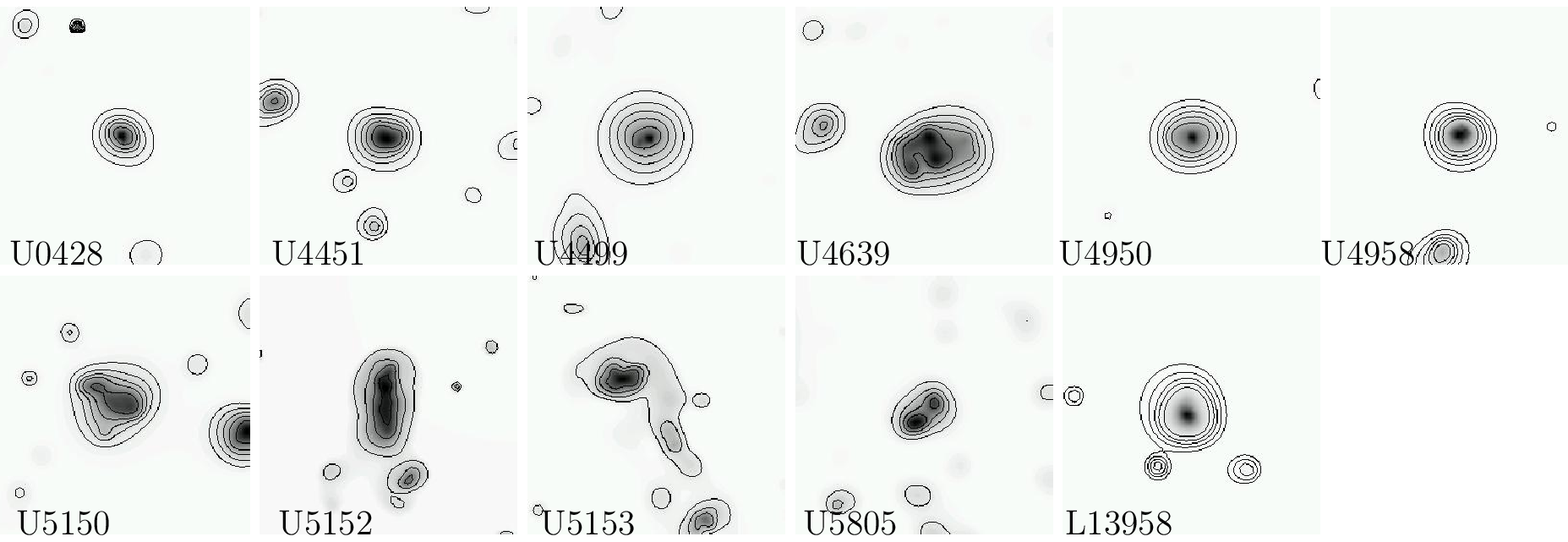}}
\caption{
\footnotesize
The plot shows the HST/NICMOS H-band 3''$\times$3'' images 
centered on 10 ULIRGs and one LIRG of our sample. Labels correspond
to the object IDs.
\label{morphnicmos}}
\end{figure*}
In the case of LIRGs, the straight and robust average are
(0.2$\pm$0.1)$\times$\,$10^{43}$\,erg/s and
(0.0$\pm$0.1)$\times$\,$10^{43}$\,erg/s, respectively.
We  conclude that for the $z$\,$\sim$\,1
sample, the majority of the galaxies do not harbor hidden AGN
and their emission  is dominated by star formation.

\subsection{Morphologies \label{morph}}

LIRGs and ULIRGs shine at infrared luminosities of $10^{44} -
10^{46}$\,ergs/s. It is not yet clear what physical processes light up
a galaxy in the infrared and if major mergers are the dominant
mechanism at $z\sim1-2$, as it is the case for local ULIRGs
\citep{sanders96,veilleux02}.  The simplest way to address this
question is to examine the morphologies of our sources using the high
spatial resolution images from {\it HST}.

Figures~\ref{morphu} and~\ref{morphl} show image cutouts for ULIRGs
and LIRGs, respectively, in the {\it HST} F435W ($B$), F606W ($V$) ,
F775W ($i$), F850LP ($z$) bands and the ground based $J$ and $K$
images.  Each cutout has a size of $10^{''}$\,$\times$\,$10^{''}$.
These two figures illustrate the steep changes of the spectral energy
distribution from $B$-$K$, and clearly indicate that many of our
sources have extremely red optical-to-NIR colors.  To better show
colors and morphologies, we present three-color images of the galaxies
which have been observed with ACS in the F435W, F606W and F850LP
bands.  The angular size of the images is 3$^{''}$ and 4$^{''}$ for
ULIRGs and LIRGs, respectively, which corresponds to 25.2\,kpc and
31.2\,kpc at the median redshift of 1.9 and 0.9, respectively.  To
obtain the three-color images, we filtered each ACS image with a
wavelet transform to get rid of the background noise and to enhance
the faint and extended morphological features. The technique used to
filter the signal is a 2-dimension generalization of the method
described in detail in \citet{fadda98}. To each band we assigned 
the same weight.

Many extended sources, such as ULIRG4367, 4451, 5652 and 5829, are
also very red, suggesting that dust obscuration could be in play.
Only about 4 sources (428,4950,4958,5877) have very compact or nearly
unresolved morphologies.  It is interesting to note that two of these
compact morphology sources (ULIRG4958 and 5877) show very blue compact
cores in Figure~\ref{colormorph}, and are AGN dominated systems with
weak or no PAH emission shown in Figure~\ref{spec}.
This correspondence between the rest-frame optical compact
morphologies and the mid-IR AGN dominant spectra has also been found
among a large sample of 24\um\ selected ULIRGs in the \spitz First
Look Survey \citep{michel09,dasyra08}. 

To better quantify morphological differences between our sample of
galaxies and other galaxies in the field which are faint infrared
emitters, we computed the Gini coefficient and M$_{20}$ for the galaxies in
our sample and control samples of faint infrared galaxies in the same
redshift ranges. The Gini coefficient, G, is a statistic to measure
the distribution of flux within the galaxy and M$_{20}$ is the second-order
moment of the brightest 20\% of the galaxy's flux. These two
nonparametric statistics have been introduced by \citet{lotz04} to
quantify galaxy morphology. Normal galaxies form a sequence in the
G$-$M$_{20}$ plot and, in the local Universe, ULIRGs lie above this
sequence.  To compute these statistics, we followed the method
described in \citet{lotz04} using our wavelet-filtered images to
reduce the role of noise in the computation.  We computed G and M$_{20}$
using the F850LP and F606W images for ULIRGs at $z\sim 2$ and LIRGs at
$z\sim 1$, respectively, to be roughly at the same 3100{\rm \,\AA}
rest-frame wavelength. We note that also the pixel size of the
galaxies at $z\sim 1$ and $z\sim 2$ correspond to a similar physical
scale (0.24 and 0.25~kpc, respectively, with our adopted cosmology),
so that we can compare the values computed for the two redshift
ranges. To compute the errors on the coefficients, we proceeded using
Monte Carlo simulations. It is, in fact, impossible to have a direct
error propagation since the computation of the coefficients involves
filtering of the image, selection of regions, sorting of values and
search of source center. Therefore, for each source, we created 50
synthetic images by adding a random realization of a noise image
(assuming a Gaussian distribution with dispersion equal to the rms of
the image) to the wavelet-filtered image. After adding the original 
zero, we rerun the code to measure the two coefficients on the 
synthetic images and estimated the errors assuming that they are 
equal to the dispersion of the values computed from the synthetic images.
The measured errors are $0.014\pm0.011$ and $0.04\pm0.04$ for the G and M$_{20}$
coefficients, respectively.

In Figure~\ref{fig:gini}, we report the G and M$_{20}$ values for the
galaxies of our sample for which the signal-to-noise was sufficient to
compute them and values for a control sample of galaxies with faint
infrared emission (less than 20$\mu$Jy at 24$\mu$m) in the redshift
ranges 0.8--1 and 1.8--2 where we find most of the galaxies of our
sample. The typical 1-$\sigma$ error is reported in the lower right
corner. A grey shaded region highlights the G-M$_{20}$ sequence which,
locally, is typically populated by ``normal'' galaxies while ULIRGs
are found above that sequence. The three regions of the plot defined
by \citet{lotz08} populated by merger, early-type and late-type
galaxies are marked with solid lines. The G-M$_{20}$ sequence found by
\citet{lotz08} in the Groth strip at $0.2<z<1.2$ in the rest-frame
blue band is well reproduced by our study, although at a bluer
rest-frame wavelength. We immediately notice that the separation
between infrared-active and infrared-quiet galaxies is not so clear as
for normal and IR-luminous local galaxies. Most of the galaxies of our
sample follow the same G-M$_{20}$ sequence as the majority of the
infrared-quiet galaxies.  \citet{lotz08} found that the percentage of
24$\mu$m sources at $0.4<z<1.0$ with $L_{IR}>10^{11}L_{\sun}$ are disk
galaxies and only $\sim$15\% are classified as major merger
candidates. At the 3100{\rm \,\AA} rest-frame, our study shows that
$z\sim 1$ LIRGs have a higher merger rate (37$^{+22}_{-16}$\%) and
that only 21$^{+15}_{-10}$\% of the ULIRGs at $z\sim 2$ can be
classified as mergers.  We can therefore conclude that the
morphologies of most of the IR-luminous galaxies at $z\sim 1$ and
$z\sim 2$ do not differ in most of the cases from those of ``normal''
infrared-quiet galaxies.  Repeating the analysis for the LIRGs of our
sample in the blue rest-frame, i.e. using the F850LP band ACS images,
only two galaxies populate the ``merger'' region for a percentage of
10$^{+14}_{-7}$\%, which is perfectly compatible with the results from
\citet{lotz08}.  This result is not surprising since the rest-frame
3100{\rm \,\AA} light is more sensitive to massive, young stars so
that late-type galaxies appear more patchy and have no prominent
optical bulge with respect to optical images (see,
e.g. \citet{kuchinski99}). In the case of ULIRGs, we have {\it HST}
NICMOS images in the H band, approximately corresponding to the blue
rest-frame at $z\sim 1.9$, for a subset of 10 ULIRGs (see
Figure~\ref{morphnicmos}). The Gini coefficients have a typical
difference of 0.02 with respect those computed in the ACS images. All
these sources are classified as ``normal'' galaxies in the G-M$_{20}$
plot. Unfortunately, we do not have NICMOS coverage for the three
ULIRGs which fall in the merger region of the G-M$_{20}$ plot.

\section{Summary}

The study presented in this paper aims to characterize the mid-IR
spectral properties of very faint 24\um\ sources with $S_{24\mu
  m}$\,$\sim$\, 0.15\,--\,0.45\,mJy, specifically targeting LIRGs at
$z$\,$\sim$\,1 and ULIRGs at $z$\,$\sim$\,2. This sample contains the
faintest sources observed with IRS, reaching the limit of the
instrument capability.  These 24\um\ faint LIRGs and ULIRGs are
infrared luminous, sampling the typical luminosities of infrared
luminosity functions at $z$\,$\sim$\,1 \&\ 2. Furthermore, they are
similar to normal galaxies selected by optical/near-IR surveys in the
sense that their stellar masses are $\simlt$\,$M^*$, the turn-over of
the mass functions ($\sim$\,2$\times$\,$10^{11}M_\odot$).  The mid-IR
spectra, in combination with X-ray, IR and optical observations, have
allowed us to determine physical properties of high-$z$ LIRGs and
ULIRGs, putting them in the context of larger galaxy populations
selected at optical/near-IR wavelength.  The major results of this
work can be summarized as follows:

$\bullet\ $ The vast majority of our sample (45 out of 48) has IRS
spectra with clear PAH emission and/or silicate absorption, allowing
secure redshift measurements. In particular, a third of the
$z$\,$\sim$\,2 ULIRGs are DRGs (galaxies with $(J-K)$\,$\simgt$\,2.3)
for which redshifts are very difficult to determine from optical or
near-IR spectroscopy.  Our finding of DRGs being dusty star forming
galaxies, rather than passively evolving old stellar population is
consistent with previous studies by \citep{labbe05,wuyts07,
  papovich06}.

$\bullet\ $ The fraction of AGN harbored by the sources in our sample
is much smaller than that of other samples of more luminous
distant ULIRGs in literature.  From the analysis of our mid-IR
sources, we estimate that their total IR emission is mainly powered by
star formation. AGN emission dominates the energy output of LIRGs and
ULIRGs only for 5\% and 12\% of them, respectively.

$\bullet\ $ A quantitative analysis of the morphology based on ACS
images reveals that most of the IR-luminous galaxies of our sample do
not differ significantly from the majority of the IR-quiet galaxies at
the same redshift range. At the 3100{\rm \,\AA} rest-frame,
37$^{+22}_{-16}$\% of the $z\sim1$ LIRGs and 21$^{+15}_{-10}$\% of the
$z\sim2$ ULIRGs have Gini coefficients higher than those of normal
IR-quiet galaxies. In the blue rest-frame, 10$^{+14}_{-7}$\% of the
$z\sim1$ LIRGs fall in the merger region of the Gini-M$_{20}$ plot
which is compatible with the 15\% found by
\citet{lotz08} among infrared galaxies with $L_{IR}>10^{11}L_{\sun}$
at $0.4<z<1.0$.

$\bullet\ $ We found a maximum value of $\sim$\,1\um\ for $EW_{6.2\mu
  m}$ for all infrared galaxies, independent of redshift and
luminosity.  This ceiling in 6.2\um\ PAH strength is due to the fact
that, for pure starbursts, the strength of the PAH line and of the
underlying continuum are both proportional to the luminosity due to
star-forming activity.  At $L_{24\mu m}$\,$\simgt$\,$10^{11}L_{\sun}$,
there is a clear anti-correlation between the rest-frame
24\um\ luminosity and 6.2\um\ PAH equivalent width. The 6.2\um\ PAH
equivalent width of the $z$\,$\simgt$1 sources is systematically
greater than that of the local luminous galaxies in the same
luminosity range. This is likely due to an AGN contribution in
high-redshift galaxies lower than that of local IR luminous galaxies.

$\bullet\ $ All our $z$\,$\sim$\,2 ULIRGs are BzK galaxies with
infrared excess, according to \citet{daddi04} definitions. Stacking
all the sources whose infrared flux is not dominated by AGN emission,
we provide quantitative constraints on the dust obscured AGN emission
at the rest-frame 6\um .  By subtracting from the spectra around the
6.2\um\ PAH feature the scaled averaged starburst spectrum, we
measured the average AGN powered continuum at 6\um, estimating an
average intrinsic X-ray AGN luminosity of $L_{2-10\,keV} = (0.1\pm
0.6)\times 10^{43}$\,erg/s, a value substantially lower than that
inferred from the X-ray stacking analysis in \citet{daddi07}. A
similar analysis on the LIRG sample suggests that their emission is
mostly powered by star formation.

In conclusion, our mid-IR spectra of LIRGs and ULIRGs at
$z$\,$\sim$\,1 and 2 reveal that the majority of these galaxies have
strong PAH emission and are starburst dominated.  The Compton thick
AGN contribution to bolometric luminosity for this sample selected
with $S_{24\mu m}$\,$\sim$\,0.15\,--\,0.5\,mJy is small.  The
so-called ``IR excess'' for these faint 24\um\ selected ULIRGs at
$z$\,$\sim$\,2 is mainly due to the strong PAH contribution to the
rest-frame mid-IR luminosity which leads to over-estimations of
$L_{IR}$ and to under-estimation of the UV dust extinction.

\acknowledgements This research made use of Tiny Tim/Spitzer,
developed by John Krist for the Spitzer Science Center. The Center is
managed by the California Institute of Technology under a contract
with NASA.  We are grateful to Haojing Yan for providing us the
reduced NICMOS H-band stamp images for some of our sources.  We thank
P. Capak for helpful discussions. We are grateful to the anonymous
referee for the detailed reading of the manuscript and for the many
ideas and suggestions to improve the quality of the paper. Support for
this work was provided by NASA through an award issued by
JPL/Caltech. S.Wuyts acknowledges support from the W. M. Keck Foundation.

\appendix

\begin{figure}[!t]
\includegraphics[width=\columnwidth]{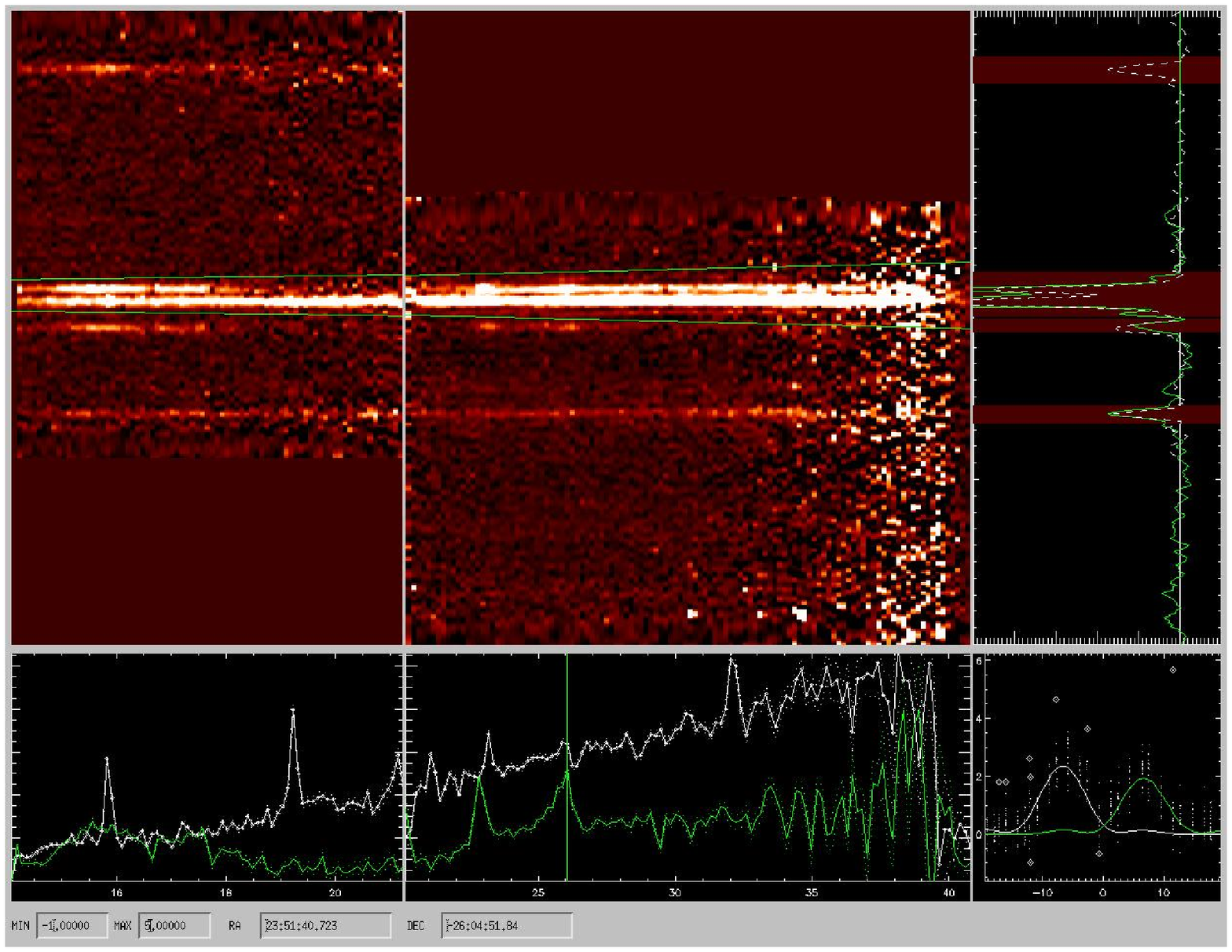}
\caption{\footnotesize
Analysis of an IRS observation through the GUI of the IrsLow code.
In this case, the target is a double source. At least other three
serendipitous sources are visible. The two images show the 2nd and 1st
orders after coadding all the frames (including the parallel observations).
In the right upper panel, the profiles of the averaged rows of the
two orders are displayed. The shaded regions have been masked to 
compute better background and rms images. Spectra of the two sources
between the green lines on the images have been extracted and are shown
in the lower left panels. Finally, the PSF spectral fit at the wavelength
corresponding to the vertical line is shown in the right-bottom window.
Each point corresponds to a pixel in the original frames. Circled points
are masked for the fit. The lower boxes contains the limits of the image
and coordinates of the extracted spectrum.
\label{fig:irslow}}
\end{figure}

\section{IrsLow, a software package for low-resolution IRS spectra}

For the analysis of the spectra presented in this paper, we developed
a new software package called ``IrsLow''. This software allows one to
exploit in an optimal way the redundancy of the data taken with the 
IRS spectrograph and to extract spectra in a very flexible way.
The software and a tutorial are available
online\footnote{web.ipac.caltech.edu/staff/fadda/IrsLow}. 
The code requires as input the basic calibrated data (BCD) produced by the
\spitz pipeline and uses the full redundancy of the data to reject bad
pixels and compute the residual background of the observations.  Once
the bad pixels are masked and the background subtracted from each
individual image, it coadds all the images of each order at different
positions.  The parallel data of non-primary orders are coadded to
identify and extract spectra of serendipitous sources.  The spectra
are extracted using PSF fitting and the code allows the simultaneous
extraction of two spatially close spectra. An example output from this
code is shown in Figure~\ref{fig:irslow}.
\begin{figure}[!t]
\includegraphics[width=\columnwidth]{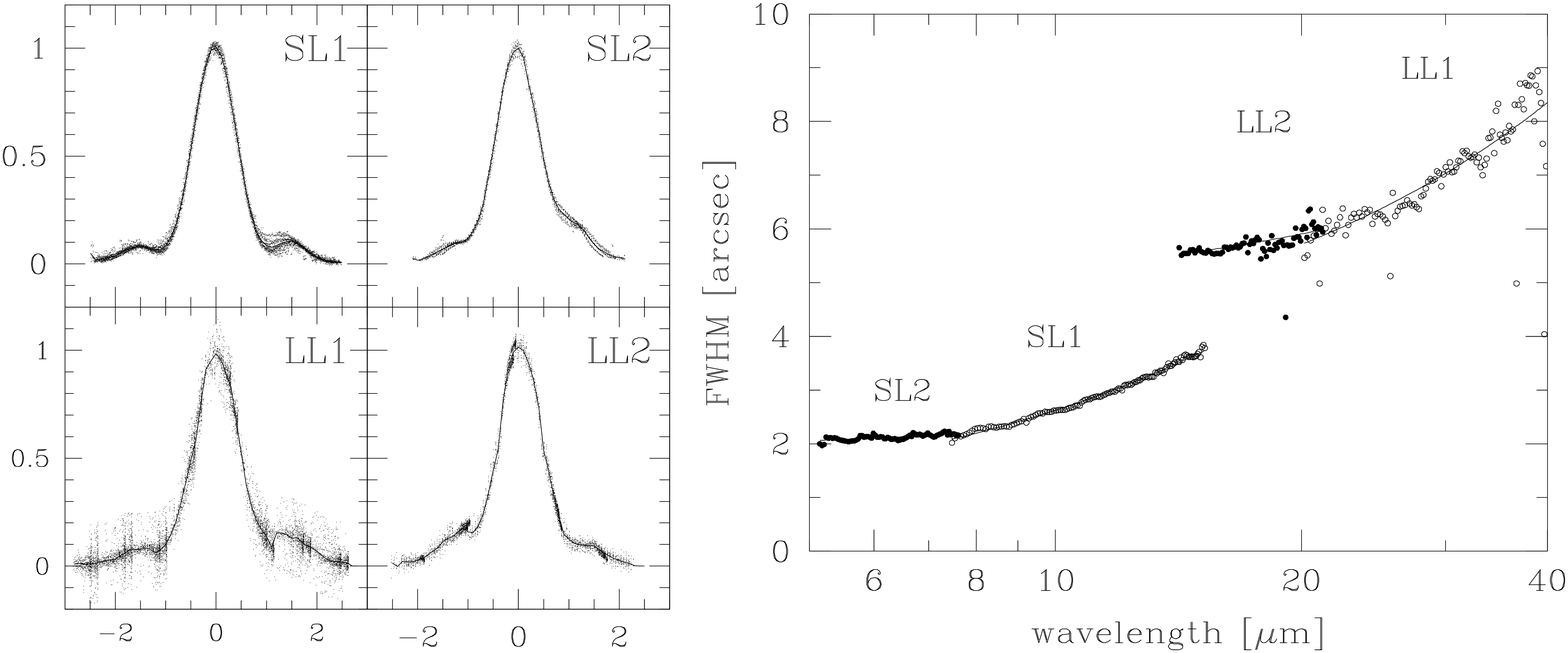}
\caption{ \footnotesize Spatial PSF profiles (left) and FWHM as a
function of the wavelength (right) in the different orders computed
with calibration stars. Amplitude and FWHM of the PSF profiles are
normalized to one. These PSFs are used for the optimal extraction
with IrsLow.
\label{fig:psf}}
\end{figure}

The optimal background subtraction is done iteratively with manually
masking off all spectra.  Uncertainty and background images are
obtained from the stacked BCD after the rejection of pixels affected
by cosmic rays or sources. A biweight estimator \citep{tukey58} is
used to compute robust estimates of average and dispersion also with a
low number ($\sim$\,10) of frames.  After extracting the spectrum,
each PSF fit for a single wavelength position can be visualized to
delete other possible outliers and therefore improve the fit. Pixels
with highly varying responsivity ({\em i.e.rogue} pixels in the IRS
jargon) are identified by comparing their RMS values to those of the
neighboring pixels. The final image is obtained by fitting a B-spline
at each wavelength to all the flux values from different pixels at
their relative spatial positions. The fit is repeated three times,
discarding 7-$\sigma$ and 4-$\sigma$ outliers in the second and third
iteration, respectively.  We derive PSF profiles using the IRS
calibration stars. The average profiles are obtained by normalizing
the flux of the spectra and scaling the FWHM as a function of the
wavelength.  The scaling of the FWHM is shown in the same
Figure~\ref{fig:psf}. Finally, the extracted spectra are saved in
ASCII files with names equal to their spatial coordinates. The flux
and wavelength calibration files are the same as what is included in
the IRS software SPICE. The calibration version is automatically
selected by the software according to the information stored in the
BCD headers.  In fact, the calibration version is linked to the
version of the pipeline used to obtain the BCD files.

One unique feature of our IRS reduction software is how it handles
confused spectra from two close-by sources.  As shown in
Figure~\ref{fig:irslow}, the software allows the optimal extraction of
two close sources at the same time by using our derived PSF profiles,
shown in Figure~\ref{fig:psf}.  
\begin{figure}[!t]
\includegraphics[width=\columnwidth]{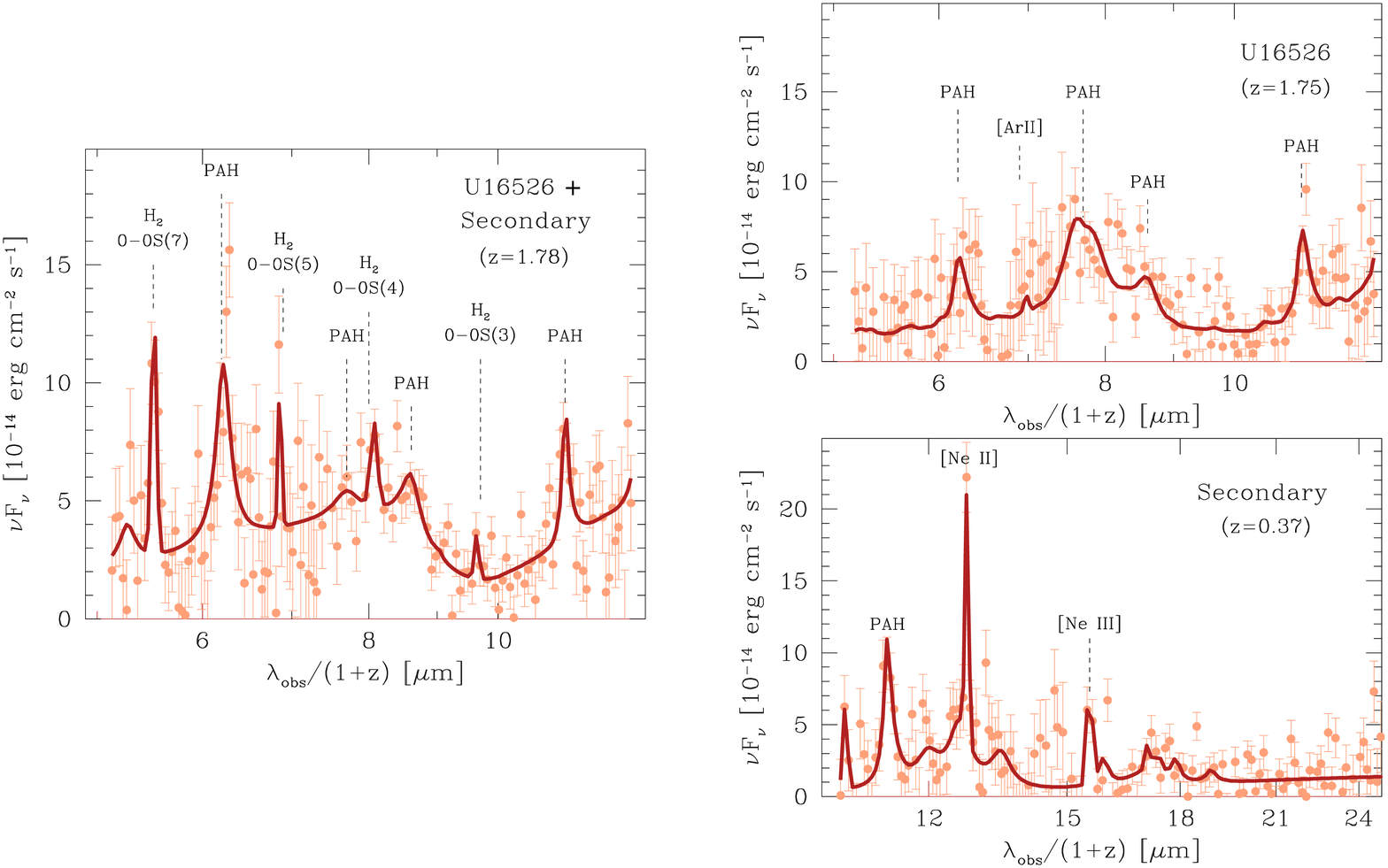}
\caption{ \footnotesize Spectral extraction of ULIRG 16526 considering
  one object (left) and fitting two objects (right). If only a single
  spectrum is extracted, spectral features from a nearby low-redshift
  object (at $z=0.37$) could be wrongly interpreted as strong H$_2$
  and 6.2~$\mu$m PAH features.
\label{fig:u16526}}
\end{figure}
This feature has been particularly useful in the case of the source
U16526. The observed IRS spectrum for U16526 is, in fact, a blend of
two spectra at $z$\,=\,0.37 and $z$\,=\,1.77 (photometric redshifts
from the FIREWORKS catalog), spatially separated by only $6^{''}$.  This
corresponds to the spatial resolution of the IRS spectrograph in the
wavelength range used in the observation. When displaying this spectrum
with our software, one can see only that the major features of the
two sources are slighlty spatially displaced although the continuum
of the two sources is blended. In our first extraction, we erroneously
considered one single source and interpreted some strong emission
features as H$_2$ lines. Then, after examining the 24$\mu$m image
of the source, we realized that the spectrum was a blend of spectra
from two different sources at different redshifts. The unique deblending
capability of our software allowed us to separate the spectra of these
two confused sources.  In Figure~\ref{fig:u16526}, the left panel
shows the total spectrum, which is the sum of two spectra, one at
$z$\,=\,0.37 and one at $z$\,=1.75, illustrated on the top right
and bottom right, respectively. At this point it was clear that 
the H$_2$ lines in the blended spectrum were in fact only strong
emission features of the $z$\,=\,0.37 source.

\clearpage

\begin{deluxetable}{lcccccc}
\singlespace
\tablecolumns{7} 
\tablewidth{0in}
\tablecaption{The Sample Targets and IRS Observation Log \label{obslog}}
\tabletypesize{\footnotesize}
\tablehead{
\colhead{ID} &
\colhead{IAU ID} &
\colhead{RA} & 
\colhead{DEC} & 
\colhead{SL\tablenotemark{b} 1st} &
\colhead{LL\tablenotemark{b} 2nd} &
\colhead{LL\tablenotemark{b} 1st}  \\
  &  &\multicolumn{2}{c}{(J2000)} & \multicolumn{3}{c}{(seconds $\times$ repeats)}
}
\startdata
U428   &   SST24 J03:32:43.45-27:49:01.8 &  3:32:43.450 & -27:49:01.82&     0 &  120x6  &   120x9  \\
U4367  &   SST24 J03:32:43.78-27:52:31.0 &  3:32:43.781 & -27:52:31.07&     0 &  120x15 &   120x22 \\
U4451  &   SST24 J03:32:14.46-27:52:33.7 &  3:32:14.465 & -27:52:33.77&     0 &  120x15 &   120x22 \\
U4499  &   SST24 J03:32:17.19-27:50:37.1 &  3:32:17.190 & -27:50:37.10&     0 &  120x15 &   120x22 \\
U4631  &   SST24 J03:32:40.24-27:49:49.6 &  3:32:40.243 & -27:49:49.64&     0 &  120x9  &   120x13 \\
U4639  &   SST24 J03:32:40.75-27:49:25.8 &  3:32:40.756 & -27:49:25.87&     0 &  120x15 &   120x22 \\
U4642  &   SST24 J03:32:31.52-27:48:53.7 &  3:32:31.529 & -27:48:53.72&     0 &  120x9  &   120x13 \\
U4812  &   SST24 J03:32:47.58-27:44:52.3 &  3:32:47.585 & -27:44:52.31&     0 &  120x6  &   120x9  \\
U4950  &   SST24 J03:32:25.68-27:43:05.6 &  3:32:25.683 & -27:43:05.65&     0 &  120x3  &   120x5  \\
U4958  &   SST24 J03:32:23.43-27:42:55.0 &  3:32:23.436 & -27:42:55.03&     0 &  120x9  &   120x13 \\
U5050  &   SST24 J03:32:39.03-27:44:20.4 &  3:32:39.034 & -27:44:20.46&     0 &  120x15 &   120x22 \\
U5059  &   SST24 J03:32:23.71-27:44:11.6 &  3:32:23.714 & -27:44:11.69&     0 &  120x9  &   120x13 \\
U5150  &   SST24 J03:32:13.87-27:43:12.3 &  3:32:13.877 & -27:43:12.35&     0 &  120x6  &   120x9  \\
U5152  &   SST24 J03:32:12.54-27:43:05.9 &  3:32:12.543 & -27:43:05.99&     0 &  120x6  &   120x9  \\
U5153  &   SST24 J03:32:12.15-27:42:49.9 &  3:32:12.150 & -27:42:49.94&     0 &  120x15 &   120x22 \\
U5632  &   SST24 J03:32:40.05-27:47:55.1 &  3:32:40.051 & -27:47:55.11&     0 &  120x6  &   60x16  \\
U5652  &   SST24 J03:32:17.44-27:50:03.1 &  3:32:17.444 & -27:50:03.12&     0 &  120x6  &   60x16  \\
U5775  &   SST24 J03:32:26.01-27:47:51.5 &  3:32:26.011 & -27:47:51.56&     0 &  120x15 &   120x22 \\
U5795  &   SST24 J03:32:18.75-27:46:26.9 &  3:32:18.750 & -27:46:26.90&     0 &  120x6  &   120x9  \\
U5801  &   SST24 J03:32:17.58-27:45:51.7 &  3:32:17.587 & -27:45:51.78&     0 &  120x15 &   120x22 \\
U5805  &   SST24 J03:32:22.56-27:45:39.0 &  3:32:22.561 & -27:45:39.00&     0 &  120x15 &   120x22 \\
U5829  &   SST24 J03:32:13.62-27:47:53.9 &  3:32:13.628 & -27:47:53.97&     0 &  120x15 &   120x22 \\
U5877  &   SST24 J03:32:20.04-27:44:47.2 &  3:32:20.049 & -27:44:47.20&     0 &  120x6  &   120x14 \\
U16526 &   SST24 J03:32:41.87-27:52:45.0 &  3:32:41.870 & -27:52:45.06&     0 &  120x6  &   120x9  \\
L3832   &   SST24 J03:32:30.18-27:57:00.7 &  3:32:30.189 & -27:57:00.75& 60x32 &  120x14 &        0 \\
L3945   &   SST24 J03:33:04.38-27:48:55.5 &  3:33:04.381 & -27:48:55.50& 60x32 &  120x14 &        0 \\
L4020   &   SST24 J03:33:15.65-27:46:07.5 &  3:33:15.655 & -27:46:07.57& 60x32 &  120x14 &        0 \\
L4177   &   SST24 J03:32:48.48-27:54:16.0 &  3:32:48.489 & -27:54:16.04& 60x32 &  120x14 &        0 \\
L4419   &   SST24 J03:32:16.87-27:54:18.4 &  3:32:16.871 & -27:54:18.43& 60x32 &  120x14 &        0 \\
L4900   &   SST24 J03:32:32.88-27:41:24.1 &  3:32:32.886 & -27:41:24.12& 60x32 &  120x14 &        0 \\
L5134   &   SST24 J03:32:42.96-27:46:50.0 &  3:32:42.962 & -27:46:50.05& 60x32 &  120x14 &        0 \\
L5367   &   SST24 J03:31:35.23-27:49:58.4 &  3:31:35.237 & -27:49:58.41& 60x32 &  120x14 &        0 \\
L5420   &   SST24 J03:32:05.99-27:45:07.3 &  3:32:05.990 & -27:45:07.36& 60x16 &   60x16 &        0 \\
L5511   &   SST24 J03:31:48.19-27:45:34.9 &  3:31:48.191 & -27:45:34.91& 60x16 &   60x16 &        0 \\
L5630   &   SST24 J03:32:42.28-27:47:45.9 &  3:32:42.287 & -27:47:45.99& 60x32 &  120x14 &        0 \\
L5659   &   SST24 J03:32:18.70-27:49:19.5 &  3:32:18.701 & -27:49:19.55& 60x32 &  120x14 &        0 \\
L5876   &   SST24 J03:32:20.71-27:44:53.5 &  3:32:20.714 & -27:44:53.53& 60x32 &  120x14 &        0 \\
L6211   &   SST24 J03:32:42.70-27:39:27.1 &  3:32:42.708 & -27:39:27.10& 120x6 &   120x5 &        0 \\
L6221   &   SST24 J03:31:17.46-27:47:55.0 &  3:31:17.463 & -27:47:55.09& 60x32 &  120x14 &        0 \\
L6473   &   SST24 J03:31:49.02-27:39:45.1 &  3:31:49.026 & -27:39:45.19& 120x6 &   120x5 &        0 \\
L6658   &   SST24 J03:32:05.48-27:36:43.9 &  3:32:05.486 & -27:36:43.96& 120x6 &   120x5 &        0 \\
L7079   &   SST24 J03:31:21.47-27:41:47.0 &  3:31:21.474 & -27:41:47.05& 60x32 &  120x14 &        0 \\
L13667  &   SST24 J03:33:27.79-28:02:50.9 &  3:33:27.794 & -28:02:50.90& 120x6 &   120x5 &        0 \\
L13958  &   SST24 J03:32:25.41-27:46:16.8 &  3:32:25.412 & -27:46:16.81& 60x16 &   60x16 &        0 \\
L14143  &   SST24 J03:31:16.36-27:40:33.5 &  3:31:16.365 & -27:40:33.57& 60x16 &   60x16 &        0 \\
L14228  &   SST24 J03:31:29.99-27:35:22.0 &  3:31:29.999 & -27:35:22.04& 60x32 &  120x14 &        0 \\
L15906  &   SST24 J03:32:45.91-27:52:19.6 &  3:32:45.914 & -27:52:19.64& 60x32 &  120x14 &        0 \\
L16005  &   SST24 J03:31:51.50-27:41:57.1 &  3:31:51.505 & -27:41:57.16& 60x32 &  120x14 &        0 \\
\enddata
\tablenotetext{a}{SL and LL stand for Short-Low and Long-Low modules. SL 1st order, LL 2nd order, and LL 1st order cover the ranges $7-14\mu m$,  $14-21\mu m$, and $21-35\mu m$, respectively.}
\end{deluxetable}

\begin{deluxetable}{l rr rr rr rrc ccccc}
\singlespace
\tablecolumns{15} 
\tablewidth{0in}
\tablecaption{The Sample Target Fluxes and Redshifts \label{targets_data}}
\tabletypesize{\footnotesize}
\tablehead{
\colhead{ID} &
\multicolumn{2}{c}{$S_{8\mu m}$} & 
\multicolumn{2}{c}{$S_{16\mu m}$} &
\multicolumn{2}{c}{$S_{24\mu m}$} &
\multicolumn{2}{c}{$S_{70\mu m}$} &
\colhead{z$_{IRS}$} & 
\colhead{z$_{opt}$} &
\colhead{$^*$} &
\colhead{$\log (L_{24\mu m})$} & 
\colhead{$M_{*}$} & 
\colhead{$EW_{6.2\mu m}$}
 \\
&
\colhead{[$\mu$Jy]} &
\colhead{SN} &  
\colhead{[$\mu$Jy]}  &
\colhead{SN} &  
\colhead{[$\mu$Jy]} &
\colhead{SN} &  
\colhead{[$m$Jy]} &
\colhead{SN} &  
& 
& 
&  
\colhead{[L$_{\sun}$]} &  
\colhead{[10$^{10}$M$_{\sun}$]} &  
\colhead{[$\mu$m]} 
}
\startdata
 U428 &  15.9&  68&    65&  24& 287& 38&   2.43&  3& 1.783 &(1.664)&a&11.47&  4.09& 1.39$\pm$0.34\\
 U4367&  22.2&  79&   101&  12& 169& 20&$<$1.55&   & 1.624 &(1.762)&a&11.23& 13.10& 0.54$\pm$0.23\\
 U4451&  11.1&  38&      &    & 225& 28&$<$1.53&   & 1.875 &(1.684)&a&11.40&  3.47& 0.83$\pm$0.50\\ 
 U4499&  16.3&  77&$<$ 35&    & 182& 23&$<$1.45&   & 1.956 &(1.909)&a&11.58& 11.90& $>$0.84    \\
 U4631&  15.4&  60&$<$ 21&    & 269& 33&$<$1.40&   & 1.841 & 1.896 &1&11.39&  9.61& $>$0.92    \\
 U4639&  13.5&  57&$<$ 14&    & 216& 27&$<$1.45&   & 2.112 & 2.130 &5&11.58&  3.57& $>$2.61    \\
 U4642&  12.3&  60&    65&   7& 268& 32&   2.14&  3& 1.898 &(1.748)&a&11.82&  2.50& 1.33$\pm$0.41\\
 U4812&  15.9&  42&$<$ 53&    & 317& 41&   3.75& 13& 1.930 & 1.910 &1&12.00&  6.63& $>$1.04   \\
 U4950& 163.3& 598&   453&  54& 547& 67&$<$1.68&   & 2.312 & 2.291 &4&11.79& 10.41& 0.12$\pm$0.06\\
 U4958&  16.4&  62&   109&  11& 262& 33&$<$1.67&   & 2.118 & 2.145 &5&11.51&  8.50&      \\
 U5050&  24.6&  97&    51&   7& 194& 22&$<$1.50&   & 1.938 &(1.726)&a&11.40& 18.93& 0.60$\pm$0.34\\
 U5059&  18.1&  64&$<$ 37&    & 253& 31&$<$1.53&   & 1.769 &(1.543)&a&11.49&  6.59& $>$1.61    \\
 U5150&  13.1&  35&    65&   6& 286& 37&   2.98& 12& 1.898 &(1.738)&a&11.92&  2.17& 0.77$\pm$0.47\\
 U5152&  13.1&  43&    66&   7& 305& 38&$<$1.77&   & 1.794 &(1.888)&a&11.44&  2.75& 0.50$\pm$0.34\\
 U5153&  19.2&  68&$<$ 34&    & 183& 21&$<$1.71&   & 2.442 &(2.030)&b&11.73&      & $>$0.41    \\
 U5632&  19.2&  86&    83&  35& 421& 52&$<$1.38&   & 2.016 & 1.998 &1&11.75&  6.79& 0.99$\pm$0.22\\
 U5652&  30.3& 153&   154&  17& 316& 44&   3.24&  8& 1.618 & 1.616 &1&11.68& 15.73& 1.21$\pm$0.23\\
 U5775&   9.6&  51&$<$ 31&    & 141& 16&   2.64& 11& 1.897 &(1.779)&a&11.61&  5.30& $>$0.72    \\
 U5795&  18.7&  63&$<$ 31&    & 264& 33&$<$1.43&   & 1.703 &(1.524)&a&11.36&  5.63& $>$2.16    \\
 U5801&  16.2&  57&$<$ 35&    & 216& 24&$<$1.45&   & 1.841 &(1.642)&a&11.25& 13.68& $>$0.83    \\
 U5805&   8.0&  25&    68&  11& 170& 19&$<$1.45&   & 2.073 &(2.093)&a&11.74&  1.16& 0.41$\pm$0.18\\
 U5829&  15.4&  69&    57&  12& 208& 26&$<$1.40&   & 1.742 &(1.597)&a&11.32&  9.21& 0.98$\pm$0.31\\
 U5877&  31.7& 120&   179&  23& 360& 48&   3.88& 19& 1.886 &(1.708)&a&11.77&  3.54& 0.17$\pm$0.10\\
 U16526& 20.6&  66&    69&  18& 306& 45&   5.17& 13& 1.749 &(1.718)&a&12.00&  4.05& 0.53$\pm$0.35\\
 L3832&  23.5&  61&      &    & 207& 28&$<$1.90&   & 0.767 &(0.763)&b&10.53&  4.39& 0.81$\pm$0.46\\ 
 L3945&  19.0&  18&      &    & 367& 39&   3.81& 11&       &(0.950)&b&11.08&  0.37& 0.20$\pm$0.05\\
 L4020&  20.8&  30&      &    & 332& 38&   4.22& 21& 0.826 &(0.779)&b&10.90&  3.10& 1.13$\pm$0.13\\ 
 L4177&  20.5&  70&   204&  22& 208& 27&$<$1.78&   & 0.842 & 0.840 &3&10.62&  3.71& 0.91$\pm$0.26\\ 
 L4419&  46.8& 159&      &    & 259& 38&   4.06& 13& 0.974 &(0.995)&b&11.08& 11.23& 0.46$\pm$0.23\\ 
 L4900&  26.8&  95&   267&  32& 213& 31&$<$1.82&   & 1.047 & 1.045 &5&10.91&  6.75& 1.11$\pm$0.05\\ 
 L5134&  22.6& 106&   269& 110& 229& 28&$<$1.42&   & 1.039 & 1.036 &1&10.84&  5.02& 1.04$\pm$0.32\\  
 L5367&  24.3&  27&      &    & 252& 26&$<$1.92&   & 0.974 &(0.868)&b&10.86&  5.79& 1.06$\pm$0.07\\  
 L5420&  36.7& 135&   450&  48& 381& 46&   2.76&  5& 1.068 & 1.068 &3&11.15&  4.42& 0.91$\pm$0.20\\ 
 L5511&  38.6&  50&      &    & 401& 54&   2.22&  5&       &( 2.08)&c&11.87& 25.90& 0.48$\pm$0.23\\
 L5630&  22.8& 104&   277& 149& 286& 36&   2.06&  3& 0.997 & 0.996 &3&10.93&  2.27& 1.24$\pm$0.16\\
 L5659&  28.6& 129&   285&  25& 231& 30&   2.41&  9& 1.044 & 1.038 &1&11.00&  4.57& 1.13$\pm$0.53\\
 L5876&  29.2& 103&   310&  41& 330& 43&   2.54& 15& 0.971 & 0.969 &1&10.98&  7.16& 1.11$\pm$0.04\\
 L6211&  31.5&  84&      &    & 457& 62&$<$1.88&   & 1.843 &(1.840)&c&11.70&      & 1.56$\pm$0.22\\
 L6221&  19.1&  14&      &    & 239& 12&$<$2.87&   & 1.012 &(0.989)&b&11.02&      & 1.04$\pm$0.26\\ 
 L6473&  32.3&  44&      &    & 484& 63&   2.78&  8& 0.816 & 0.811 &1&10.83&      & 1.10$\pm$0.28\\
 L6658&  45.2&  60&      &    & 467& 59&$<$1.87&   & 0.969 &(0.963)&b&10.94& 12.54& 1.33$\pm$0.24\\
 L7079&  23.3&  20&      &    & 289& 27&   3.21&  8& 0.955 &(0.884)&b&11.00&      & 1.01$\pm$0.01\\
L13667&  33.8&  30&      &    & 496& 64&$<$2.65&   & 0.936 &(0.924)&b&11.00&      & 1.57$\pm$0.01\\
L13958&  21.8&  83&   292&  39& 376& 47&   3.31& 13& 0.891 & 0.896 &2&10.96&  2.10& 1.99$\pm$0.24\\
L14143&  34.8&  22&      &    & 427& 35&   4.71&  8& 1.043 &(0.965)&b&11.29&      & 0.89$\pm$0.08\\
L14228&  23.8&  23&      &    & 313& 35&$<$2.12&   & 0.953 &(0.961)&b&10.89&  3.02& 1.55$\pm$0.21\\
L15906&  18.4&  63&   193&  23& 262& 41&   3.23&  8& 0.976 &(1.045)&a&11.02&  0.44& 1.13$\pm$0.17\\
L16005&  24.2&  32&      &    & 253& 42&   3.29&  9& 0.865 &(0.984)&b&10.86&  1.59& 1.60$\pm$0.06\\
\enddata
\tablenotetext{*}{Spectroscopic redshifts from: 
(1) \citet{vanz08,vanz06}, 
(2) \citet{mig05},  
(3) \citet{lefev04}, 
(4) \citet{szokoly04}, 
(5) \citet{popesso08}, 
Photometric redshifts from: 
(a) \citet{wuyts08}, 
(b) \citet{karina04},
(c) this paper.}
\end{deluxetable}

\clearpage

\end{document}